\documentclass[prx,twocolumn,superscriptaddress,amsmath,amssymb]{revtex4-2}
\usepackage[colorlinks=true,citecolor=blue,linkcolor=blue,breaklinks=true]{hyperref}

\usepackage{color,graphicx,soul}
\usepackage{bm}
\usepackage{amsmath}
\usepackage{amssymb}
\usepackage{caption}
\usepackage[usenames,dvipsnames]{xcolor}
\usepackage{verbatim}
\usepackage{float}
\usepackage[justification=Justified]{caption}
\usepackage{booktabs}

\begin{document}
		
		\title{Measurement-induced enhancement of the $X$-parity lifetime in a Majorana tetron}
			
		\author{Katharina Laubscher}
		\affiliation{Condensed Matter Theory Center and Joint Quantum Institute, Department of Physics, University of Maryland, College Park, MD 20742, USA}
        
		\author{Jay D. Sau}
		\affiliation{Condensed Matter Theory Center and Joint Quantum Institute, Department of Physics, University of Maryland, College Park, MD 20742, USA}

        \author{Sankar Das Sarma}
		\affiliation{Condensed Matter Theory Center and Joint Quantum Institute, Department of Physics, University of Maryland, College Park, MD 20742, USA}
		
		\date{\today}
		\begin{abstract}
A recent experiment [arXiv:2507.08795] has reported measurements of the interwire $X$-parity lifetime in a Majorana tetron device via the quantum capacitance of an auxiliary readout dot. Motivated by these measurements, we theoretically study the interwire parity lifetime in a minimal model of the coupled dot-tetron system subject to electrostatic noise. We show that coupling to the measurement dot can itself strongly modify the parity dynamics by generating an energy splitting that shifts noise-induced transitions to finite frequency and thereby stabilizes the measured states against low-frequency fluctuations. For representative noise strengths motivated by disordered nanowire devices, we find $X$-parity switching lifetimes of order microseconds in the measurement configuration even when the corresponding bare-tetron coherence time is only tens of nanoseconds long. We analyze both the dispersive regime, where the dot can be integrated out and the dynamics reduces to an effective two-level system, and the near-resonant regime, where all four low-energy states participate. In the latter case, rapid transitions organize the spectrum into two long-lived manifolds that approximately correspond to sectors of the combined dot-tetron $X$-loop parity, yielding effective two-state switching despite the underlying multilevel structure. Our results show that measurement-induced stabilization can substantially enhance the observed $X$-parity switching lifetime, such that the lifetime measured under continuous readout can greatly exceed the intrinsic idle $T_2^*$ coherence time of the isolated Majorana tetron qubit.
\end{abstract}
		
		\maketitle

\section{Introduction}
\label{sec:intro}

Majorana zero modes (MZMs) are exotic zero-energy quasiparticle excitations predicted to occur at the boundaries of topological superconductors~\cite{kitaev2001unpaired}. Their non-Abelian exchange statistics make them promising building blocks for topological quantum computation, where quantum information can be encoded and processed in a way that is intrinsically protected against a broad class of local perturbations~\cite{nayak2008nonabelian,sarma2015majorana}. Considerable interest has focused on spin-orbit-coupled semiconductor nanowires proximity-coupled to conventional superconductors, which can enter a topological superconducting phase when a magnetic field is applied along the wire~\cite{lutchyn2010majorana,sau2010generic,sau2010nonabelian,oreg2010helical}. The relative simplicity and tunability of this platform have motivated extensive experimental efforts toward the detection and realization of MZMs in hybrid nanowires~\cite{dassarma2023search,kouwenhoven2025perspective}.

Building on this approach, Microsoft Quantum has developed a large industrial-scale program aimed at realizing measurement-based Majorana qubits in semiconductor--superconductor hybrid devices~\cite{Aasen2025}. Recent work on InAs-Al hybrid nanowires reported local and nonlocal transport signatures consistent with topological superconductivity~\cite{microsoftquantum2023inasal}, followed by reports of interferometric single-shot fermion-parity measurements in individual hybrid nanowires~\cite{aghaee2025interferometric}. Here, the quantity to be read out is the fermion parity of the nonlocal fermionic degree of freedom defined by a particular pair of MZMs. Since this information cannot be accessed by probing either Majorana individually, the two MZMs are coupled through an interferometric loop~\cite{akhmerov2009electrically,fu2010electron} containing an auxiliary quantum dot. The resulting dot-Majorana hybridization depends on both their joint fermion parity and an interferometric phase $\phi$ controlled by the magnetic flux through the loop, producing a parity- and phase-dependent energy spectrum and hence quantum capacitance. The latter can be monitored by radio-frequency reflectometry, allowing time-resolved detection of switching between the two readout configurations. The observed switching is interpreted as fermion-parity switching, with the corresponding characteristic switching time serving as a measure of parity lifetime. This approach has subsequently been incorporated into a tetron device~\cite{karzig2017scalable,plugge2017majorana} containing two parallel hybrid nanowires and distinct interferometers designed to implement Pauli-$X$ and $Z$ measurements~\cite{aghaee2025distinctlifetimesxz}. Repeated measurements in this device yielded markedly different switching times for the two loops, with an $X$-loop timescale of order $10~\mu{\rm s}$ and a much longer $Z$-loop timescale. More recently, related measurements in InAs-Pb tetron devices have reported $Z$-parity lifetimes reaching tens of seconds~\cite{aghaee202620secondparitylifetime}. Independently of the Microsoft efforts, related quantum-capacitance-based MZM parity measurements have also been reported in gate-defined minimal Kitaev chains, together with coherent manipulation of the associated low-energy parity states~\cite{zatelli2026majoranaparityqubitcoupled,vanLoo2026}.

In this work, we focus on the $X$-loop measurement in a Majorana nanowire tetron and study how the coupling to the measurement dot itself affects the observed parity lifetime. Using a minimal model consisting of four MZMs coupled to a readout dot with a single fermionic level, see Fig.~\ref{fig:tetron}, we calculate the $X$-parity lifetimes in the presence of fluctuations of both the readout-dot potential and the residual MZM splittings. Our main finding is that the coupling to the measurement dot can strongly modify the parity dynamics by inducing an energy splitting between the two parity states, which shifts noise-induced transitions to finite frequency and thereby suppresses transitions driven by low-frequency noise. For representative noise strengths motivated by disordered nanowire devices, we find that the $X$-parity lifetimes in the measurement configuration can reach the microsecond scale even when the corresponding isolated-tetron coherence time—which characterizes the free-induction decay of the same $X$-parity degree of freedom in the bare tetron—is more than an order of magnitude shorter. A long $X$-parity switching time observed during continuous readout can therefore result from stabilization by the measurement coupling itself and should not be taken as a direct indication that the idle coherence time of the isolated MZM tetron qubit is long.

An important ingredient of our analysis is the inclusion of fluctuations of the residual MZM splittings. Such fluctuations, which may be screened by the superconductor, are strongly suppressed in the ideal limit of well-separated, exponentially protected MZMs~\cite{Knapp2018,mishmash2020dephasing} and are therefore neglected in the theoretical modeling of Ref.~\cite{aghaee2025distinctlifetimesxz}. However, whether present devices have reached this regime remains an open question. Indeed, microscopic studies suggest that disorder may remain nonnegligible even in relatively recent devices and, together with the necessarily finite wire length, can substantially restrict the regime of exponentially protected MZMs~\cite{dassarma2023density,dassarma2023spectral,Pan2026,Taylor2026}. Furthermore, capacitance and $Z$-parity signatures with qualitative similarities to those reported experimentally have been predicted for overlapping MZMs in disordered wires~\cite{Sau2025,stanescu2025fermionparityquantumcapacitance}. Once the residual MZM splitting is sensitive to the local electrostatic environment, time-dependent charge fluctuations can in turn generate appreciable MZM-splitting noise~\cite{Sau2025}. Importantly, even a small static splitting at a selected operating point does not by itself imply weak sensitivity of this splitting to electrostatic fluctuations. We therefore include the MZM-splitting noise channel alongside fluctuations of the readout-dot potential. This allows for a regime where a long $X$-parity switching time under continuous measurement can coexist with a much shorter idle $T_2^*$ coherence time of the isolated tetron.

We analyze the effect of measurement-induced stabilization both in the dispersive regime, where the measurement dot can be integrated out and the low-energy dynamics reduces to an effective two-level system, and closer to charge resonance, where all four low-energy states of the dot-tetron system can participate. In the latter regime, rapid transitions organize the spectrum into two long-lived manifolds that approximately correspond to sectors of the combined dot-tetron $X$-loop parity, yielding effective two-state switching despite the underlying multilevel structure. The relation between the measured states and the computational states of an idle tetron is then less direct, as the dot is strongly hybridized with the tetron throughout the measurement. How the measurement dot is ultimately decoupled from the system when the measurement is turned off is therefore an important part of the protocol. This issue is particularly relevant for measurement-based Majorana quantum computation, where qubit operations rely on fast and selective parity measurements along different noncommuting axes~\cite{Aasen2025}.

The remainder of this paper is organized as follows. In Sec.~\ref{sec:model}, we introduce the coupled dot-tetron model and its effective description in the dispersive regime. In Sec.~\ref{sec:measurement_lifetime}, we calculate the noise-induced $X$-parity lifetime in the measurement configuration, which we then compare with the $X$-parity dynamics of the isolated tetron in Sec.~\ref{sec:bare_lifetime}. In Sec.~\ref{sec:detuning}, we extend our analysis beyond the dispersive limit and discuss the multilevel dynamics and the emergence of two long-lived manifolds near charge resonance. In Sec.~\ref{sec:measurement_basis}, we discuss the relation between the measured loop parity and the tetron $X$ parity. We conclude in Sec.~\ref{sec:conclusions}.

\section{Model}
\label{sec:model}

\begin{figure}[tb]
	\centering
	\includegraphics[width=0.9\columnwidth]{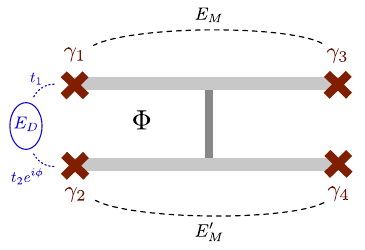}
	\caption{Schematic of the dot-tetron model in the $X$-measurement configuration. Two topological nanowires (light gray) host Majorana zero modes $\gamma_1$, $\gamma_3$ (upper wire) and $\gamma_2$, $\gamma_4$ (lower wire) with residual intrawire hybridizations $E_M$ and $E'_M$. An auxiliary quantum dot with detuning $E_D$ is tunnel coupled to $\gamma_1$ and $\gamma_2$ with amplitudes $t_1$ and $t_2e^{i\phi}$, respectively. The interferometric phase $\phi$ is controlled by the magnetic flux $\Phi$ through the measurement loop. The vertical dark gray line indicates the superconducting backbone connecting the two nanowires.}
	\label{fig:tetron}
\end{figure}

We consider a minimal low-energy model of the Majorana tetron of Ref.~\cite{aghaee2025distinctlifetimesxz} in the $X$ measurement configuration. The device consists of two topological nanowires hosting four Majorana zero modes (MZMs), see Fig.~\ref{fig:tetron}, where we label the Majorana operators such that $\gamma_1,\gamma_3$
belong to one nanowire and $\gamma_2,\gamma_4$ to the other.
To facilitate measurement of the interwire parity $X=i\gamma_1\gamma_2$, an auxiliary quantum dot is further coupled to $\gamma_1$ and $\gamma_2$. The coupled dot-tetron system can be described by the minimal Hamiltonian
\begin{align}
    H
    &=
    iE_M\gamma_1\gamma_3+iE'_M\gamma_2\gamma_4
    +E_{\rm D}\left(d^\dagger d-\frac{1}{2}\right)
    \nonumber\\&\quad+\left(t_1\gamma_1 d+t_2e^{i\phi}\gamma_2 d
    +{\rm H.c.}\right),
    \label{eq:Hfull}
\end{align}
where $E_{\rm D}$ denotes the dot detuning and $\phi$ is the flux-dependent phase accumulated around the measurement loop.

For a sufficiently detuned quantum dot,
$|E_{\rm D}|\gg |t_1|,|t_2|$, the dot can be integrated out
perturbatively. Up to parity-independent energy shifts, this generates
an effective coupling $t_\mathrm{eff}(\phi)$ between the two MZMs participating in the $X$
measurement,
\begin{equation}
    H_{\rm eff}
    =
    it_{\rm eff}(\phi)\gamma_1\gamma_2
    +iE_M\gamma_1\gamma_3
    +iE'_M\gamma_2\gamma_4,
    \label{eq:Heff}
\end{equation}
where
\begin{equation}
    t_{\rm eff}(\phi)
    \simeq
    \frac{2t_1t_2}{E_{\rm D}}\sin\phi
    \label{eq:teff}
\end{equation}
for $E_{\rm D}>0$. This regime allows a clean separation of the quantum dot from the qubit and a corresponding non-destructive measurement. Unfortunately, the detuning also suppresses the capacitance signal and therefore increases the measurement time.

Next, we can express Eq.~\eqref{eq:Heff} as a two-level Hamiltonian by introducing the fermionic operators $a=(\gamma_1+i\gamma_3)/2$, $b=(\gamma_2+i\gamma_4)/2$. Equation~\eqref{eq:Heff} then becomes, up to an overall energy shift,
\begin{equation}
    H_{\rm eff}
    =
    2E_Ma^\dagger a
    +2E'_Mb^\dagger b
    +
    \left(
        \Delta a^\dagger b^\dagger
        +t a^\dagger b
        +{\rm H.c.}
    \right)
    \label{eq:Hefffermion}
\end{equation}
with $\Delta=t=it_{\rm eff}(\phi)$. Since the total tetron parity is conserved, the four-dimensional
Hilbert space separates into two two-dimensional parity sectors $p=\pm 1$. The
even-parity sector $p=+1$ is spanned by
\begin{equation}
    |0\rangle_{\rm +}=|n_a=0,n_b=0\rangle,
    \quad
    |1\rangle_{\rm +}=|n_a=1,n_b=1\rangle,
    \label{eq:evenbasis}
\end{equation}
while the odd-parity sector $p=-1$ is spanned by
\begin{equation}
    |0\rangle_{\rm -}=|n_a=1,n_b=0\rangle,
    \quad
    |1\rangle_{\rm -}=|n_a=0,n_b=1\rangle.
    \label{eq:oddbasis}
\end{equation}
In the even sector, the pairing term proportional to $\Delta$ provides an off-diagonal coupling between
the two states, whereas the hopping term proportional to $t$ vanishes. In the odd sector, the opposite occurs. Since $\Delta=t=it_{\rm eff}(\phi)$,
an appropriate choice of phase for the basis states allows the off-diagonal coupling to be chosen real in either sector. Both parity sectors can therefore be written in the common form
\begin{equation}
    H_p
    =
    t_{\rm eff}(\phi)\sigma_x
    +
    \left(E_M+pE'_M\right)\sigma_z.
    \label{eq:Hp}
\end{equation}
Thus, the coupling to the quantum dot generates an effective
field along the $x$ direction, whereas the residual intrawire MZM splittings generate an effective field along $z$. Interestingly, for equal intrawire splittings $E_M=E'_M$, the static $z$ component vanishes in the odd-parity sector ($p=-1$) even when the individual MZM splittings remain finite.

Due to noise, which is unavoidably present in any experimental semiconductor system, the effective couplings introduced above will generally fluctuate in time:
\begin{align}
    t_{\rm eff}(t)
    &=t_{\rm eff}^0+\delta t_{\rm eff}(t),\\
    E_M(t)
    &=E_M^0+\delta E_M(t),\label{eq:EMnoise}\\
    E'_M(t)
    &=E_M^{\prime0}+\delta E'_M(t)\label{eq:EMprimenoise}.
\end{align}
Fluctuations in $t_{\rm eff}$ can arise, for example, from fluctuations of the quantum dot detuning or the dot-MZM tunnel couplings, whereas fluctuations in $E_M, E_M'$ arise from fluctuations of the intrawire MZM
splittings, e.g., due to fluctuations of the electrostatic potentials in the nanowires. The Hamiltonian in either parity sector then takes the
form
\begin{equation}
    H_p(t)
    =
    \left[B_x^0+\delta B_x(t)\right]\sigma_x
    +
    \left[B_{z,p}^0+\delta B_{z,p}(t)\right]\sigma_z,
    \label{eq:Hnoisegeneral}
\end{equation}
where
\begin{align}
    &B_x^0
    =t_{\rm eff}^0,\qquad\quad\qquad\,
    \delta B_x(t)
    =\delta t_{\rm eff}(t),
    \label{eq:Bxnoise}\\
    &B_{z,p}^0
    =E_M^0+pE_M^{\prime0},\quad
    \delta B_{z,p}(t)
    =\delta E_M(t)+p\,\delta E'_M(t).
    \label{eq:Bznoise}
\end{align}

In the following, we distinguish between the `measurement configuration' considered in Ref.~\cite{aghaee2025distinctlifetimesxz}, where the coupling to the dot is on, and the `isolated' configuration, in which the tetron is decoupled from all measurement devices. In the measurement configuration, the
coupling to the auxiliary dot produces a finite
$B_x^0=t_{\rm eff}(\phi)$, which is assumed to be large compared to the MZM splittings in the regime close to maximal measurement visibility $\phi\sim\pi/2$, such that $|B_x^0|\gg |B_{z,p}^0|$. The eigenstates of the measurement Hamiltonian are then approximately aligned with the $X=i\gamma_1\gamma_2$ basis of the MZM two-level system, up to a small rotation induced by the finite MZM splittings. The lifetime extracted from transitions between these states can therefore approximately be associated with an $X$-parity lifetime. Importantly, however, the finite $B_x^0$ responsible for this approximate alignment is itself generated by the measurement apparatus.

For an isolated tetron, by contrast, the auxiliary quantum dot is
disconnected. Both the static measurement-induced coupling and its
fluctuations then vanish, \mbox{$B_x^0\rightarrow0$}, $\delta B_x(t)\rightarrow0$. The bare tetron is therefore
described by
\begin{equation}
    H_{p,\rm bare}(t)
    =
    [B_{z,p}^0+\delta B_{z,p}(t)]\sigma_z.
    \label{eq:Hbare}
\end{equation}
As we show below, removing the measurement-induced splitting
qualitatively changes the sensitivity of the interwire parity to
low-frequency noise, so that the free-induction coherence time of the bare Majorana qubit can be substantially shorter than the $X$-parity lifetime extracted in the measurement configuration.

We note that we focus on electrostatic fluctuations of the effective low-energy couplings as the relevant noise mechanisms throughout this work, while we neglect quasiparticle poisoning. This choice is motivated by the interpretation of the experiment in Ref.~\cite{aghaee2025distinctlifetimesxz}, where the substantially longer $Z$-loop switching time is associated with external quasiparticle poisoning, while the shorter $X$-loop lifetime is attributed to dynamics within a fixed-parity sector. We therefore treat the total tetron parity as fixed and focus on fluctuations of the effective low-energy couplings within a given parity sector.

The effective two-level description introduced above provides a simple setting in which the physical origin of the measurement-induced lifetime enhancement can be identified transparently. In Secs.~\ref{sec:measurement_lifetime} and \ref{sec:bare_lifetime}, we use this description to contrast the relaxation of the measurement eigenstates with the intrinsic dynamics of the bare tetron. However, since the experiment of Ref.~\cite{aghaee2025distinctlifetimesxz} operates closer to dot resonance, the dispersive approximation need not be quantitatively accurate. In Sec.~\ref{sec:detuning}, we therefore return to the full dot-tetron Hamiltonian of Eq.~(\ref{eq:Hfull}) and show that the resulting qualitative picture persists beyond the large-detuning limit.

\section{Lifetime in the measurement configuration}
\label{sec:measurement_lifetime}

We first consider the lifetime of the two-level system in the
measurement configuration, where the auxiliary quantum dot is coupled to the tetron. The relevant time-dependent Hamiltonian is given in Eq.~(\ref{eq:Hnoisegeneral}). The static part of this Hamiltonian, $ H_0=B_x^0\sigma_x+B_z^0\sigma_z$, has a level splitting
\begin{equation}
    \hbar\Omega=2B_0,
    \qquad
    B_0=\sqrt{(B_x^0)^2+(B_z^0)^2},
    \label{eq:Omega}
\end{equation}
where we suppress the parity label $p$ for notational simplicity for now. We introduce an angle $\vartheta$ such that $\sin\vartheta=B_x^0/B_0$, $\cos\vartheta=B_z^0/B_0$, and define Pauli operators parallel and perpendicular to the static
effective field,
\begin{align}
\sigma_\parallel&=\sin\vartheta\,\sigma_x+
\cos\vartheta\,\sigma_z,
\label{eq:sigmaparallel}\\
\sigma_\perp&=\cos\vartheta\,\sigma_x-
\sin\vartheta\,\sigma_z.
\label{eq:sigmaperp}
\end{align}
In this basis, Eq.~\eqref{eq:Hnoisegeneral} becomes
\begin{equation}
    H(t)=\left[B_0+\delta B_\parallel(t)\right]\sigma_\parallel
    +\delta B_\perp(t)\sigma_\perp,
    \label{eq:Hrotated}
\end{equation}
where
\begin{align}
\delta B_\parallel(t)&=\sin\vartheta\,\delta B_x(t)
+\cos\vartheta\,\delta B_z(t),
\label{eq:Bparallel}\\
\delta B_\perp(t)
&=\cos\vartheta\,\delta B_x(t)-\sin\vartheta\,\delta B_z(t).
\label{eq:Bperp}
\end{align}
The longitudinal noise $\delta B_\parallel$ produces fluctuations of the instantaneous level splitting and therefore contributes to
dephasing. By contrast, the transverse component $\delta B_\perp$
induces transitions between the eigenstates of the measurement
Hamiltonian and determines their relaxation time.

We consider two statistically independent sources of electrostatic noise. First, fluctuations of the intrawire MZM splittings generate
$\delta B_z(t)$. Second, fluctuations of the quantum-dot detuning,
\begin{equation}
E_D(t)=E_D+\delta E_D(t)\label{eq:dotnoise}
\end{equation}
modify the measurement-induced coupling. Since $B_x^0=(2t_1t_2/E_D)\sin\phi$, weak fluctuations of $E_D$ give
\begin{equation}
\delta B_x(t)=-\frac{B_x^0}{E_D}\delta E_D(t)
\label{eq:deltaBx_ED}
\end{equation}
to leading order. While we focus on dot-detuning noise here, other microscopic sources of fluctuations in $B_x$ could in principle be treated analogously.

We characterize the two noise sources by the spectral densities
\begin{align}
S_{zz}(\omega)&=\frac{1}{2\pi}\int_{-\infty}^{\infty}dt\,\langle
\delta B_z(t)\delta B_z(0)\rangle e^{i\omega t},\label{eq:SBz}\\
S_D(\omega)&=\frac{1}{2\pi}\int_{-\infty}^{\infty}dt\,
\langle\delta E_D(t)\delta E_D(0)\rangle
e^{i\omega t}.
\label{eq:SD}
\end{align}
Assuming that $\delta B_z$ and $\delta E_D$ are statistically independent, their contributions to the transverse noise add,
\begin{equation}
S_\perp(\omega)=\frac{(B_x^0)^2}{B_0^2}S_{zz}(\omega)
+\frac{(B_x^0)^2(B_z^0)^2}{E_D^2B_0^2}
S_D(\omega).\label{eq:Sperp_general}
\end{equation}

For sufficiently weak noise and a well-resolved level splitting,
transitions between the two eigenstates can be determined using
Fermi's golden rule. This treatment requires the level splitting to
remain well resolved on the relaxation time scale,
$\Omega T_1\gg 1$. The downward and upward transition rates are then
\begin{align}
\Gamma_\downarrow=\frac{2\pi}{\hbar^2}S_\perp(+\Omega),\qquad
\Gamma_\uparrow=\frac{2\pi}{\hbar^2}S_\perp(-\Omega).
\end{align}
The conventional relaxation time $T_1$, which characterizes the
relaxation of the population difference toward its stationary value,
is therefore
\begin{equation}
\frac{1}{T_1}=\Gamma_\downarrow+\Gamma_\uparrow
=\frac{2\pi}{\hbar^2}\left[
S_\perp(+\Omega)+S_\perp(-\Omega)\right].    
\label{eq:T1general}
\end{equation}
We note that our definition differs by a factor of two from the
interwire-parity lifetime used in Ref.~\cite{aghaee2025distinctlifetimesxz}, where
$\tau_X=2/(\Gamma_\downarrow+\Gamma_\uparrow)=2T_1^X$. We further assume $\hbar\Omega\ll k_{\rm B}T$, such that the upward and downward transition rates can be approximated as equal, $S_\perp(+\Omega)\simeq S_\perp(-\Omega)$. This is consistent with the approximately equal dwell times observed experimentally in Ref.~\cite{aghaee2025distinctlifetimesxz} and yields
\begin{align}
\frac{1}{T_1}&=\frac{4\pi}{\hbar^2}S_\perp(\Omega).
\label{eq:T1classical}
\end{align}

We now focus on the regime relevant to the $X$-loop measurement.
The dispersive treatment above requires $|E_D| \gg |t_1|,|t_2|$, while identifying the measurement eigenstates with the interwire-parity eigenstates simultaneously requires $|B_x^0|=2|t_1t_2\sin\phi|/|E_D| \gg |B_z^0|$. In this regime, the measurement axis is approximately aligned with the $X$ axis and we denote the corresponding relaxation time by $T_1^X$. Assuming the system is initially polarized along the measurement axis, the population difference decays as
\begin{equation}
\langle \sigma_\parallel(t)\rangle=e^{-t/T_1^X}.
\label{eq:X_decay_measurement}
\end{equation}
Since $\sigma_\parallel\simeq\sigma_x$ in this regime, this decay
corresponds approximately to relaxation of the interwire parity.

Equations~\eqref{eq:Sperp_general} and~\eqref{eq:T1classical} already capture the central effect of the measurement apparatus on the observed lifetime. Coupling to the auxiliary dot generates a finite splitting $\hbar\Omega$ between the measurement eigenstates, such that relaxation samples the noise spectral densities at the finite frequency $\Omega$. For noise concentrated predominantly at low frequencies, this can strongly suppress transitions compared with the corresponding bare
tetron. Fluctuations of the MZM splittings contribute directly through $S_{zz}(\Omega)$, while dot-detuning fluctuations are additionally suppressed by the factor $(B_z^0/E_D)^2$ in the approximately $X$-like regime.

Electrostatic charge noise in semiconductor devices is commonly observed to exhibit an approximately $1/f^\alpha$ spectrum over broad frequency ranges, with $\alpha$ typically of order unity, although both the exponent and the crossover between different frequency regimes are device dependent. We therefore use $1/f$ spectra as a simple representative model for the electrostatic fluctuations underlying both noise mechanisms,
\begin{equation}
S_{zz}(\omega)=\frac{A_z}{|\omega|},
\qquad
S_D(\omega)=\frac{A_D}{|\omega|}.
\label{eq:1f_both}
\end{equation}
Using Eqs.~\eqref{eq:Sperp_general} and~\eqref{eq:T1classical}, the relaxation rate within the dispersive model is then
\begin{equation}
\frac{1}{T_{1}}=\frac{2\pi}{\hbar}
\frac{(B_x^0)^2}{B_{0}^3}
\left[A_z+\left(\frac{B_{z}^0}{E_D}
\right)^2 A_D \right].
\label{eq:T1_1f_general}
\end{equation}
In the $X$-like limit $|B_x^0|\gg|B_z^0|$, the prefactor $(B_x^0)^2/B_0^3$ reduces to $1/|B_x^0|$, making it explicit that the finite splitting generated by the measurement apparatus suppresses relaxation driven by low-frequency \(1/f\) noise.

While the $1/|B_x^0|$ scaling follows specifically from the $1/f$ form assumed above, the underlying stabilization mechanism is more general. For a fixed transverse noise coupling, it only requires the relevant noise spectral density to decrease over the frequency range accessed by the measurement-induced splitting. We note that the charge-noise measurements used to motivate our noise
amplitudes probe frequencies well below the qubit transition frequencies relevant here, such that extending the $1/f$ form to these frequencies necessarily involves an extrapolation. Similar extrapolations have previously been used to estimate the effective noise strengths relevant for Majorana readout experiments~\cite{aghaee2025interferometric,Boutin2025}. The absolute lifetimes obtained from our $1/f$ model should therefore be regarded as illustrative examples. At higher frequencies, the charge-noise spectrum itself may cross over to a different form or approach a finite noise floor, and additional relaxation channels may become relevant. Such contributions would limit the lifetime enhancement once the low-frequency noise has been sufficiently suppressed. At the same time, the long switching lifetimes observed experimentally place constraints on the overall strength of these additional finite-frequency relaxation channels. Our central conclusion therefore does not rely on a strict $1/f$ spectrum, but rather on the relevant transverse noise being reduced at the measurement-induced transition frequency compared with the low-frequency fluctuations responsible for bare-tetron dephasing.

\begin{figure*}[tb]
	\centering
	\includegraphics[width=0.85\textwidth]{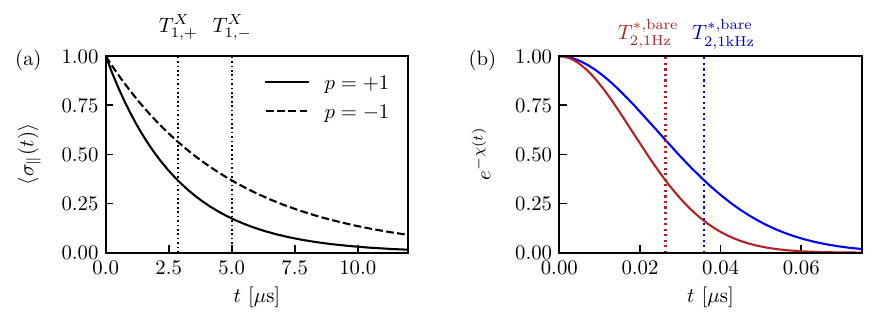}
	\caption{Comparison of the parity dynamics in the measurement and isolated-tetron configurations. (a) Relaxation in the measurement configuration for $t_1=2~\mu{\rm eV}$, $t_2=2.5~\mu{\rm eV}$, $E_D=15~\mu{\rm eV}$, and $\phi=\pi/2$, corresponding to $B_x^0\approx0.67~\mu{\rm eV}$. We take intrawire MZM splittings $E_M=0.05~\mu{\rm eV}$, $E_M'=0.15~\mu{\rm eV}$, such that $B_{z,+}^0=0.2~\mu{\rm eV}$ and $|B_{z,-}^0|=0.1~\mu{\rm eV}$. Both MZM-splitting noise with $A_z=10^{-17}~{\rm eV}^2$ and dot-detuning noise with $A_D=10^{-13}~{\rm eV}^2$ are included. Solid and dashed black curves correspond to $p=+1$ and $p=-1$, respectively, and the vertical dotted lines indicate the corresponding relaxation times $T_{1,+}^X$ and $T_{1,-}^X$. Since $|B_x^0|\gg|B_{z,p}^0|$, the measurement eigenstates remain close to the $X$ eigenstates. (b) Decay envelope of the bare interwire parity after the measurement dot is disconnected, for the same MZM-splitting noise strength $A_z$ and two representative cutoffs $f_\mathrm{ir}=\omega_{\rm ir}/2\pi=1~{\rm Hz}$ (red) and $f_\mathrm{ir}=\omega_{\rm ir}/2\pi=1~{\rm kHz}$ (blue). The vertical dotted lines indicate the characteristic free-induction coherence times $T_{2}^{*,\mathrm{bare}}$, defined by $e^{-\chi(T_{2}^{*,\mathrm{bare}})}=1/e$. For the representative parameters considered here, the bare-tetron free-induction coherence time is a few tens of ns, whereas the relaxation time in the measurement configuration is a few $\mu{\rm s}$.}
\label{fig:lifetime}
\end{figure*}

An important special case occurs in the $p=-1$ tetron-parity sector for equal intrawire MZM splittings,
$E_M=E'_M$. This leads to $B_{z,-}^0=E_M-E'_M=0$, where we have restored the parity label $p$ for clarity. The leading dot-noise contribution in Eq.~\eqref{eq:T1_1f_general} then vanishes. At this order, both the static measurement-induced coupling and the fluctuations generated by $\delta E_D$ point along the same effective-field axis, so dot-detuning noise is purely longitudinal. This cancellation does not, however, persist to higher order in the large-$E_D$ expansion. As shown explicitly in Appendix~\ref{app:dot_tetron}, virtual dot processes generate a small component of the static effective field perpendicular to the leading measurement axis. Since this correction scales as $E_D^{-2}$, while the leading measurement-induced field scales as $E_D^{-1}$, fluctuations of $E_D$ change the direction of the total effective field and thereby acquire a transverse component. For $E_M=E'_M$, the resulting leading relaxation rate due to dot-detuning noise is then (see Appendix~\ref{app:dot_tetron})
\begin{equation}
\Gamma_D^{(-)}=\frac{4\pi}{\hbar^2}S_D(\Omega)
\left|K_{+-}^{(-)}\right|^2
\label{eq:T1_dot_higher}
\end{equation}
with
\begin{equation}
\left|K_{+-}^{(-)}\right|^2=\frac{4E_M^2}{E_D^6}
\left[(t_1^2-t_2^2)^2+4t_1^2t_2^2\cos^2\phi\right],
\label{eq:Kminus_main}
\end{equation}
provided that the leading measurement-induced field dominates
the higher-order transverse correction,
$|\mathbf B_\perp|\ll |B_\parallel|$ in the notation of
Appendix~\ref{app:dot_tetron}. The same large-$E_D$ result was obtained in Ref.~\cite{aghaee2025distinctlifetimesxz}. This contribution is parametrically suppressed at large $E_D$ and vanishes at the perfectly symmetric operating point $t_1=t_2$ and $\phi=\pi/2$. Away from this point, tunnel-coupling asymmetry or a phase offset generates a finite relaxation rate. The
full derivation of Eq.~\eqref{eq:Kminus_main} is given in Appendix~\ref{app:dot_tetron}.

To obtain a quantitative estimate of the parity lifetime, we set $A_D=10^{-13}\,\mathrm{eV}^2$ for fluctuations of the dot detuning~\footnote{Adapting the noise-amplitude estimate of Ref.~\cite{Boutin2025} to our convention for the noise spectral density yields $A_D\sim (1~\mu{\rm eV})^2/2\pi\sim 1.6\times10^{-13}~{\rm eV}^2$.}. Furthermore, we use $A_z=10^{-17}\,\mathrm{eV}^2$ as an order-of-magnitude estimate motivated by the electrostatic sensitivity obtained in microscopic calculations of disordered Majorana nanowires~\footnote{We estimate $A_z$ by combining a representative dimensionless charge sensitivity $D\equiv |\partial \varepsilon_M/\partial\mu|\sim0.1$ of the Majorana quasiparticle energy $\varepsilon_M=2E_M$ to the local electrostatic potential $\mu$, inferred from the decoherence rates reported in Ref.~\cite{Sau2025}, with the electric-field noise model of Ref.~\cite{Knapp2018}. In our convention, the intrawire splitting $E_M$ is half the corresponding BdG excitation energy, giving $A_M\simeq\alpha_E(eDd_{\rm eff})^2/4$ for a single wire, where $\alpha_E$ is the amplitude of the electric-field noise spectrum and $d_{\rm eff}$ is an effective electrostatic length scale converting electric-field fluctuations into potential fluctuations. Assuming comparable independent fluctuations in the two wires then gives $A_z=A_M+A'_M\simeq\alpha_E(eDd_{\rm eff})^2/2$. Using $\alpha_E\sim10~({\rm V/m})^2$, $D\sim0.1$, and $d_{\rm eff}\sim10-20~{\rm nm}$ gives $A_z\sim5\times10^{-18}-2\times 10^{-17}~{\rm eV}^2$, motivating the representative value $A_z=10^{-17}~{\rm eV}^2$ used here. Since $d_{\rm eff}$ and the microscopic charge sensitivity are device dependent, this should be regarded only as an order of magnitude estimate. For comparison, the decoherence rate reported for the ideal clean-wire case in the same work corresponds to an effective charge sensitivity $D\sim10^{-4}$, illustrating that the electrostatic sensitivity can increase strongly away from the ideal clean-wire regime.}. This value should be viewed as a representative value for a device that has not yet reached the strongly exponentially protected limit rather than as a universal MZM-noise strength.

In Fig.~\ref{fig:lifetime}(a), we illustrate the resulting relaxation dynamics in the measurement configuration for the two parity sectors $p=\pm 1$. We set $t_1=2~\mu{\rm eV}$, $t_2=2.5~\mu{\rm eV}$, $E_D=15~\mu{\rm eV}$, and $\phi=\pi/2$, corresponding to $B_{x}^0\approx0.67~\mu{\rm eV}$, together with intrawire MZM splittings $E_M=0.05~\mu{\rm eV}$, $E_M'=0.15~\mu{\rm eV}$. The resulting static splittings are $B_{z,+}^0=0.2~\mu{\rm eV}$ and $|B_{z,-}^0|=0.1~\mu{\rm eV}$, respectively. For these representative parameters, we obtain relaxation times on the order of a few $\mu{\rm s}$. As we will show in the following section, this time scale can be substantially longer than the free-induction coherence time of the isolated tetron.

\section{$X$-parity dynamics of the bare tetron}
\label{sec:bare_lifetime}

We now contrast the lifetime obtained in the measurement configuration
with the intrinsic $X$-parity dynamics of an isolated tetron. When the
auxiliary quantum dot is disconnected, the measurement-induced
coupling and its fluctuations vanish, $B_x^0=\delta B_x(t)=0$, while
the static intrawire MZM splittings and their fluctuations generally
remain finite. The bare-tetron Hamiltonian is then $H_{\rm bare}(t)=\left[B_z^0+\delta B_z(t)\right]\sigma_z$ as given in Eq.~(\ref{eq:Hbare}), where we again suppress the parity-sector label for simplicity.

Unlike in the measurement configuration, there is now no
measurement-induced splitting along the $x$ direction that suppresses the effect of low-frequency fluctuations of the MZM splittings. To determine the intrinsic $X$-parity dynamics, we consider an initial state with $X=\sigma_x=+1$. Since the Hamiltonian in Eq.~\eqref{eq:Hbare} is proportional to $\sigma_z$ at all times, the time evolution can be evaluated exactly for a given realization of the noise. The accumulated relative phase between the two $\sigma_z$ eigenstates can be written as
\begin{equation}
    \Phi(t)=\frac{2B_z^0t}{\hbar}+\varphi(t),
    \quad
    \varphi(t)=\frac{2}{\hbar}
    \int_0^t d\tau\,\delta B_z(\tau).
    \label{eq:phase_bare}
\end{equation}
The corresponding $X$ expectation value for a fixed noise realization is
\begin{equation}
    \langle\sigma_x(t)\rangle_{\delta B}=\cos[\Phi(t)].
    \label{eq:X_single_realization}
\end{equation}
For Gaussian noise with zero mean, averaging over noise realizations
gives
\begin{equation}
    \langle\sigma_x(t)\rangle=
    \cos\left(\frac{2B_z^0t}{\hbar}\right)
    \exp\left[-\frac{1}{2}
    \left\langle\varphi^2(t)\right\rangle
    \right].
    \label{eq:X_gaussian}
\end{equation}
Thus, a finite static MZM splitting produces coherent oscillations of the interwire parity but does not modify the noise-induced decay
envelope.

Using Eq.~\eqref{eq:SBz}, we obtain
\begin{align}
    \left\langle\varphi^2(t)\right\rangle
    &=
    \frac{4}{\hbar^2}
    \int_0^t d\tau
    \int_0^t d\tau'\,
    \left\langle
        \delta B_z(\tau)\delta B_z(\tau')
    \right\rangle
    \nonumber\\
    &=
    \frac{4}{\hbar^2}
    \int_{-\infty}^{\infty}d\omega\,
    S_{zz}(\omega)
    \frac{4\sin^2(\omega t/2)}{\omega^2}.
    \label{eq:phase_variance_general}
\end{align}
Introducing the decoherence function
\begin{equation}
    \chi(t)=\frac{2}{\hbar^2}
    \int_{-\infty}^{\infty}d\omega\,
    S_{zz}(\omega)
    \frac{4\sin^2(\omega t/2)}{\omega^2},
    \label{eq:chi_general}
\end{equation}
we then have
\begin{equation}
    \langle\sigma_x(t)\rangle=
    \cos\left(\frac{2B_z^0t}{\hbar}\right)e^{-\chi(t)}.
    \label{eq:X_chi}
\end{equation}

The decay considered here is therefore the free-induction dephasing of the idle qubit, which essentially corresponds to its $T_2^*$ coherence time. Both this quantity and the $T_1^X$ introduced for the measurement configuration characterize the decay of the $X$ parity, but the conventional terminology differs because coupling to the measurement dot changes the orientation of the effective qubit field. In the $X$-like measurement regime, fluctuations of the MZM splittings are predominantly transverse to the qubit eigenbasis and induce relaxation, whereas in the bare tetron they are longitudinal and dephase an initial $X$ superposition. Thus, the different terminology reflects the operating Hamiltonian rather than a difference in the underlying observable.

Because the $1/f$ spectrum diverges at low frequency, its contribution to the dephasing depends on how the infrared part of the noise is treated. Here, we consider a fixed infrared cutoff $\omega_{\rm ir}$. For $\omega_{\rm ir}t\ll1$, the dominant contribution arises from frequencies $\omega_{\rm ir}\lesssim\omega\lesssim1/t$, for which
    $4\sin^2(\omega t/2)/\omega^2
    \simeq t^2$.
Equation~\eqref{eq:chi_general} then gives, to logarithmic accuracy,
\begin{equation}
    \chi(t)
    \simeq
    \frac{4A_z t^2}{\hbar^2}
    \ln\left(\frac{1}{\omega_{\rm ir}t}\right).
    \label{eq:chi_1f}
\end{equation}
The decay envelope $e^{-\chi(t)}$ is therefore approximately Gaussian, with an additional weak logarithmic dependence on the infrared cutoff.

Figure~\ref{fig:lifetime}(b) shows this decay envelope for two representative cutoffs $f_\mathrm{ir}=\omega_{\rm ir}/2\pi=1~{\rm Hz},\ 1~{\rm kHz}$. In contrast to the measurement configuration shown in Fig.~\ref{fig:lifetime}(a), the free-induction envelope decays on a time scale that is about two orders of magnitude shorter for the same MZM-splitting noise strength. This large difference arises because coupling to the measurement dot generates a finite splitting $2|B_x^0|$ in panel (a), which shifts the relevant noise sensitivity to the finite frequency
$\Omega\simeq2|B_x^0|/\hbar$. Once the dot is disconnected, this measurement-induced stabilization is absent and the bare-tetron $X$ coherence is directly exposed to the low-frequency part of the $1/f$ noise. The strong separation of timescales does not depend on the exact choice of infrared cutoff.

Two general remarks about this comparison are useful. First, we note that the coupling to the measurement dot introduces a tradeoff: while the resulting finite splitting suppresses transitions driven by low-frequency noise, fluctuations of the dot detuning provide an additional source of noise that is absent for the isolated tetron. Nevertheless, we find that the splitting-induced stabilization persists over a broad range of experimentally motivated parameters. In particular, increasing the dot-detuning noise strength used in Fig.~\ref{fig:lifetime}(a) by an order of magnitude still yields measurement-configuration lifetimes that exceed the bare-tetron coherence times by about an order of magnitude, keeping all other parameters fixed. The resulting lifetimes remain of order $\sim 1~\mu$s and therefore broadly compatible with the $X$-loop switching times observed experimentally~\cite{aghaee2025distinctlifetimesxz}; even larger effective noise strengths would instead drive the modeled lifetimes below this scale.

Second, the bare-tetron coherence time depends strongly on the magnitude of the MZM-splitting noise. In the clean, long-wire limit, this noise is expected to be exponentially suppressed, leading to substantially longer coherence times. In contrast, we have estimated our noise amplitude using a numerical charge matrix element estimate from Ref.~\cite{Sau2025}, where disorder is explicitly included. Ultimately, the few tens of ns bare-tetron coherence times in Fig.~\ref{fig:lifetime}(b) should not be viewed as universal. Rather, our main point is that the measurement-induced splitting can lead to microsecond-scale lifetimes even in disordered devices with comparably strong MZM noise, for which the corresponding bare-tetron coherence times can be much shorter.

\begin{figure*}[tb]
	\centering
	\includegraphics[width=\textwidth]{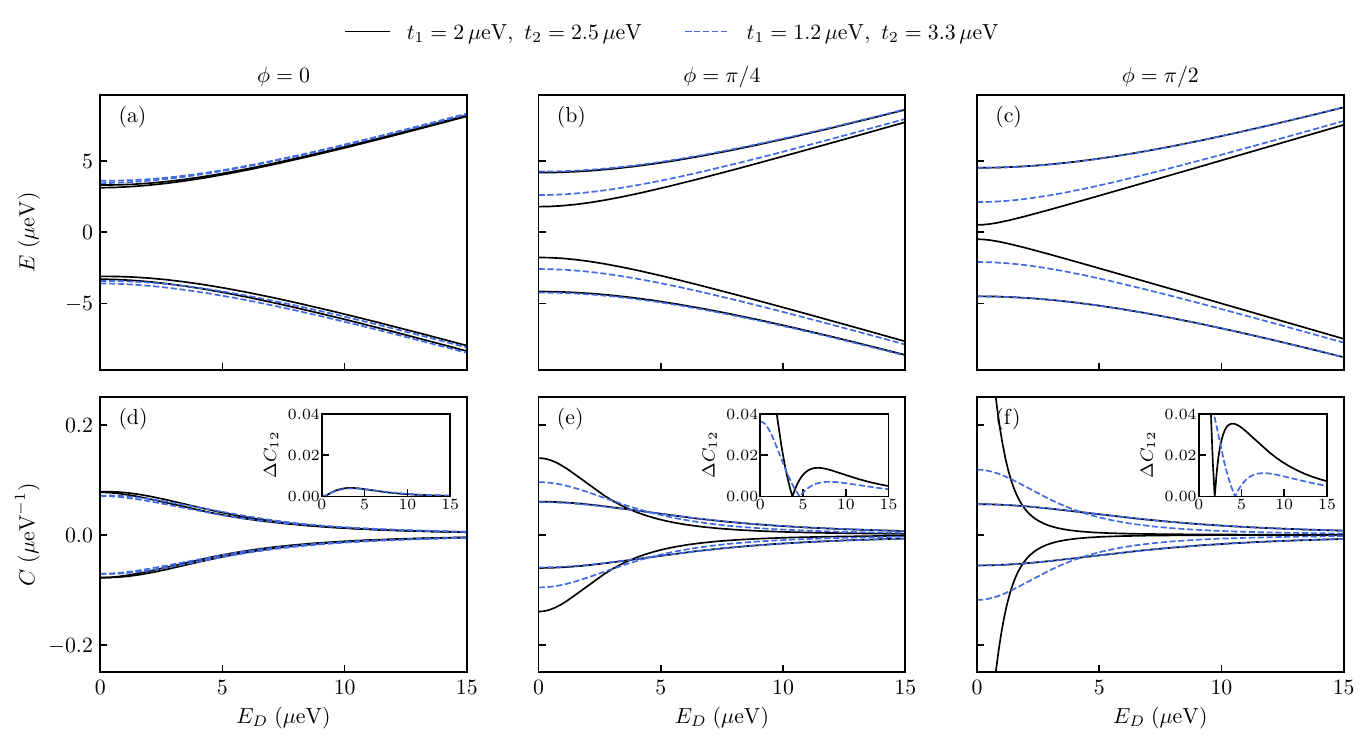}
	\caption{Spectrum and quantum capacitance response of the full dot--tetron model. (a)-(c) Eigenenergies $E$ and (d)-(f) corresponding detuning curvatures $C=\partial^2E/\partial E_D^2$ as functions of the dot detuning $E_D$, for $\phi=0,\pi/4,\pi/2$, respectively. The curvatures $C$ are proportional to the state-dependent quantum capacitance. We show the $\mathcal P=+1$ sector for $E_M=0.05~\mu{\rm eV}$ and $E_M'=0.15~\mu{\rm eV}$. The solid black curves correspond to $(t_1,t_2)=(2~\mu{\rm eV},2.5~\mu{\rm eV})$ and the dashed blue curves to $(t_1,t_2)=(1.2~\mu{\rm eV},3.3~\mu{\rm eV})$. The insets in panels (d)-(f) show the difference in capacitance $\Delta C_{12}=|C_{1}-C_2|$ between the two lowest-energy eigenstates.}
\label{fig:spectra}
\end{figure*}

\section{Relaxation beyond the large-detuning limit}
\label{sec:detuning}

The analytical results of Sec.~III were obtained in the dispersive limit $|E_D|\gg |t_1|,|t_2|,|E_M|,|E_M'|$, where the auxiliary measurement dot can be integrated out perturbatively. This limit is particularly useful for identifying the different relaxation mechanisms and their parametric dependence. However, for the purpose of enhancing the capacitance signal, the $X$-loop measurements of Ref.~\cite{aghaee2025distinctlifetimesxz} are performed closer to the quantum dot-nanowire resonance, such that the relevant dot detuning need not lie deep in the dispersive regime. We therefore now return to the full dot-tetron Hamiltonian of Eq.~\eqref{eq:Hfull} and evaluate the relaxation rates in this full model.

Although tunneling between the dot and the tetron does not conserve the tetron parity $p$ introduced in Sec.~\ref{sec:model}, the total parity of the combined dot-tetron system remains conserved. We denote this total parity by $\mathcal P=\pm1$. For a fixed dot occupation $n_d=d^\dagger d$, it is related to the tetron parity by $\mathcal P=(-1)^{n_d}p$. The full eight-dimensional Hilbert space therefore separates into two four-dimensional sectors of fixed $\mathcal P$.

To represent the Hamiltonian within either sector, we introduce two
sets of Pauli matrices $\boldsymbol{\sigma}$ and
$\boldsymbol{\tau}$, where $\boldsymbol{\sigma}$ acts in MZM space as
\begin{equation}
X=i\gamma_1\gamma_2=\sigma_x,
\qquad
i\gamma_1\gamma_3=\sigma_z,
\qquad
i\gamma_2\gamma_3=\sigma_y
\end{equation}
and $\boldsymbol{\tau}$ acts in dot space with $d^\dagger d=n_d=(1+\tau_z)/2$ describing the dot occupation.
This can be realized by
\begin{align}
&\gamma_1=\pi_x\sigma_y,\qquad
\gamma_2=-\pi_x\sigma_z,\qquad\nonumber\\
&\gamma_3=\pi_x\sigma_x,\qquad
\gamma_4=-\pi_y\tau_z,\qquad\nonumber\\
&d=\pi_y\tau_-,
\label{eq:full_pauli_identities}
\end{align}
where $\pi_z=\mathcal P$ acts in total-parity space.
Equation~\eqref{eq:Hfull} then becomes, within a fixed
$\mathcal P$ sector,
\begin{align}
H_{\mathcal P}={}&\frac{E_D}{2}\tau_z+E_M\sigma_z
-\mathcal P E_M'\sigma_z\tau_z\nonumber\\&+
\mathcal P\left[\left(-it_1\sigma_y+it_2e^{-i\phi}\sigma_z\right)\tau_+
+\mathrm{H.c.}\right].
\label{eq:Hfull_pauli}
\end{align}

For each $\mathcal P$, we diagonalize Eq.~\eqref{eq:Hfull_pauli} exactly and denote its four eigenstates
and eigenenergies by
\begin{equation}
H_{\mathcal P}|n,\mathcal P\rangle=
E_{n,\mathcal P}|n,\mathcal P\rangle,
\qquad n=1,\ldots,4,
\end{equation}
ordered such that
$E_{1,\mathcal P}\le E_{2,\mathcal P}\le
E_{3,\mathcal P}\le E_{4,\mathcal P}$.
The two lowest states continuously approach the low-energy
measurement doublet considered in Sec.~\ref{sec:measurement_lifetime} in the dispersive limit. In addition to the energy spectrum, it is useful to calculate the curvature of each eigenstate
\begin{equation}
\mathcal C_{n,\mathcal P}=\frac{\partial^2 E_{n,\mathcal P}}{\partial E_D^2},
\label{eq:curvature}
\end{equation}
which is proportional to the static quantum-capacitance contribution of state $|n,\mathcal P\rangle$. By focusing on the static contribution, we neglect finite-frequency readout effects as well as readout-induced changes in state occupations. As such, the resulting capacitance differences should be understood as approximate estimates of the measured signal. Figure~\ref{fig:spectra} shows the energy spectrum and state-resolved curvatures of the coupled dot-tetron system as a function of $E_D$ for three representative values of flux $\phi=0$, $\pi/4$, $\pi/2$ and two different sets of tunneling amplitudes $(t_1,t_2)=(2~\mu\mathrm{eV},2.5~\mu\mathrm{eV})$ and $(t_1,t_2)=(1.2~\mu\mathrm{eV},3.3~\mu\mathrm{eV})$.

We next evaluate the noise-induced transition rates directly within the full four-state spectrum. As in Sec.~\ref{sec:measurement_lifetime}, we consider independent fluctuations of the two intrawire MZM splittings and the dot detuning as given in Eqs.~(\ref{eq:EMnoise}), (\ref{eq:EMprimenoise}),  and (\ref{eq:dotnoise}), respectively.
The corresponding noise Hamiltonian can be written as
\begin{equation}
\delta H_{\mathcal P}(t)=
\delta E_D(t)\mathcal O_D+\delta E_M(t)\mathcal O_M+\delta E_M'(t)\mathcal O_{M'},
\label{eq:full_noise}
\end{equation}
with
\begin{equation}
\mathcal O_D=\frac{\tau_z}{2},\qquad
\mathcal O_M=\sigma_z,\qquad
\mathcal O_{M'}=-\mathcal P\,\sigma_z\tau_z.
\label{eq:full_noise_operators}
\end{equation}

We denote the spectral densities of the three statistically
independent noise sources by $S_D(\omega)$, $S_M(\omega)$, and $S_{M'}(\omega)$, using the convention introduced in Sec.~\ref{sec:measurement_lifetime}. Fermi's golden rule then gives the transition rate
\begin{align}
\Gamma_{n\rightarrow m}^{(\mathcal P)}=
\frac{2\pi}{\hbar^2}\bigg[&S_D(\omega_{nm})\left|
\langle m,\mathcal P|\mathcal O_D|n,\mathcal P\rangle
\right|^2\nonumber\\&+S_M(\omega_{nm})\left|
\langle m,\mathcal P|\mathcal O_M|n,\mathcal P\rangle
\right|^2\nonumber\\&+S_{M'}(\omega_{nm})\left|
\langle m,\mathcal P|\mathcal O_{M'}|n,\mathcal P\rangle
\right|^2\bigg],
\label{eq:full_transition_rate}
\end{align}
where $\omega_{nm}=(E_{n,\mathcal P}-E_{m,\mathcal P})/\hbar$.
Equation~\eqref{eq:full_transition_rate} can be evaluated for arbitrary $E_D$, provided the noise remains sufficiently weak for the golden-rule treatment to apply. For the $1/f$ noise considered throughout this work, we write
$S_\lambda(\omega)=A_\lambda/|\omega|$, where $A_D$ is the dot-detuning noise amplitude introduced in Sec.~\ref{sec:measurement_lifetime}, while $A_M$ and $A_M'$ characterize fluctuations of the two intrawire MZM splittings $E_M$ and $E'_M$, respectively. Since $\delta B_{z,p}=\delta E_M+p\,\delta E'_M$ and the two MZM noise sources are assumed to be statistically independent, their spectral densities add, such that we can relate them to the splitting noise parameter from Sec.~\ref{sec:measurement_lifetime} as $A_z=A_M+A_{M'}$. In the numerical calculations below, we choose $A_M=A_{M}'=A_z/2$. With these definitions, Eq.~\eqref{eq:full_transition_rate}
takes the simple form
\begin{align}
\Gamma_{n\rightarrow m}^{(\mathcal P)}=\frac{2\pi}{\hbar^2}
\sum_{\lambda=D,M,M'}\frac{A_\lambda\left|\langle m,\mathcal P|\mathcal O_\lambda
|n,\mathcal P\rangle\right|^2}{|\omega_{nm}|},
\label{eq:full_transition_rate_1f}
\end{align}
whenever the corresponding transition energy is small compared with temperature, such that upward and downward rates can be treated as approximately equal.

\subsection{Two-level regime at large detuning}

We first consider the large-detuning regime, in which the two lowest eigenstates are well separated from the remaining states. In the dispersive limit, these two states continuously approach the measurement doublet considered in Sec.~\ref{sec:measurement_lifetime}. If transitions involving the upper states are sufficiently suppressed, the dynamics can therefore be reduced to the lowest doublet. The relevant thermal energy hierarchy is
\begin{equation}
    E_{2,\mathcal P}-E_{1,\mathcal P}\ll k_B T
    \ll E_{3,\mathcal P}-E_{1,\mathcal P},
    E_{4,\mathcal P}-E_{1,\mathcal P}.
    \label{eq:two_level_hierarchy}
\end{equation}
However, we note that the accuracy of the two-level approximation also depends on the corresponding transition matrix elements and on the frequency dependence of the noise spectrum, so that transitions to the upper pair can remain relevant even when these states are only weakly thermally occupied.

When transitions involving states $3$ and $4$ can be neglected, the relaxation dynamics is governed by the rates within the lowest doublet. For
$E_{2,\mathcal P}-E_{1,\mathcal P}\ll k_BT$, upward and downward rates are approximately equal, and
\begin{equation}
    \frac{1}{T_{1,\mathcal P}^{(2)}}=
    \Gamma_{1\rightarrow2}^{(\mathcal P)}
    +\Gamma_{2\rightarrow1}^{(\mathcal P)}
    \simeq 2\Gamma_{1\rightarrow2}^{(\mathcal P)}.
    \label{eq:T1_lowest_doublet}
\end{equation}
In this limit, the full model continuously approaches the effective dispersive description developed in Sec.~\ref{sec:measurement_lifetime}.

In Appendix~\ref{app:comparison}, we compare the lowest-doublet
relaxation time calculated from the full dot-tetron Hamiltonian
with the analytical dispersive expressions of Sec.~\ref{sec:measurement_lifetime}. Furthermore, we also evaluate the total escape rates from the two lowest eigenstates to assess when transitions to the upper states can be neglected.

\begin{figure*}[tb]
	\centering
	\includegraphics[width=0.95\textwidth]{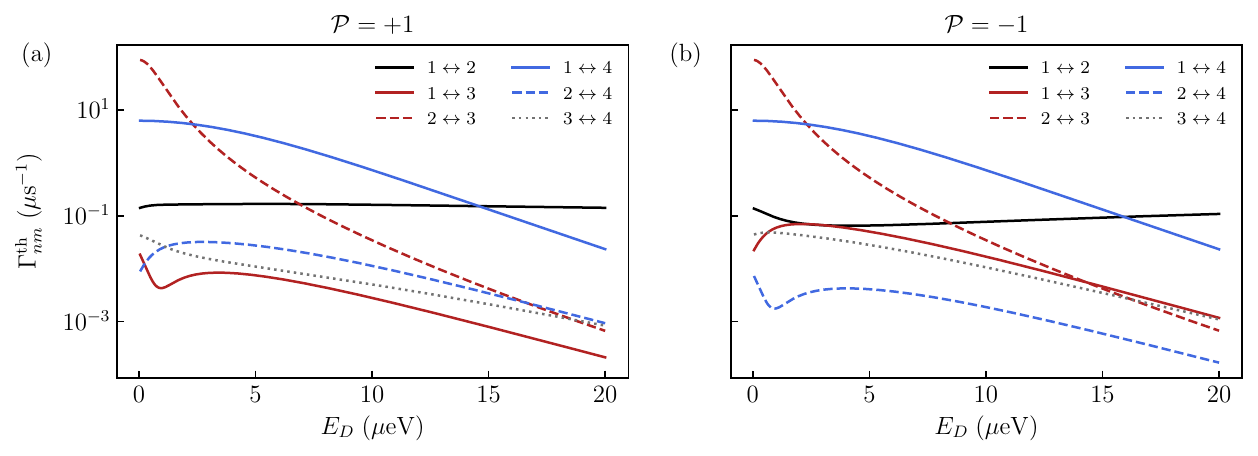}
	\caption{Thermally weighted transition rates in the full four-state dot–tetron model. The equilibrium event rates $\Gamma_{nm}^{\rm th}$, defined in Eq.~(\ref{eq:thermal_event_rate}), are shown as functions of dot detuning $E_D$ at $\phi=\pi/2$ for total parity (a) $\mathcal P=+1$ and (b) $\mathcal P=-1$. Near resonance, the dominant $1\leftrightarrow4$ and $2\leftrightarrow3$ transitions motivate the two-manifold description $A=\{1,4\}$, $B=\{2,3\}$. With increasing detuning, transition traffic involving the upper states is suppressed and the lowest-doublet channel becomes dominant. Parameters are $t_1=2~\mu\mathrm{eV}$, $t_2=2.5~\mu\mathrm{eV}$, $E_M=0.05~\mu{\rm eV}$, $E_M'=0.15~\mu\mathrm{eV}$, and $T=50~\mathrm{mK}$. The independent noise sources have amplitudes $A_D=10^{-13}~\mathrm{eV}^2$ and $A_M=A_M'=0.5\times10^{-17}~\mathrm{eV}^2$, with spectral densities given by Eq.~(\ref{eq:thermal_1f}).}
\label{fig:multilevel}
\end{figure*}

\subsection{Multilevel dynamics near resonance}

The dynamics becomes qualitatively different as the measurement dot is brought closer to resonance. In this regime, thermal occupation of the upper states can no longer generally be neglected, and there is no obvious reason to restrict the dynamics to the lowest doublet. Since the four eigenstates have different capacitance responses, as illustrated in Fig.~\ref{fig:spectra}, one might then expect up to four branches in the capacitance signal rather than just two as observed in Ref.~\cite{aghaee2025distinctlifetimesxz}. Whether these branches can be resolved, however, also depends on the transition rates between the states. For now, we therefore retain all pairwise transitions within a fixed-$\mathcal P$ sector and examine whether their dynamics can give rise to two effective readout levels.

For transition energies comparable to or larger than $k_BT$, the symmetric high-temperature form of the $1/f$ spectrum used above must be supplemented by the detailed-balance asymmetry between absorption and emission. Following Ref.~\cite{Boutin2025}, we use
\begin{equation}
    S_\lambda(\omega)
    =
    \frac{A_\lambda}{|\omega|}
    \frac{2}
    {1+\exp[-\hbar\omega/k_BT]},
    \label{eq:thermal_1f}
\end{equation}
where our convention $\omega_{nm}=(E_{n,\mathcal P}-E_{m,\mathcal P})/\hbar$ implies that $\omega_{nm}>0$ corresponds to a downward transition. Equation~\eqref{eq:thermal_1f} satisfies the detailed balance relation
\begin{equation}
    \frac{S_\lambda(-\omega)}
         {S_\lambda(\omega)}=e^{-\hbar\omega/k_BT}
    \qquad (\omega>0).
\end{equation}

The corresponding equilibrium occupations are
\begin{equation}
P_{n,\mathcal P}^{\rm eq}=
\frac{e^{-E_{n,\mathcal P}/k_BT}}
         {\sum_m e^{-E_{m,\mathcal P}/k_BT}}.
\end{equation}
To visualize which transitions dominate the steady-state dynamics, we define the thermally weighted rates associated with the undirected link $n\leftrightarrow m$,
\begin{equation}
\Gamma_{nm}^{\rm th}=
P_{n,\mathcal P}^{\rm eq}
\Gamma_{n\rightarrow m}^{(\mathcal P)}
+P_{m,\mathcal P}^{\rm eq}
\Gamma_{m\rightarrow n}^{(\mathcal P)}.
\label{eq:thermal_event_rate}
\end{equation}
Detailed balance makes the two contributions on the right-hand side equal. Thus, $\Gamma_{nm}^{\rm th}$ measures the equilibrium frequency of transitions between a given pair of states while accounting for the reduced occupation of higher-energy states.

Figure~\ref{fig:multilevel} shows all six rates
$\Gamma_{nm}^{\rm th}$ as a function of $E_D$ at
$\phi=\pi/2$. At large detuning, transitions involving states $3$ and $4$ become suppressed and
$\Gamma_{12}^{\rm th}$ becomes the dominant low-energy rate, consistent with the two-level description discussed above. Closer to resonance, however, a different hierarchy emerges: the $1\leftrightarrow4$ and $2\leftrightarrow3$ rates become much larger than the remaining four rates. The four-state system therefore separates dynamically into two rapidly mixed manifolds, $A=\{1,4\}$ and $B=\{2,3\}$. Transitions within each manifold occur rapidly, whereas transitions between the two manifolds are substantially slower for the parameters considered.

The manifolds can be characterized via an approximate conserved quantity of the full dot-tetron Hamiltonian. In the absence of intrawire MZM splittings, $E_M=E'_M=0$, the Hamiltonian conserves the fermion parity of the $X$-measurement loop, $\Pi_X=(-1)^{n_d}X=-\sigma_x\tau_z$, where the last equality uses Eqs.~(\ref{eq:full_pauli_identities}). The four-dimensional Hilbert space then separates into two two-dimensional sectors with $\Pi_X=\pm1$, which are exactly the manifolds $A$ and $B$. Dot-detuning noise preserves $\Pi_X$ and therefore drives transitions within these sectors, whereas finite MZM splittings weakly break this conservation law and hybridize the two sectors. At moderate detuning, dot-detuning noise produces rapid transitions within each approximate $\Pi_X$ sector. At large $|E_D|$, the increasing level separation, reduced charge hybridization, and thermal suppression of the upper states progressively freeze out these transitions, recovering the effective lowest-doublet description.

The strong separation of timescales at small to moderate $E_D$ can also have direct consequences for the capacitance readout. If transitions within a manifold are fast compared with the measurement integration time, the detector does not resolve the individual capacitances of its two states. Instead, it probes their thermally averaged response. Provided that switching between the manifolds remains slow enough to be resolved, the measured signal can therefore consist of two effective capacitance levels. Defining the conditional thermal weights within the two manifolds,
\begin{equation}
    p_{n|A}=\frac{e^{-E_n/k_BT}}
         {\sum_{m\in A}e^{-E_m/k_BT}},
    \qquad n\in A,
\end{equation}
and analogously for $B$, the corresponding average curvatures are
\begin{equation}
    \overline C_A=
    \sum_{n\in A}p_{n|A}C_n,
    \qquad\overline C_B=
    \sum_{n\in B}p_{n|B}C_n.
    \label{eq:manifold_capacitances}
\end{equation}
Here we have suppressed the parity index $\mathcal{P}$ and evaluate all averages in a fixed parity sector. The curvature difference relevant for distinguishing the two manifolds is therefore
\begin{equation}
    \Delta C_{\rm manif}
    =
    |\overline C_A-\overline C_B|.
    \label{eq:manifold_capacitance_diff}
\end{equation}

\begin{figure*}[tb]
	\centering
	\includegraphics[width=0.9\textwidth]{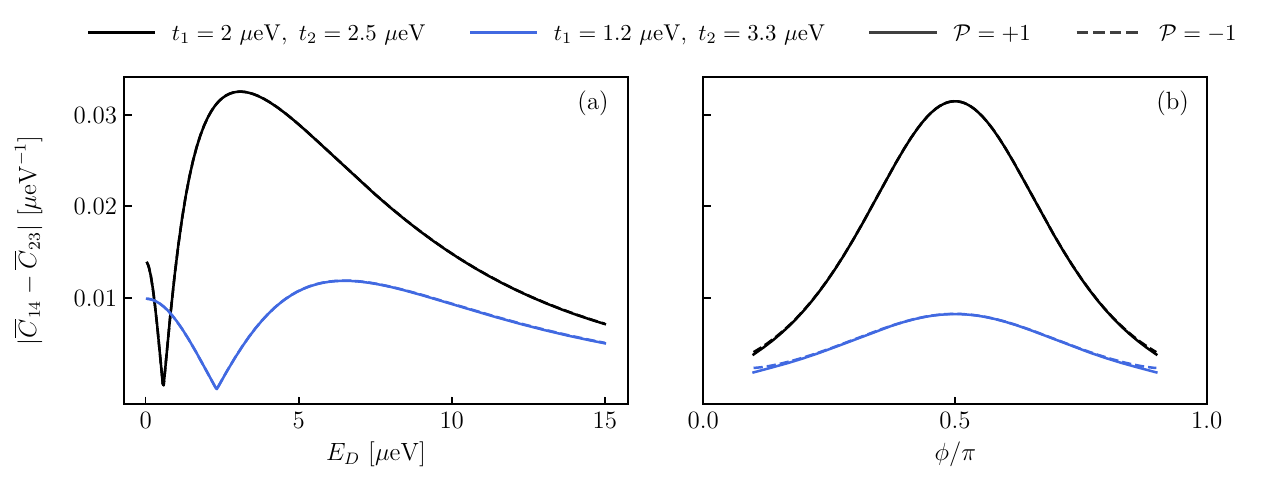}
	\caption{Thermally averaged curvature difference between the manifolds $A=\{1,4\}$ and $B=\{2,3\}$. The magnitude $\Delta C_{\rm manif}=|\overline C_A-\overline C_B|$, defined in Eq.~(\ref{eq:manifold_capacitance_diff}), is shown as a function of (a) dot detuning $E_D$ at $\phi=\pi/2$ and (b) phase $\phi$ at $E_D=4~\mu\mathrm{eV}$. Solid and dashed curves denote $\mathcal P=+1$ and $\mathcal P=-1$, respectively, and nearly overlap. Black curves are for $t_1=2~\mu\mathrm{eV}$, $t_2=2.5~\mu\mathrm{eV}$, while blue curves are for $t_1=1.2~\mu\mathrm{eV}$, $t_2=3.3~\mu\mathrm{eV}$. The other parameters are $E_M=0.05~\mu{\rm eV}$, $E'_M=0.15~\mu\mathrm{eV}$, and $T=50~\mathrm{mK}$ for all curves.}
\label{fig:manifold_capacitance}
\end{figure*}

Figure~\ref{fig:manifold_capacitance} shows that the two thermally averaged manifolds retain a finite capacitance contrast over the same near-resonant parameter regime in which the rate hierarchy of Fig.~\ref{fig:multilevel} is observed. At the representative point $E_D=4\,\mu\mathrm{eV}$, $\phi=\pi/2$, for the nearly balanced case $t_1=2~\mu\mathrm{eV}$, $t_2=2.5~\mu\mathrm{eV}$ and with the remaining parameters as given in the caption of Fig.~\ref{fig:manifold_capacitance}, we obtain $\Delta C_{\rm manif}\simeq0.03\,\mu\mathrm{eV}^{-1}$. Using the dimensionless gate lever arm $\alpha=0.15$ quoted in Ref.~\cite{aghaee2025distinctlifetimesxz}, we estimate the corresponding quantum-capacitance difference as $\Delta C_Q\simeq(e\alpha)^2\Delta C_{\rm manif}\simeq110\,\mathrm{aF}$, which is comparable to the experimentally reported signal and should therefore remain resolvable under similar readout conditions. Thus, rapid transitions within the two manifolds provide a possible explanation for how a four-state spectrum can give rise to a bimodal capacitance response for reasonable parameters.

The same averaging can provide an effective description of the slow switching dynamics. Assuming that each manifold reaches internal thermal equilibrium on a timescale short compared with transitions between the manifolds, the effective intermanifold transition rates are
\begin{align}
    \Gamma_{A\rightarrow B}&=\sum_{n\in A}
    p_{n|A}\sum_{m\in B}\Gamma_{n\rightarrow m},
    \\\Gamma_{B\rightarrow A}&=
    \sum_{n\in B} p_{n|B}\sum_{m\in A}
    \Gamma_{n\rightarrow m}.
    \label{eq:manifold_rates}
\end{align}
The corresponding dwell times are then    $\tau_A=1/\Gamma_{A\rightarrow B}$, $\tau_B=1/\Gamma_{B\rightarrow A}$, and we define the manifold lifetime as
\begin{equation}
    T_X^{\rm manif}=\frac{1}
    {\Gamma_{A\rightarrow B}+\Gamma_{B\rightarrow A}}.
    \label{eq:manifold_lifetime}
\end{equation}
We again note that, as in the two-state discussion, the lifetime definition in Ref.~\cite{aghaee2025distinctlifetimesxz} contains an additional factor of two compared to ours. For a more direct comparison, we also introduce $\tau_X^\mathrm{manif}=2T_X^{\rm manif}$.

\begin{figure}[tb]
	\centering
	\includegraphics[width=\columnwidth]{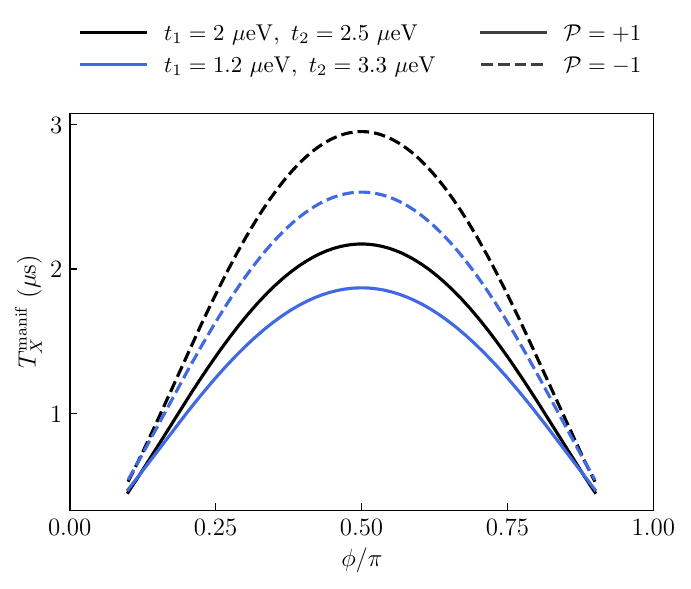}
	\caption{Phase dependence of the coarse-grained manifold-switching time defined in Eq.~(\ref{eq:manifold_lifetime}). Black curves correspond to $(t_1,t_2)=(2~\mu\mathrm{eV},2.5~\mu\mathrm{eV})$, while blue curves correspond to $(t_1,t_2)=(1.2~\mu\mathrm{eV},3.3~\mu\mathrm{eV})$; solid and dashed lines denote total parity $\mathcal P=+1$ and $\mathcal P=-1$, respectively. Other parameters are $E_D=4~\mu\mathrm{eV}$, $E_M=0.05~\mu{\rm eV}$, $E_M'=0.15~\mu\mathrm{eV}$, $T=50~\mathrm{mK}$, $A_D=10^{-13}~\mathrm{eV}^2$, and $A_M=A_M'=0.5\times10^{-17}~\mathrm{eV}^2$. Enhanced switching times occur around $\phi=\pi/2$. We note again that, in the lifetime convention of Ref.~\cite{aghaee2025distinctlifetimesxz}, the values shown here would be larger by a factor of two, $\tau_X^\mathrm{manif}=2T_X^{\rm manif}$.}
\label{fig:manifold_lifetime}
\end{figure}

Figure~\ref{fig:manifold_lifetime} shows the resulting manifold lifetimes as functions of the interferometric phase. The lifetimes are of order a few microseconds and are maximal around $\phi=\pi/2$, recovering the characteristic scale and phase dependence of the experimental switching dynamics at a qualitative level. For our representative parameters, Eq.~(\ref{eq:manifold_lifetime}) gives lifetimes $T_X^{\rm manif}$ on the order of a few $\mu$s. The manifold picture therefore preserves the main qualitative behavior seen in the doublet regime. However, we note that the parity and tunnel asymmetry dependence of the lifetimes obtained in the manifold regime is less pronounced than in the doublet regime.

Importantly, these microsecond switching times remain much longer than the intrinsic free-induction coherence times $\sim$ tens of $\mathrm{ns}$ obtained for the bare tetron in Sec.~\ref{sec:bare_lifetime} at the same MZM splitting noise strength. The enhancement can thus remain substantial even when all four states participate in the dynamics. The central conclusion of Secs.~\ref{sec:measurement_lifetime} and \ref{sec:bare_lifetime} therefore survives beyond the lowest-doublet approximation: coupling to the measurement dot can strongly suppress switching between the readout configurations, and the lifetime measured in this configuration need not reflect the intrinsic $X$-parity lifetime of the isolated tetron.

Finally, we note that the two-manifold description breaks down in the immediate vicinity of $\phi=0,\,\pi$, see the phase-dependent transition rates presented in Appendix~\ref{app:phi_dependence}. Neither the conditional manifold capacitances nor the manifold lifetimes can then be reliably interpreted as properties of two long-lived readout configurations. We have therefore restricted the phase dependence in Figs.~\ref{fig:manifold_capacitance}(b) and \ref{fig:manifold_lifetime} to an interval $0.1\pi\leq\phi\leq0.9\pi$, where the rate hierarchy supports the manifold picture for both tunnel-coupling pairs considered.

\section{Relation between the measured loop parity and $X$}
\label{sec:measurement_basis}

Above, we have focused on two complementary descriptions of the $X$-loop measurement. In the dispersive regime considered in Sec.~\ref{sec:measurement_lifetime}, the measurement dot can be integrated out and the low-energy dynamics reduces to an effective two-level system. Closer to dot resonance, on the other hand, all four low-energy states can participate in the dynamics, while rapid transitions organize them into the two effective manifolds discussed in Sec.~\ref{sec:detuning}. It is now interesting to discuss how the quantity distinguished by the measurement in these two regimes is related to the bare interwire parity $X$.

As discussed in Sec.~\ref{sec:detuning}, in the absence of intrawire MZM splittings, the full dot-tetron Hamiltonian conserves the $X$-loop parity $\Pi_X=(-1)^{n_d}X$. The four-dimensional Hilbert space then separates into two two-dimensional sectors with $\Pi_X=\pm1$, which correspond to the two manifolds $A$ and $B$ introduced in Sec.~\ref{sec:detuning}. Finite intrawire MZM splittings weakly break this conservation law and mix the two sectors. In the regime where this mixing is small, the two manifolds $A$ and $B$ can be viewed as approximate $X$-loop parity sectors.

The $X$-loop parity $\Pi_X$ is not identical to the bare interwire parity $X$ when the dot occupation fluctuates. In particular, tunneling between the dot and the tetron can change $X$ while preserving $\Pi_X$. Thus, close to resonance, where the eigenstates contain appreciable admixtures of both dot-charge configurations, the slow switching observed between the two manifolds reflects a switching of the $X$-loop parity rather than a direct switching of the bare $X$ parity. Only when the dot occupation is effectively fixed, $\Pi_X$ reduces to $X$ (up to an overall sign).

This identification can become less accurate for two related reasons that are relevant in both the multilevel and dispersive regimes. First, finite intrawire MZM splittings break the conservation of $\Pi_X$. Close to resonance, sufficiently large splittings therefore mix the two approximate $\Pi_X$ sectors and make the identification of the manifolds $A$ and $B$ with definite $X$-loop parity less precise. In the dispersive regime, the same effect appears as a residual splitting $B_z^0$ transverse to the measurement axis $B_x^0$, which rotates the measurement eigenstates away from the bare $X$ basis. The identification of the measured switching time with an $X$-parity lifetime is therefore most direct when $|B_x^0|\gg|B_z^0|$, which is the case we have focused on in Sec.~\ref{sec:measurement_lifetime}. If the two splittings become comparable, the switching between energy eigenstates should no longer be identified directly with switching of $X$.

Second, the strength of the measurement-induced coupling depends strongly on the interferometric phase. The operating point $\phi\simeq\pi/2$ maximizes the loop-induced coupling and also lies in the regime where the two-manifold structure discussed in Sec.~\ref{sec:detuning} is most clearly developed. If the effective phase is displaced substantially from this point, the manifold hierarchy can weaken or disappear altogether, as occurs near $\phi=0,\pi$. In the dispersive limit, the same phase dependence is reflected in $B_x^0(\phi)=t_{\rm eff}(\phi)\propto\sin\phi$, so that moving away from $\phi=\pi/2$ reduces the measurement-induced splitting relative to the residual MZM splittings.

We note that these effects do not qualitatively change the central stabilization mechanism discussed in Secs.~\ref{sec:measurement_lifetime}-\ref{sec:detuning}: a finite measurement-induced splitting can strongly suppress switching driven by low-frequency noise.  They do, however, affect how the resulting switching lifetime should be interpreted. In particular, for large residual MZM splittings, switching between the measured states can no longer be straightforwardly identified with switching of an $X$ or $X$-loop parity.

Finally, the relation between the measured $X$-loop parity in the multilevel regime and the bare $X$ parity also depends on how the measurement dot is decoupled following readout. If the dot is first detuned sufficiently far from charge resonance while remaining tunnel coupled, rapid relaxation within a fixed $\Pi_X$ manifold can drive the system toward a definite dot occupation. Once $n_d$ is effectively fixed, $\Pi_X=(-1)^{n_d}X$ maps onto the bare $X$ parity. However, for large-capacitance dots, detuning the dot after each measurement operation may be challenging in practice. By contrast, suppressing the tunnel coupling directly while remaining close to resonance need not provide a unique mapping between $\Pi_X$ and the bare $X$ parity. In the multilevel regime, both charge-hybridized states within a given $\Pi_X$ manifold can participate in the dynamics and evolve toward different dot occupations, corresponding to opposite values of $X$ for the same loop parity. A sudden switch-off can generally leave the dot occupation indefinite and need not prepare a definite bare $X$ parity. These considerations suggest that a controlled detuning step may be important for faithfully converting an $X$-loop measurement outcome into the parity state of the isolated tetron, provided that the protocol preserves the approximate loop-parity sector and avoids unwanted qubit rotation during decoupling.

\section{Conclusions}
\label{sec:conclusions}

In this work, we have studied the lifetime of the interwire $X$-parity in a Majorana tetron motivated by the recent experimental measurement of parity switching through the quantum capacitance of an auxiliary readout dot~\cite{aghaee2025distinctlifetimesxz}. Using a minimal model of the coupled dot–tetron system in the presence of electrostatic noise, we have analyzed both the dispersive regime, where the dot can be integrated out and the low-energy dynamics reduces to an effective two-level system, and the near-resonant regime, where all four low-energy states participate in the dynamics.

Our main result is that the coupling to the measurement dot can substantially modify the parity lifetime by shifting the noise-induced transitions to finite frequency, thereby strongly suppressing switching driven by low-frequency noise. This distinguishes the switching lifetime in the measurement configuration from a conventional coherence time such as $T_2^*$, which is usually viewed as intrinsic to the qubit. For representative MZM-splitting noise motivated by disordered nanowire devices, we find that switching times in the measurement configuration can remain of order microseconds even when the corresponding bare-tetron free-induction coherence time is only tens of nanoseconds. This stabilization persists beyond the dispersive limit. Close to dot resonance, the four-state system can instead separate dynamically into two long-lived manifolds connected by slow switching, providing an effective two-state description despite the underlying multilevel spectrum. The two manifolds approximately correspond to sectors of the combined dot–tetron $X$-loop parity when the intrawire MZM splittings are sufficiently small.

Our results also show that identifying the states whose switching is measured with the computational states of an idle Majorana qubit is not automatic. In the dispersive regime, this connection is relatively direct when the dot occupation is fixed and the measurement-induced coupling dominates the residual MZM splittings. Near resonance, however, the observed configurations are manifolds of charge-hybridized dot-tetron states and are more naturally associated with the $X$-loop parity than with $X$ alone. From the perspective of a measurement-based qubit architecture, it is therefore important to understand how the coupling to the readout dot is turned on and off. The ramp between the idle and measurement configurations determines how the well-defined qubit states of the isolated tetron map onto the eigenstates or dynamical manifolds of the coupled dot-tetron system. In particular, a protocol that first detunes the dot into a definite charge configuration before suppressing the tunnel coupling could provide a natural route for mapping the measured loop parity back onto the idle $X$ parity.

Our analysis assumes that the wire degrees of freedom relevant at low energies are adequately described by four MZMs, while allowing their residual hybridization to remain sensitive to electrostatic fluctuations. We thus consider the favorable scenario in which MZMs are present but the device has not yet reached a strongly exponentially protected regime. Additional low-energy states that may arise from disorder or smooth potential variations~\cite{bagrets2012class,kells2012nearzeroenergy,prada2012transport,Reeg2018,vuik2019reproducing, pan2020physical,dassarma2021disorderinduced,pan2021quantized,prada2020andreev} are not included in this description. Even within this favorable scenario, long parity-switching lifetimes during readout need not imply comparably long coherence times in the isolated tetron.

Whether the performance of a Majorana tetron is ultimately limited primarily by measurement-based operations or by idle coherence depends on the relative importance of readout-dot noise and fluctuations of the residual MZM splittings. If the time-dependent noise is dominated by fluctuations of the readout-dot potential while the MZM splittings remain small and essentially static, dot noise drives unwanted transitions during measurement, whereas the idle tetron can remain comparatively coherent. In particular, very long idle coherence times are expected for ideal MZMs in the exponentially localized regime. However, our results show that the opposite hierarchy is possible for non-ideal MZMs in disordered nanowires. Here, disorder can make the residual splitting sensitive to local potential variations, which converts electrostatic noise into fluctuations of the MZM splitting. These fluctuations can be significant even when the static splittings themselves are small at a particular operating point. Theoretical studies of closely related semiconductor-superconductor nanowire devices indicate that substantial disorder may be present even in relatively recent generations of Majorana nanowire devices~\cite{dassarma2023density,dassarma2023spectral}, which can generate appreciable MZM-splitting noise~\cite{Sau2025}. In this regime, fluctuations of the residual MZM splittings could become an important limitation on the idle $T_2^*$ even when the continuously measured $X$-parity switching time remains long due to measurement-induced stabilization.

Improving materials quality and reducing electrostatic disorder should consequently remain a central goal even when long switching lifetimes are observed under continuous readout. Reduced disorder generally improves MZM localization and suppresses unwanted hybridization, and also reduces the susceptibility of the idle qubit to electrostatic fluctuations. Furthermore, complementary readout schemes that reduce the measurement-induced coupling while retaining sensitivity to the qubit parity could provide a more direct probe of the intrinsic idle coherence time. Measurements of parity dynamics in different samples and under different dot couplings and detunings may ultimately help disentangle noise originating in the measurement circuit from noise intrinsic to the Majorana system, helping to clarify which timescale limits a tetron-based qubit in practice.

\section*{Acknowledgments}%
This work is supported by the Laboratory for Physical Sciences through the Condensed Matter Theory Center. We thank Gijs de Lange and Anton Akhmerov for insightful discussions that motivated this project.

\appendix

\section{Higher-order dot-noise relaxation in the odd-parity sector}
\label{app:dot_tetron}

In Sec.~\ref{sec:measurement_lifetime}, we showed that for equal intrawire MZM splittings, $E_M=E_M'$, the leading dot-detuning-noise contribution to the relaxation rate vanishes in the odd tetron-parity sector $p=-1$. Within the leading-order dispersive description, this follows from $B_{z,-}=0$, such that both the static measurement-induced coupling and its fluctuations generated by $\delta E_D$ point along the same effective-field axis. Here, we derive the leading nonzero correction to this result arising from higher-order virtual occupation of the measurement dot. We assume $E_D>0$ and work in the dispersive regime $E_D\gg |t_1|,|t_2|,|E_M|$, with $E_M=E_M'$, and retain only fluctuations of the dot detuning.

In the full dot-tetron Hamiltonian, the low-energy subspace of the
$\mathcal{P}=-1$ sector consists predominantly of the two states with an empty quantum dot. This low-energy manifold is coupled by tunneling to two virtual states with an occupied quantum dot with energies $E_D+2E_M$ and $E_D-2E_M$. After subtracting an irrelevant overall energy shift, we can write the Hamiltonian as $H=H_0+H_T$ with
\begin{equation}
H_0 =
\begin{pmatrix}
0 & 0 \\
0 & D
\end{pmatrix},
\qquad
H_T =
\begin{pmatrix}
0 & V^\dagger \\
V & 0
\end{pmatrix},
\label{eq:app_blockH}
\end{equation}
where $D=E_D\openone+2E_M\sigma_z$
for $E_M=E_M'$ and the tunneling matrix between the low- and high-energy subspaces is
\begin{equation}
V=\begin{pmatrix}
-t_2e^{i\phi} & t_1\\
t_1 & t_2e^{i\phi}
\end{pmatrix}=
t_1\sigma_x-t_2e^{i\phi}\sigma_z.
\end{equation}

We now perform a Schrieffer-Wolff transformation
$\widetilde H=e^SHe^{-S}$ with an anti-Hermitian generator $S$ chosen such that the coupling between the two subspaces vanishes to first order. For the block structure above, a convenient choice is
\begin{equation}
S=\begin{pmatrix}
0 & -V^\dagger D^{-1}\\
D^{-1}V & 0
\end{pmatrix},
\end{equation}
which satisfies $H_T+[S,H_0]=0$. Projecting the transformed Hamiltonian onto the low-energy subspace then gives the standard second-order result
\begin{equation}
H_{-,\mathrm{eff}}=-V^\dagger D^{-1}V.
\label{eq:app_SW_result}
\end{equation}
For $|E_M|\ll E_D$, we have
\begin{equation}
D^{-1}\simeq\frac{1}{E_D}\openone-
\frac{2E_M}{E_D^2}\sigma_z.
\label{eq:app_Dinv}
\end{equation}
Thus, we find
\begin{equation}
H_{-,\mathrm{eff}}\simeq-\frac{1}{E_D}M_0
+\frac{2E_M}{E_D^2}M_1,
\label{eq:app_Heff_next}
\end{equation}
where
\begin{align}
M_0&=(t_1^2+t_2^2)\openone
-2t_1t_2\sin\phi\,\sigma_y,\\
M_1&=-2t_1t_2\cos\phi\,\sigma_x
+(t_2^2-t_1^2)\sigma_z.
\end{align}
The first term in Eq.~\eqref{eq:app_Heff_next} reproduces the measurement-induced coupling
$2t_1t_2\sin\phi/E_D$, where the physical interwire-parity operator is represented by $\sigma_y$ in the occupation basis used here. This differs from the $\sigma_x$ convention adopted in the main text only by a relative phase choice for the two basis states; throughout, $X$ denotes the physical Majorana bilinear $i\gamma_1\gamma_2$. The second term in Eq.~\eqref{eq:app_Heff_next} introduces a small component of the
static effective field perpendicular to the leading measurement axis. We can thus write the effective field as
$\mathbf B = B_{\parallel}\,\hat{\mathbf y} + \mathbf B_{\perp}$,
where $B_{\parallel}=2t_1t_2\sin\phi /E_D$ is the leading
measurement-induced contribution and
\begin{equation}
\mathbf B_{\perp}=2E_M\left[-2t_1t_2\cos\phi\,\hat{\mathbf x}
+(t_2^2-t_1^2)\hat{\mathbf z}\right]/E_D^2
\end{equation}
is generated by the first
correction in Eq.~\eqref{eq:app_Heff_next}.

Fluctuations of the dot detuning generate a noise field
\begin{equation}
\mathbf K=\frac{\partial \mathbf B}{\partial E_D}
=-\frac{B_{\parallel}}{E_D}\hat{\mathbf y}
-\frac{2\mathbf B_{\perp}}{E_D}.
\end{equation}
The component of $\mathbf K$ transverse to the full static field $\mathbf{B}$ is
%
$\mathbf K_{\perp}=-\mathbf B_{\perp}/E_D$
%
to leading order in $|\mathbf B_{\perp}|/|B_{\parallel}|$. Thus, the relevant matrix element is
\begin{align}
|K_{+-}^{(-)}|^2&=|\langle +|\mathbf{K}_\perp\cdot\boldsymbol{\sigma}|-\rangle|^2\nonumber\\
	&=
	\frac{4E_M^2}{E_D^6}
	\left[
	(t_1^2-t_2^2)^2
	+4t_1^2t_2^2\cos^2\phi
	\right].
\label{eq:app_odd_rate}
\end{align}
This shows explicitly that the leading-order cancellation that led to $|K_{+-}^{(-)}|^2=0$ at lowest order is now lifted by tunnel-coupling asymmetry or by detuning the interferometer phase away from the point of maximal visibility.

To compare with Ref.~\cite{aghaee2025distinctlifetimesxz}, we introduce
\begin{equation}
t=\frac{t_1+t_2}{2},
\qquad
x=\frac{t_1-t_2}{2},
\qquad
\delta\phi=\phi-\frac{\pi}{2}.
\end{equation}
For $|x|\ll t$ and $|\delta\phi|\ll1$,
Eq.~\eqref{eq:app_odd_rate} becomes
\begin{equation}
|K_{+-}^{(-)}|^2
\simeq
\frac{16E_M^2t^2}{E_D^6}
\left(
4x^2+t^2\delta\phi^2
\right).
\label{eq:app_odd_Aghaee}
\end{equation}
Equation (\ref{eq:app_odd_Aghaee}) reproduces the large-$E_D$ result of Ref.~\cite{aghaee2025distinctlifetimesxz} after accounting for the different convention used for the noise spectral density.

Thus, in the dot-noise-only model, the simple two-level theory captures the leading transition rate when the two MZM splittings add. When they cancel, the leading projected theory predicts no relaxation, while the first correction due to virtual dot occupation restores the flux- and interferometer-asymmetry dependence found in the full dot-tetron calculation.

\begin{figure*}[tb]
	\centering
	\includegraphics[width=\textwidth]{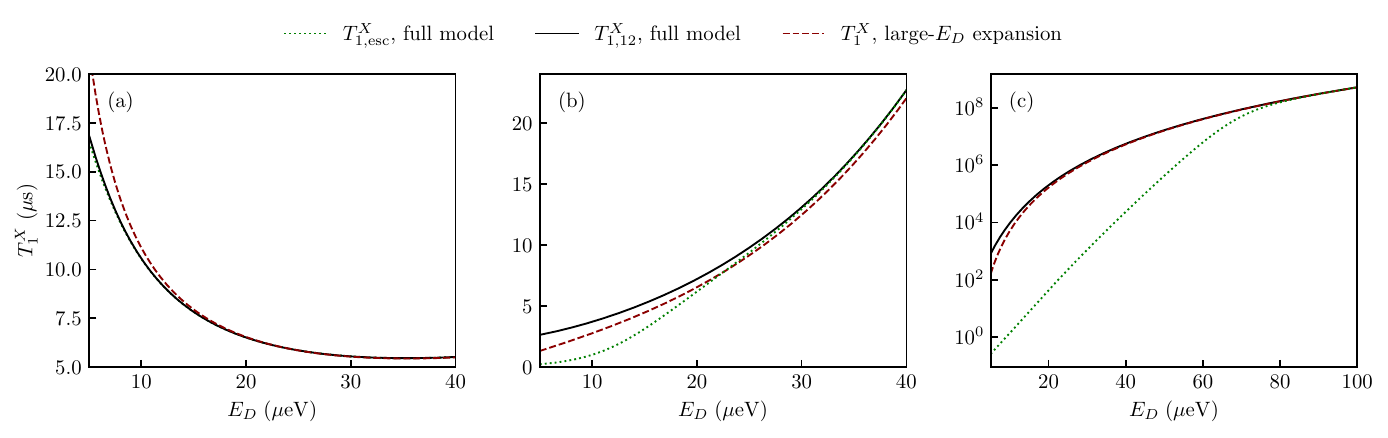}
	\caption{Comparison of the lowest-doublet relaxation time calculated using the full dot-tetron Hamiltonian (black solid curves), the corresponding analytical large-$E_D$ expressions (red dashed curves), and the escape-based timescale defined in Eq.~(\ref{eq:T1_escape}) (green dotted curves). (a) MZM-splitting noise only, with $A_M=A'_M=0.5\times10^{-17}\,\mathrm{eV}^2$ and $A_D=0$. The analytical result is the first term in Eq.~(\ref{eq:T1_1f_general}). (b) Dot-detuning noise only, with $A_D=10^{-13}\,\mathrm{eV}^2$ and $A_M=A'_M=0$. The analytical result is the second term in Eq.~(\ref{eq:T1_1f_general}). In (a) and (b), we have set $E_M=0.05\,\mu\mathrm{eV}$, $E'_M=0.15\,\mu\mathrm{eV}$, and $P=+1$. (c) Dot-detuning noise only, with the same noise amplitudes as in (b), but $E_M=E'_M=0.1\,\mu\mathrm{eV}$ and $P=-1$. Here the leading lowest-doublet contribution vanishes, and the analytical curve uses Eq.~(\ref{eq:T1_dot_higher}). All panels use $t_1=2\,\mu\mathrm{eV}$, $t_2=2.5\,\mu\mathrm{eV}$, $\phi=\pi/2$, and $T=50\,\mathrm{mK}$. The delayed convergence of the green and black curves in (c) shows that transitions to the upper states remain relevant relative to the strongly suppressed lowest-doublet rate over a wider detuning range.}
\label{fig:lifetimes_large_ED}
\end{figure*}

\section{Comparison between full dot-tetron model and large-$E_D$ approximation}
\label{app:comparison}

In the large-$E_D$ regime, the two lowest eigenstates of the full dot-tetron Hamiltonian reduce to the effective measurement doublet considered in Sec.~\ref{sec:measurement_lifetime}. In Fig.~\ref{fig:lifetimes_large_ED}, we plot the relaxation time defined in Eq.~(\ref{eq:T1_lowest_doublet}) obtained using the exact transition matrix elements and eigenenergies of the full four-state Hamiltonian, see the black solid curves, and compare it to the corresponding analytical dispersive results derived in Sec.~\ref{sec:measurement_lifetime}, see the red dashed curves. The agreement at large $E_D$ confirms that the projected two-level description correctly reproduces the transition rate within the lowest measurement doublet.

A separate question is whether transitions from the lowest
doublet to the upper two states can be neglected. To assess
this, we evaluate the total escape rates from each of the two
lowest eigenstates,
\begin{equation}
    \Gamma_n^{\rm esc}=\sum_{m\neq n}\Gamma_{n\to m},\qquad n=1,2,
\end{equation}
where the finite-temperature directional rates include the
detailed-balance factors introduced in Sec.~\ref{sec:detuning}. We then introduce the escape-based relaxation scale
\begin{equation}
    \frac{1}{T_{1,\rm esc}^X}=
    \Gamma_1^{\rm esc}+\Gamma_2^{\rm esc}.
    \label{eq:T1_escape}
\end{equation}
This quantity differs from $1/T^X_{1,12}$ by the sum of the
transition rates from states $1$ and $2$ to states $3$ and $4$.
Consequently, $T^X_{1,\mathrm{esc}}\leq T^X_{1,12}$, with
equality when these additional channels are absent.

In Fig.~\ref{fig:lifetimes_large_ED}, we show $T^X_{1,\mathrm{esc}}$ as green dotted curves. Their convergence toward the black curves indicates that excitation of the upper states becomes negligible relative to transitions within the lowest doublet. We emphasize that $T^X_{1,\mathrm{esc}}$ is a diagnostic of
the validity of restricting the dynamics to the lowest
doublet, rather than a relaxation time of the full four-state
system. In particular, an excursion into an upper state can
be followed by a return to the same readout manifold and need
not produce a switch between readout configurations.
Determining the observable switching dynamics generally
requires the full transition-rate matrix.

\begin{figure*}[tb]
	\centering
	\includegraphics[width=\textwidth]{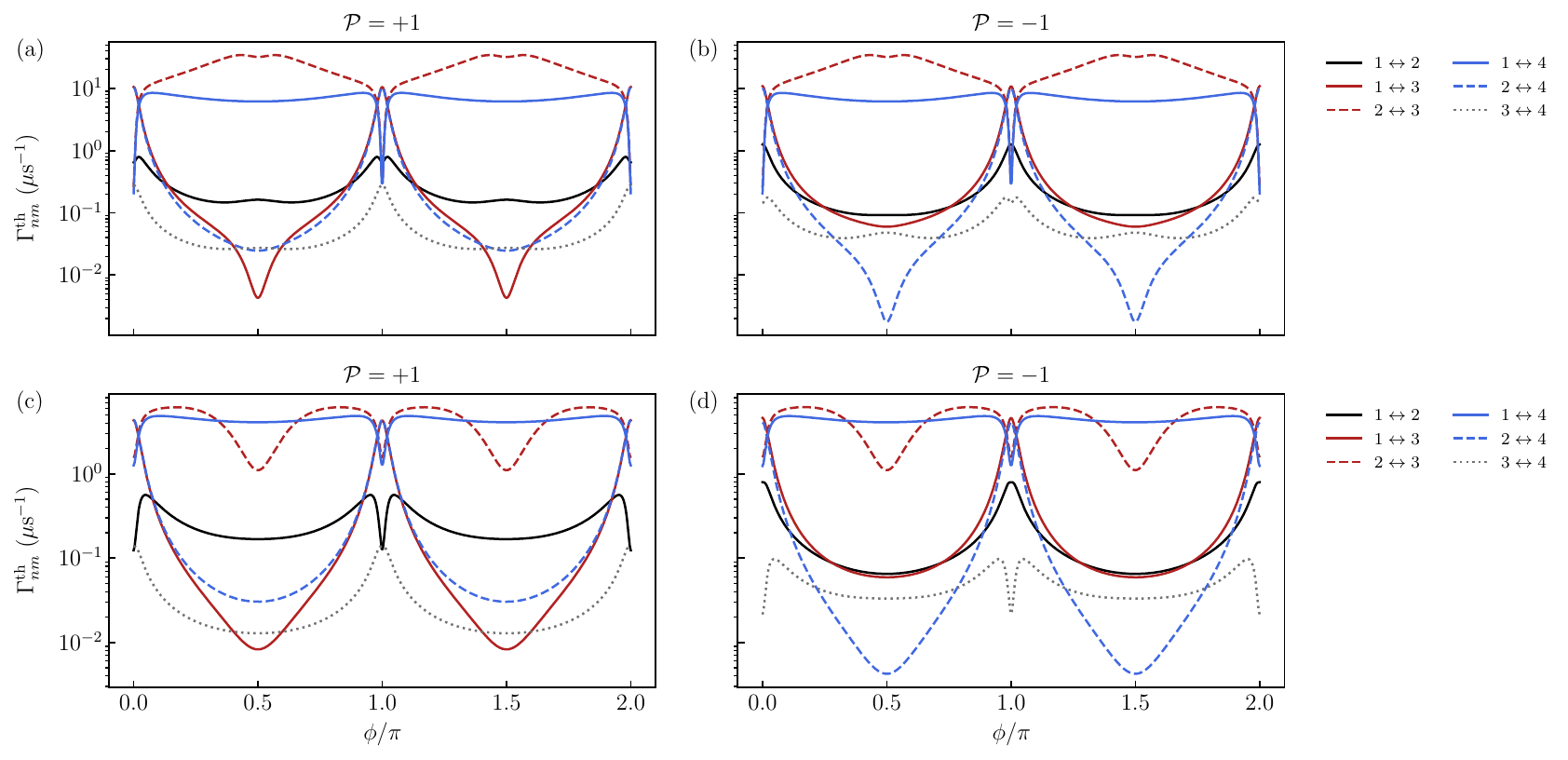}
	\caption{Equilibrium transition rates $\Gamma_{nm}^{\rm th}$, defined in Eq.~(\ref{eq:thermal_event_rate}), as a function of interferometer phase $\phi$ for total parity (a,c) $\mathcal P=+1$ and (b,d) $\mathcal P=-1$. The dot detuning is (a,b) $E_D=1~\mu$eV and (c,d) $E_D=4~\mu$eV. The two-manifold description $A=\{1,4\}$, $B=\{2,3\}$ discussed in the main text remains valid in a relatively large range around $\phi=\pi/2$, whereas it breaks down in the immediate vicinity of $\phi=0,\,\pi$. For small dot detunings, an alternative manifold picture corresponding to $A'=\{1,3\}$ and $B'=\{2,4\}$ emerges in the latter regions. The other system parameters are the same as in Fig.~\ref{fig:multilevel}.}
\label{fig:multilevel_phi}
\end{figure*}

For the generic cases shown in Figs.~\ref{fig:lifetimes_large_ED}(a) and \ref{fig:lifetimes_large_ED}(b), the
escape-based timescale approaches the lowest-doublet result
relatively rapidly as $E_D$ increases. Once the upper two states are energetically separated from the measurement doublet, their contribution to the total escape rate becomes negligible, and $T_{1,\rm esc}^X$ closely follows the lowest-doublet result.  The situation is different in the equal-splitting odd-parity case shown in Fig.~\ref{fig:lifetimes_large_ED}(c). There, the leading dot-noise matrix element within the lowest doublet vanishes, leaving only the small higher-order contribution derived in Appendix~\ref{app:dot_tetron}. Transitions to the upper states can therefore remain significant relative to the lowest-doublet rate even when they are strongly suppressed in absolute magnitude. The escape-based timescale therefore only approaches the lowest-doublet timescales at substantially larger detuning.

\section{$\phi$-dependence of multilevel transition rates}
\label{app:phi_dependence}

In Fig.~\ref{fig:multilevel_phi}, we show the equilibrium transition rates defined in Eq.~(\ref{eq:thermal_event_rate}) as a function of the interferometer phase $\phi$. The two-manifold picture $A=\{1,4\}$, $B=\{2,3\}$ discussed in the main text remains valid over a relatively broad range around $\phi=\pi/2$, whereas it breaks down in the immediate vicinity of $\phi=0,\pi$. The lifetime calculation presented in the main text is therefore not intended to describe these narrow phase regions.

The transition-rate hierarchy near $\phi=0,\pi$ can, however, reorganize in a different way. For sufficiently small dot detuning, the dominant transitions can instead group the states into the alternative manifolds $A'=\{1,3\}$ and $B'=\{2,4\}$, as illustrated in Figs.~\ref{fig:multilevel_phi}(a) and \ref{fig:multilevel_phi}(b). This alternative coarse graining is not generically realized throughout parameter space, but it shows that the breakdown of the manifold structure used in the main text does not necessarily imply the absence of an effective two-state dynamics.

\begin{figure}[tb]
	\centering
	\includegraphics[width=\columnwidth]{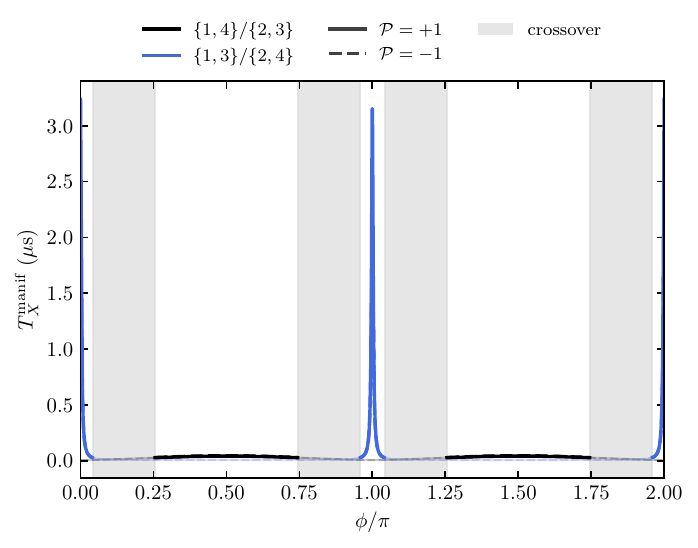}
	\caption{Effective switching lifetimes obtained from the two possible manifold decompositions as a function of interferometer phase for the representative parameters indicated in the figure. Black curves correspond to the manifold structure $A=\{1,4\}$, $B=\{2,3\}$ used in the main text, while blue curves correspond to the alternative structure $A'=\{1,3\}$, $B'=\{2,4\}$. Solid and dashed lines denote total parity $\mathcal P=+1$ and $-1$, respectively. Gray regions indicate crossover intervals in which neither manifold hierarchy is well separated. Narrow lifetime enhancements appear near $\phi=0,\pi$, where the alternative manifold structure becomes relevant. We have used the parameters $t_1=0.5~\mu$eV, $t_2=0.75~\mu$eV, $E_D=1~\mu$eV, $E_M=E_M'=0.12~\mu$eV, $A_D=10^{-13}~{\rm eV}^2$, $A_M=A_M'=0.5\times 10^{-17}~{\rm eV}^2$, and $T=50$~mK.}
\label{fig:manifold_lifetime_alt}
\end{figure}

Figure~\ref{fig:manifold_lifetime_alt} illustrates a parameter regime in which this alternative manifold structure gives rise to narrow enhancements of the effective switching lifetime near $\phi=0,\pi$. Away from these regions, the original $A=\{1,4\}$, $B=\{2,3\}$ description again provides the appropriate coarse graining. The resulting phase dependence is therefore qualitatively different from the broad maximum around $\phi=\pi/2$ discussed in the main text and arises from a change in the underlying transition-rate hierarchy rather than from the same stabilization mechanism within a fixed pair of manifolds.

The capacitance contrast is generally reduced near $\phi=0,\pi$, but it need not vanish. For the representative parameters of Fig.~\ref{fig:manifold_lifetime_alt}, we find that the corresponding difference between the thermally averaged manifold capacitances remains of order a few tens of aF in the vicinity of the sharp lifetime features, such that the two effective configurations can in principle remain distinguishable. We emphasize that these peaks are centered at $\phi=0,\pi$, rather than at the experimentally observed locations near $\phi=\pi/2,3\pi/2$. Nevertheless, if the relation between the applied magnetic field and the effective interferometer phase includes an unaccounted offset, features occurring near $\phi=0,\pi$ in the model could appear at shifted values of the experimentally controlled field.

\bibliography{Majorana.bib}

\begin{thebibliography}{40}%
\makeatletter
\providecommand \@ifxundefined [1]{%
 \@ifx{#1\undefined}
}%
\providecommand \@ifnum [1]{%
 \ifnum #1\expandafter \@firstoftwo
 \else \expandafter \@secondoftwo
 \fi
}%
\providecommand \@ifx [1]{%
 \ifx #1\expandafter \@firstoftwo
 \else \expandafter \@secondoftwo
 \fi
}%
\providecommand \natexlab [1]{#1}%
\providecommand \enquote  [1]{``#1''}%
\providecommand \bibnamefont  [1]{#1}%
\providecommand \bibfnamefont [1]{#1}%
\providecommand \citenamefont [1]{#1}%
\providecommand \href@noop [0]{\@secondoftwo}%
\providecommand \href [0]{\begingroup \@sanitize@url \@href}%
\providecommand \@href[1]{\@@startlink{#1}\@@href}%
\providecommand \@@href[1]{\endgroup#1\@@endlink}%
\providecommand \@sanitize@url [0]{\catcode `\\12\catcode `\$12\catcode
  `\&12\catcode `\#12\catcode `\^12\catcode `\_12\catcode `\%12\relax}%
\providecommand \@@startlink[1]{}%
\providecommand \@@endlink[0]{}%
\providecommand \url  [0]{\begingroup\@sanitize@url \@url }%
\providecommand \@url [1]{\endgroup\@href {#1}{\urlprefix }}%
\providecommand \urlprefix  [0]{URL }%
\providecommand \Eprint [0]{\href }%
\providecommand \doibase [0]{https://doi.org/}%
\providecommand \selectlanguage [0]{\@gobble}%
\providecommand \bibinfo  [0]{\@secondoftwo}%
\providecommand \bibfield  [0]{\@secondoftwo}%
\providecommand \translation [1]{[#1]}%
\providecommand \BibitemOpen [0]{}%
\providecommand \bibitemStop [0]{}%
\providecommand \bibitemNoStop [0]{.\EOS\space}%
\providecommand \EOS [0]{\spacefactor3000\relax}%
\providecommand \BibitemShut  [1]{\csname bibitem#1\endcsname}%
\let\auto@bib@innerbib\@empty
\bibitem [{\citenamefont {Kitaev}(2001)}]{kitaev2001unpaired}%
  \BibitemOpen
  \bibfield  {author} {\bibinfo {author} {\bibfnamefont {A.~Y.}\ \bibnamefont
  {Kitaev}},\ }\bibfield  {title} {\bibinfo {title} {Unpaired {{Majorana}}
  fermions in quantum wires},\ }\href
  {https://doi.org/10.1070%2F1063-7869%2F44%2F10s%2Fs29} {\bibfield  {journal}
  {\bibinfo  {journal} {Phys.-Usp.}\ }\textbf {\bibinfo {volume} {44}},\
  \bibinfo {pages} {131} (\bibinfo {year} {2001})}\BibitemShut {NoStop}%
\bibitem [{\citenamefont {Nayak}\ \emph {et~al.}(2008)\citenamefont {Nayak},
  \citenamefont {Simon}, \citenamefont {Stern}, \citenamefont {Freedman},\ and\
  \citenamefont {Das~Sarma}}]{nayak2008nonabelian}%
  \BibitemOpen
  \bibfield  {author} {\bibinfo {author} {\bibfnamefont {C.}~\bibnamefont
  {Nayak}}, \bibinfo {author} {\bibfnamefont {S.~H.}\ \bibnamefont {Simon}},
  \bibinfo {author} {\bibfnamefont {A.}~\bibnamefont {Stern}}, \bibinfo
  {author} {\bibfnamefont {M.}~\bibnamefont {Freedman}},\ and\ \bibinfo
  {author} {\bibfnamefont {S.}~\bibnamefont {Das~Sarma}},\ }\bibfield  {title}
  {\bibinfo {title} {Non-abelian anyons and topological quantum computation},\
  }\href {https://doi.org/10.1103/RevModPhys.80.1083} {\bibfield  {journal}
  {\bibinfo  {journal} {Rev. Mod. Phys.}\ }\textbf {\bibinfo {volume} {80}},\
  \bibinfo {pages} {1083} (\bibinfo {year} {2008})}\BibitemShut {NoStop}%
\bibitem [{\citenamefont {Das~Sarma}\ \emph {et~al.}(2015)\citenamefont
  {Das~Sarma}, \citenamefont {Freedman},\ and\ \citenamefont
  {Nayak}}]{sarma2015majorana}%
  \BibitemOpen
  \bibfield  {author} {\bibinfo {author} {\bibfnamefont {S.}~\bibnamefont
  {Das~Sarma}}, \bibinfo {author} {\bibfnamefont {M.}~\bibnamefont
  {Freedman}},\ and\ \bibinfo {author} {\bibfnamefont {C.}~\bibnamefont
  {Nayak}},\ }\bibfield  {title} {\bibinfo {title} {Majorana zero modes and
  topological quantum computation},\ }\href
  {https://www.nature.com/articles/npjqi20151} {\bibfield  {journal} {\bibinfo
  {journal} {npj Quantum Inf.}\ }\textbf {\bibinfo {volume} {1}},\ \bibinfo
  {pages} {15001} (\bibinfo {year} {2015})}\BibitemShut {NoStop}%
\bibitem [{\citenamefont {Lutchyn}\ \emph {et~al.}(2010)\citenamefont
  {Lutchyn}, \citenamefont {Sau},\ and\ \citenamefont
  {Das~Sarma}}]{lutchyn2010majorana}%
  \BibitemOpen
  \bibfield  {author} {\bibinfo {author} {\bibfnamefont {R.~M.}\ \bibnamefont
  {Lutchyn}}, \bibinfo {author} {\bibfnamefont {J.~D.}\ \bibnamefont {Sau}},\
  and\ \bibinfo {author} {\bibfnamefont {S.}~\bibnamefont {Das~Sarma}},\
  }\bibfield  {title} {\bibinfo {title} {Majorana fermions and a topological
  phase transition in semiconductor-superconductor heterostructures},\ }\href
  {https://link.aps.org/doi/10.1103/PhysRevLett.105.077001} {\bibfield
  {journal} {\bibinfo  {journal} {Phys. Rev. Lett.}\ }\textbf {\bibinfo
  {volume} {105}},\ \bibinfo {pages} {077001} (\bibinfo {year}
  {2010})}\BibitemShut {NoStop}%
\bibitem [{\citenamefont {Sau}\ \emph {et~al.}(2010{\natexlab{a}})\citenamefont
  {Sau}, \citenamefont {Lutchyn}, \citenamefont {Tewari},\ and\ \citenamefont
  {Das~Sarma}}]{sau2010generic}%
  \BibitemOpen
  \bibfield  {author} {\bibinfo {author} {\bibfnamefont {J.~D.}\ \bibnamefont
  {Sau}}, \bibinfo {author} {\bibfnamefont {R.~M.}\ \bibnamefont {Lutchyn}},
  \bibinfo {author} {\bibfnamefont {S.}~\bibnamefont {Tewari}},\ and\ \bibinfo
  {author} {\bibfnamefont {S.}~\bibnamefont {Das~Sarma}},\ }\bibfield  {title}
  {\bibinfo {title} {Generic new platform for topological quantum computation
  using semiconductor heterostructures},\ }\href
  {https://link.aps.org/doi/10.1103/PhysRevLett.104.040502} {\bibfield
  {journal} {\bibinfo  {journal} {Phys. Rev. Lett.}\ }\textbf {\bibinfo
  {volume} {104}},\ \bibinfo {pages} {040502} (\bibinfo {year}
  {2010}{\natexlab{a}})}\BibitemShut {NoStop}%
\bibitem [{\citenamefont {Sau}\ \emph {et~al.}(2010{\natexlab{b}})\citenamefont
  {Sau}, \citenamefont {Tewari}, \citenamefont {Lutchyn}, \citenamefont
  {Stanescu},\ and\ \citenamefont {Das~Sarma}}]{sau2010nonabelian}%
  \BibitemOpen
  \bibfield  {author} {\bibinfo {author} {\bibfnamefont {J.~D.}\ \bibnamefont
  {Sau}}, \bibinfo {author} {\bibfnamefont {S.}~\bibnamefont {Tewari}},
  \bibinfo {author} {\bibfnamefont {R.~M.}\ \bibnamefont {Lutchyn}}, \bibinfo
  {author} {\bibfnamefont {T.~D.}\ \bibnamefont {Stanescu}},\ and\ \bibinfo
  {author} {\bibfnamefont {S.}~\bibnamefont {Das~Sarma}},\ }\bibfield  {title}
  {\bibinfo {title} {Non-{{Abelian}} quantum order in spin-orbit-coupled
  semiconductors: {{Search}} for topological {{Majorana}} particles in
  solid-state systems},\ }\href
  {https://link.aps.org/doi/10.1103/PhysRevB.82.214509} {\bibfield  {journal}
  {\bibinfo  {journal} {Phys. Rev. B}\ }\textbf {\bibinfo {volume} {82}},\
  \bibinfo {pages} {214509} (\bibinfo {year} {2010}{\natexlab{b}})}\BibitemShut
  {NoStop}%
\bibitem [{\citenamefont {Oreg}\ \emph {et~al.}(2010)\citenamefont {Oreg},
  \citenamefont {Refael},\ and\ \citenamefont {{von Oppen}}}]{oreg2010helical}%
  \BibitemOpen
  \bibfield  {author} {\bibinfo {author} {\bibfnamefont {Y.}~\bibnamefont
  {Oreg}}, \bibinfo {author} {\bibfnamefont {G.}~\bibnamefont {Refael}},\ and\
  \bibinfo {author} {\bibfnamefont {F.}~\bibnamefont {{von Oppen}}},\
  }\bibfield  {title} {\bibinfo {title} {Helical liquids and {Majorana} bound
  states in quantum wires},\ }\href
  {http://link.aps.org/doi/10.1103/PhysRevLett.105.177002} {\bibfield
  {journal} {\bibinfo  {journal} {Phys. Rev. Lett.}\ }\textbf {\bibinfo
  {volume} {105}},\ \bibinfo {pages} {177002} (\bibinfo {year}
  {2010})}\BibitemShut {NoStop}%
\bibitem [{\citenamefont {Das~Sarma}(2023)}]{dassarma2023search}%
  \BibitemOpen
  \bibfield  {author} {\bibinfo {author} {\bibfnamefont {S.}~\bibnamefont
  {Das~Sarma}},\ }\bibfield  {title} {\bibinfo {title} {In search of
  {{Majorana}}},\ }\href {https://www.nature.com/articles/s41567-022-01900-9}
  {\bibfield  {journal} {\bibinfo  {journal} {Nat. Phys.}\ }\textbf {\bibinfo
  {volume} {19}},\ \bibinfo {pages} {165} (\bibinfo {year} {2023})}\BibitemShut
  {NoStop}%
\bibitem [{\citenamefont {Kouwenhoven}(2025)}]{kouwenhoven2025perspective}%
  \BibitemOpen
  \bibfield  {author} {\bibinfo {author} {\bibfnamefont {L.}~\bibnamefont
  {Kouwenhoven}},\ }\bibfield  {title} {\bibinfo {title} {Perspective on
  {Majorana} bound-states in hybrid superconductor-semiconductor nanowires},\
  }\href {https://doi.org/10.1142/S0217984925400020} {\bibfield  {journal}
  {\bibinfo  {journal} {Modern Physics Letters B}\ }\textbf {\bibinfo {volume}
  {39}},\ \bibinfo {pages} {2540002} (\bibinfo {year} {2025})}\BibitemShut
  {NoStop}%
\bibitem [{\citenamefont {Aasen}\ \emph {et~al.}(2025)\citenamefont {Aasen},
  \citenamefont {Aghaee}, \citenamefont {Alam}, \citenamefont {Andrzejczuk},
  \citenamefont {Antipov}, \citenamefont {Astafev}, \citenamefont {Avilovas},
  \citenamefont {Barzegar}, \citenamefont {Bauer}, \citenamefont {Becker},
  \citenamefont {Bello-Rivas}, \citenamefont {Bhaskar}, \citenamefont
  {Bocharov}, \citenamefont {Boddapati}, \citenamefont {Bohn}, \citenamefont
  {Bommer}, \citenamefont {Bonderson}, \citenamefont {Borovsky}, \citenamefont
  {Bourdet}, \citenamefont {Boutin}, \citenamefont {Brown}, \citenamefont
  {Campbell}, \citenamefont {Casparis}, \citenamefont {Chakravarthi},
  \citenamefont {Chao}, \citenamefont {Chapman}, \citenamefont {Chatoor},
  \citenamefont {Christensen}, \citenamefont {Codd}, \citenamefont {Cole},
  \citenamefont {Cooper}, \citenamefont {Corsetti}, \citenamefont {Cui},
  \citenamefont {van Dam}, \citenamefont {Dandachi}, \citenamefont
  {Daraeizadeh}, \citenamefont {Dumitrascu}, \citenamefont {Ekefj{\"a}rd},
  \citenamefont {Fallahi}, \citenamefont {Galletti}, \citenamefont {Gardner},
  \citenamefont {Gatta}, \citenamefont {Gavranovic}, \citenamefont {Goulding},
  \citenamefont {Govender}, \citenamefont {Griggio}, \citenamefont {Grigoryan},
  \citenamefont {Grijalva}, \citenamefont {Gronin}, \citenamefont
  {Gukelberger}, \citenamefont {Haah}, \citenamefont {Hamdast}, \citenamefont
  {Hansen}, \citenamefont {Hastings}, \citenamefont {Heedt}, \citenamefont
  {Ho}, \citenamefont {Hogaboam}, \citenamefont {Holgaard}, \citenamefont
  {Van~Hoogdalem}, \citenamefont {Indrapiromkul}, \citenamefont {Ingerslev},
  \citenamefont {Ivancevic}, \citenamefont {Jablonski}, \citenamefont {Jensen},
  \citenamefont {Jhoja}, \citenamefont {Jones}, \citenamefont {Kalashnikov},
  \citenamefont {Kallaher}, \citenamefont {Kalra}, \citenamefont {Karimi},
  \citenamefont {Karzig}, \citenamefont {Kimes}, \citenamefont {Kliuchnikov},
  \citenamefont {Kloster}, \citenamefont {Knapp}, \citenamefont {Knee},
  \citenamefont {Koski}, \citenamefont {Kostamo}, \citenamefont {Kuesel},
  \citenamefont {Lackey}, \citenamefont {Laeven}, \citenamefont {Lai},
  \citenamefont {de~Lange}, \citenamefont {Larsen}, \citenamefont {Lee},
  \citenamefont {Lee}, \citenamefont {Leum}, \citenamefont {Li}, \citenamefont
  {Lindemann}, \citenamefont {Lucas}, \citenamefont {Lutchyn}, \citenamefont
  {Madsen}, \citenamefont {Madulid}, \citenamefont {Manfra}, \citenamefont
  {Markussen}, \citenamefont {Martinez}, \citenamefont {Mattila}, \citenamefont
  {Mattinson}, \citenamefont {McNeil}, \citenamefont {Mei}, \citenamefont
  {Mishmash}, \citenamefont {Mohandas}, \citenamefont {Mollgaard},
  \citenamefont {de~Moor}, \citenamefont {Morgan}, \citenamefont {Moussa},
  \citenamefont {Narla}, \citenamefont {Nayak}, \citenamefont {Nielsen},
  \citenamefont {Nielsen}, \citenamefont {Nolet}, \citenamefont {Nystrom},
  \citenamefont {O'Farrell}, \citenamefont {Otani}, \citenamefont {Paetznick},
  \citenamefont {Papon}, \citenamefont {Paz}, \citenamefont {Petersson},
  \citenamefont {Petit}, \citenamefont {Pikulin}, \citenamefont {Pons},
  \citenamefont {Quinn}, \citenamefont {Rajpalke}, \citenamefont {Ramirez},
  \citenamefont {Rasmussen}, \citenamefont {Razmadze}, \citenamefont
  {Reichardt}, \citenamefont {Ren}, \citenamefont {Reneris}, \citenamefont
  {Riccomini}, \citenamefont {Sadovskyy}, \citenamefont {Sainiemi},
  \citenamefont {Salda{\~n}a}, \citenamefont {Sanlorenzo}, \citenamefont
  {Schaal}, \citenamefont {Schmidgall}, \citenamefont {Sfiligoj}, \citenamefont
  {da~Silva}, \citenamefont {Singh}, \citenamefont {Sinha}, \citenamefont
  {Soeken}, \citenamefont {Sohr}, \citenamefont {Stankevic}, \citenamefont
  {Stek}, \citenamefont {Str{\o{}}m-Hansen}, \citenamefont {Stuppard},
  \citenamefont {Sundaram}, \citenamefont {Suominen}, \citenamefont {Suter},
  \citenamefont {Suzuki}, \citenamefont {Svore}, \citenamefont {Teicher},
  \citenamefont {Thiyagarajah}, \citenamefont {Tholapi}, \citenamefont
  {Thomas}, \citenamefont {Tom}, \citenamefont {Toomey}, \citenamefont {Tracy},
  \citenamefont {Troyer}, \citenamefont {Turley}, \citenamefont {Turner},
  \citenamefont {Upadhyay}, \citenamefont {Urban}, \citenamefont {Vaschillo},
  \citenamefont {Viazmitinov}, \citenamefont {Vogel}, \citenamefont {Wang},
  \citenamefont {Watson}, \citenamefont {Webster}, \citenamefont {Weston},
  \citenamefont {Williamson}, \citenamefont {Winkler}, \citenamefont {van
  Woerkom}, \citenamefont {W{\"u}tz}, \citenamefont {Yang}, \citenamefont {Yu},
  \citenamefont {Yucelen}, \citenamefont {Zamorano}, \citenamefont {Zeisel},
  \citenamefont {Zheng}, \citenamefont {Zilke},\ and\ \citenamefont
  {Zimmerman}}]{Aasen2025}%
  \BibitemOpen
  \bibfield  {author} {\bibinfo {author} {\bibfnamefont {D.}~\bibnamefont
  {Aasen}}, \bibinfo {author} {\bibfnamefont {M.}~\bibnamefont {Aghaee}},
  \bibinfo {author} {\bibfnamefont {Z.}~\bibnamefont {Alam}}, \bibinfo {author}
  {\bibfnamefont {M.}~\bibnamefont {Andrzejczuk}}, \bibinfo {author}
  {\bibfnamefont {A.}~\bibnamefont {Antipov}}, \bibinfo {author} {\bibfnamefont
  {M.}~\bibnamefont {Astafev}}, \bibinfo {author} {\bibfnamefont
  {L.}~\bibnamefont {Avilovas}}, \bibinfo {author} {\bibfnamefont
  {A.}~\bibnamefont {Barzegar}}, \bibinfo {author} {\bibfnamefont
  {B.}~\bibnamefont {Bauer}}, \bibinfo {author} {\bibfnamefont
  {J.}~\bibnamefont {Becker}}, \bibinfo {author} {\bibfnamefont {J.~M.}\
  \bibnamefont {Bello-Rivas}}, \bibinfo {author} {\bibfnamefont
  {U.}~\bibnamefont {Bhaskar}}, \bibinfo {author} {\bibfnamefont
  {A.}~\bibnamefont {Bocharov}}, \bibinfo {author} {\bibfnamefont
  {S.}~\bibnamefont {Boddapati}}, \bibinfo {author} {\bibfnamefont
  {D.}~\bibnamefont {Bohn}}, \bibinfo {author} {\bibfnamefont {J.}~\bibnamefont
  {Bommer}}, \bibinfo {author} {\bibfnamefont {P.}~\bibnamefont {Bonderson}},
  \bibinfo {author} {\bibfnamefont {J.}~\bibnamefont {Borovsky}}, \bibinfo
  {author} {\bibfnamefont {L.}~\bibnamefont {Bourdet}}, \bibinfo {author}
  {\bibfnamefont {S.}~\bibnamefont {Boutin}}, \bibinfo {author} {\bibfnamefont
  {T.}~\bibnamefont {Brown}}, \bibinfo {author} {\bibfnamefont
  {G.}~\bibnamefont {Campbell}}, \bibinfo {author} {\bibfnamefont
  {L.}~\bibnamefont {Casparis}}, \bibinfo {author} {\bibfnamefont
  {S.}~\bibnamefont {Chakravarthi}}, \bibinfo {author} {\bibfnamefont
  {R.}~\bibnamefont {Chao}}, \bibinfo {author} {\bibfnamefont {B.~J.}\
  \bibnamefont {Chapman}}, \bibinfo {author} {\bibfnamefont {S.}~\bibnamefont
  {Chatoor}}, \bibinfo {author} {\bibfnamefont {A.~W.}\ \bibnamefont
  {Christensen}}, \bibinfo {author} {\bibfnamefont {P.}~\bibnamefont {Codd}},
  \bibinfo {author} {\bibfnamefont {W.}~\bibnamefont {Cole}}, \bibinfo {author}
  {\bibfnamefont {P.}~\bibnamefont {Cooper}}, \bibinfo {author} {\bibfnamefont
  {F.}~\bibnamefont {Corsetti}}, \bibinfo {author} {\bibfnamefont
  {A.}~\bibnamefont {Cui}}, \bibinfo {author} {\bibfnamefont {W.}~\bibnamefont
  {van Dam}}, \bibinfo {author} {\bibfnamefont {T.~E.}\ \bibnamefont
  {Dandachi}}, \bibinfo {author} {\bibfnamefont {S.}~\bibnamefont
  {Daraeizadeh}}, \bibinfo {author} {\bibfnamefont {A.}~\bibnamefont
  {Dumitrascu}}, \bibinfo {author} {\bibfnamefont {A.}~\bibnamefont
  {Ekefj{\"a}rd}}, \bibinfo {author} {\bibfnamefont {S.}~\bibnamefont
  {Fallahi}}, \bibinfo {author} {\bibfnamefont {L.}~\bibnamefont {Galletti}},
  \bibinfo {author} {\bibfnamefont {G.}~\bibnamefont {Gardner}}, \bibinfo
  {author} {\bibfnamefont {R.}~\bibnamefont {Gatta}}, \bibinfo {author}
  {\bibfnamefont {H.}~\bibnamefont {Gavranovic}}, \bibinfo {author}
  {\bibfnamefont {M.}~\bibnamefont {Goulding}}, \bibinfo {author}
  {\bibfnamefont {D.}~\bibnamefont {Govender}}, \bibinfo {author}
  {\bibfnamefont {F.}~\bibnamefont {Griggio}}, \bibinfo {author} {\bibfnamefont
  {R.}~\bibnamefont {Grigoryan}}, \bibinfo {author} {\bibfnamefont
  {S.}~\bibnamefont {Grijalva}}, \bibinfo {author} {\bibfnamefont
  {S.}~\bibnamefont {Gronin}}, \bibinfo {author} {\bibfnamefont
  {J.}~\bibnamefont {Gukelberger}}, \bibinfo {author} {\bibfnamefont
  {J.}~\bibnamefont {Haah}}, \bibinfo {author} {\bibfnamefont {M.}~\bibnamefont
  {Hamdast}}, \bibinfo {author} {\bibfnamefont {E.~B.}\ \bibnamefont {Hansen}},
  \bibinfo {author} {\bibfnamefont {M.}~\bibnamefont {Hastings}}, \bibinfo
  {author} {\bibfnamefont {S.}~\bibnamefont {Heedt}}, \bibinfo {author}
  {\bibfnamefont {S.}~\bibnamefont {Ho}}, \bibinfo {author} {\bibfnamefont
  {J.}~\bibnamefont {Hogaboam}}, \bibinfo {author} {\bibfnamefont
  {L.}~\bibnamefont {Holgaard}}, \bibinfo {author} {\bibfnamefont
  {K.}~\bibnamefont {Van~Hoogdalem}}, \bibinfo {author} {\bibfnamefont
  {J.}~\bibnamefont {Indrapiromkul}}, \bibinfo {author} {\bibfnamefont
  {H.}~\bibnamefont {Ingerslev}}, \bibinfo {author} {\bibfnamefont
  {L.}~\bibnamefont {Ivancevic}}, \bibinfo {author} {\bibfnamefont
  {S.}~\bibnamefont {Jablonski}}, \bibinfo {author} {\bibfnamefont
  {T.}~\bibnamefont {Jensen}}, \bibinfo {author} {\bibfnamefont
  {J.}~\bibnamefont {Jhoja}}, \bibinfo {author} {\bibfnamefont
  {J.}~\bibnamefont {Jones}}, \bibinfo {author} {\bibfnamefont
  {K.}~\bibnamefont {Kalashnikov}}, \bibinfo {author} {\bibfnamefont
  {R.}~\bibnamefont {Kallaher}}, \bibinfo {author} {\bibfnamefont
  {R.}~\bibnamefont {Kalra}}, \bibinfo {author} {\bibfnamefont
  {F.}~\bibnamefont {Karimi}}, \bibinfo {author} {\bibfnamefont
  {T.}~\bibnamefont {Karzig}}, \bibinfo {author} {\bibfnamefont
  {S.}~\bibnamefont {Kimes}}, \bibinfo {author} {\bibfnamefont
  {V.}~\bibnamefont {Kliuchnikov}}, \bibinfo {author} {\bibfnamefont {M.~E.}\
  \bibnamefont {Kloster}}, \bibinfo {author} {\bibfnamefont {C.}~\bibnamefont
  {Knapp}}, \bibinfo {author} {\bibfnamefont {D.}~\bibnamefont {Knee}},
  \bibinfo {author} {\bibfnamefont {J.}~\bibnamefont {Koski}}, \bibinfo
  {author} {\bibfnamefont {P.}~\bibnamefont {Kostamo}}, \bibinfo {author}
  {\bibfnamefont {J.}~\bibnamefont {Kuesel}}, \bibinfo {author} {\bibfnamefont
  {B.}~\bibnamefont {Lackey}}, \bibinfo {author} {\bibfnamefont
  {T.}~\bibnamefont {Laeven}}, \bibinfo {author} {\bibfnamefont
  {J.}~\bibnamefont {Lai}}, \bibinfo {author} {\bibfnamefont {G.}~\bibnamefont
  {de~Lange}}, \bibinfo {author} {\bibfnamefont {T.}~\bibnamefont {Larsen}},
  \bibinfo {author} {\bibfnamefont {J.}~\bibnamefont {Lee}}, \bibinfo {author}
  {\bibfnamefont {K.}~\bibnamefont {Lee}}, \bibinfo {author} {\bibfnamefont
  {G.}~\bibnamefont {Leum}}, \bibinfo {author} {\bibfnamefont {K.}~\bibnamefont
  {Li}}, \bibinfo {author} {\bibfnamefont {T.}~\bibnamefont {Lindemann}},
  \bibinfo {author} {\bibfnamefont {M.}~\bibnamefont {Lucas}}, \bibinfo
  {author} {\bibfnamefont {R.}~\bibnamefont {Lutchyn}}, \bibinfo {author}
  {\bibfnamefont {M.~H.}\ \bibnamefont {Madsen}}, \bibinfo {author}
  {\bibfnamefont {N.}~\bibnamefont {Madulid}}, \bibinfo {author} {\bibfnamefont
  {M.}~\bibnamefont {Manfra}}, \bibinfo {author} {\bibfnamefont {S.~B.}\
  \bibnamefont {Markussen}}, \bibinfo {author} {\bibfnamefont {E.}~\bibnamefont
  {Martinez}}, \bibinfo {author} {\bibfnamefont {M.}~\bibnamefont {Mattila}},
  \bibinfo {author} {\bibfnamefont {J.}~\bibnamefont {Mattinson}}, \bibinfo
  {author} {\bibfnamefont {R.}~\bibnamefont {McNeil}}, \bibinfo {author}
  {\bibfnamefont {A.~R.}\ \bibnamefont {Mei}}, \bibinfo {author} {\bibfnamefont
  {R.~V.}\ \bibnamefont {Mishmash}}, \bibinfo {author} {\bibfnamefont
  {G.}~\bibnamefont {Mohandas}}, \bibinfo {author} {\bibfnamefont
  {C.}~\bibnamefont {Mollgaard}}, \bibinfo {author} {\bibfnamefont
  {M.}~\bibnamefont {de~Moor}}, \bibinfo {author} {\bibfnamefont
  {T.}~\bibnamefont {Morgan}}, \bibinfo {author} {\bibfnamefont
  {G.}~\bibnamefont {Moussa}}, \bibinfo {author} {\bibfnamefont
  {A.}~\bibnamefont {Narla}}, \bibinfo {author} {\bibfnamefont
  {C.}~\bibnamefont {Nayak}}, \bibinfo {author} {\bibfnamefont {J.~H.}\
  \bibnamefont {Nielsen}}, \bibinfo {author} {\bibfnamefont {W.~H.~P.}\
  \bibnamefont {Nielsen}}, \bibinfo {author} {\bibfnamefont {F.}~\bibnamefont
  {Nolet}}, \bibinfo {author} {\bibfnamefont {M.}~\bibnamefont {Nystrom}},
  \bibinfo {author} {\bibfnamefont {E.}~\bibnamefont {O'Farrell}}, \bibinfo
  {author} {\bibfnamefont {K.}~\bibnamefont {Otani}}, \bibinfo {author}
  {\bibfnamefont {A.}~\bibnamefont {Paetznick}}, \bibinfo {author}
  {\bibfnamefont {C.}~\bibnamefont {Papon}}, \bibinfo {author} {\bibfnamefont
  {A.}~\bibnamefont {Paz}}, \bibinfo {author} {\bibfnamefont {K.}~\bibnamefont
  {Petersson}}, \bibinfo {author} {\bibfnamefont {L.}~\bibnamefont {Petit}},
  \bibinfo {author} {\bibfnamefont {D.}~\bibnamefont {Pikulin}}, \bibinfo
  {author} {\bibfnamefont {D.~O.~F.}\ \bibnamefont {Pons}}, \bibinfo {author}
  {\bibfnamefont {S.}~\bibnamefont {Quinn}}, \bibinfo {author} {\bibfnamefont
  {M.}~\bibnamefont {Rajpalke}}, \bibinfo {author} {\bibfnamefont {A.~A.}\
  \bibnamefont {Ramirez}}, \bibinfo {author} {\bibfnamefont {K.}~\bibnamefont
  {Rasmussen}}, \bibinfo {author} {\bibfnamefont {D.}~\bibnamefont {Razmadze}},
  \bibinfo {author} {\bibfnamefont {B.}~\bibnamefont {Reichardt}}, \bibinfo
  {author} {\bibfnamefont {Y.}~\bibnamefont {Ren}}, \bibinfo {author}
  {\bibfnamefont {K.}~\bibnamefont {Reneris}}, \bibinfo {author} {\bibfnamefont
  {R.}~\bibnamefont {Riccomini}}, \bibinfo {author} {\bibfnamefont
  {I.}~\bibnamefont {Sadovskyy}}, \bibinfo {author} {\bibfnamefont
  {L.}~\bibnamefont {Sainiemi}}, \bibinfo {author} {\bibfnamefont {J.~C.~E.}\
  \bibnamefont {Salda{\~n}a}}, \bibinfo {author} {\bibfnamefont
  {I.}~\bibnamefont {Sanlorenzo}}, \bibinfo {author} {\bibfnamefont
  {S.}~\bibnamefont {Schaal}}, \bibinfo {author} {\bibfnamefont
  {E.}~\bibnamefont {Schmidgall}}, \bibinfo {author} {\bibfnamefont
  {C.}~\bibnamefont {Sfiligoj}}, \bibinfo {author} {\bibfnamefont {M.~P.}\
  \bibnamefont {da~Silva}}, \bibinfo {author} {\bibfnamefont {S.}~\bibnamefont
  {Singh}}, \bibinfo {author} {\bibfnamefont {S.}~\bibnamefont {Sinha}},
  \bibinfo {author} {\bibfnamefont {M.}~\bibnamefont {Soeken}}, \bibinfo
  {author} {\bibfnamefont {P.}~\bibnamefont {Sohr}}, \bibinfo {author}
  {\bibfnamefont {T.}~\bibnamefont {Stankevic}}, \bibinfo {author}
  {\bibfnamefont {L.}~\bibnamefont {Stek}}, \bibinfo {author} {\bibfnamefont
  {P.}~\bibnamefont {Str{\o{}}m-Hansen}}, \bibinfo {author} {\bibfnamefont
  {E.}~\bibnamefont {Stuppard}}, \bibinfo {author} {\bibfnamefont
  {A.}~\bibnamefont {Sundaram}}, \bibinfo {author} {\bibfnamefont
  {H.}~\bibnamefont {Suominen}}, \bibinfo {author} {\bibfnamefont
  {J.}~\bibnamefont {Suter}}, \bibinfo {author} {\bibfnamefont
  {S.}~\bibnamefont {Suzuki}}, \bibinfo {author} {\bibfnamefont
  {K.}~\bibnamefont {Svore}}, \bibinfo {author} {\bibfnamefont
  {S.}~\bibnamefont {Teicher}}, \bibinfo {author} {\bibfnamefont
  {N.}~\bibnamefont {Thiyagarajah}}, \bibinfo {author} {\bibfnamefont
  {R.}~\bibnamefont {Tholapi}}, \bibinfo {author} {\bibfnamefont
  {M.}~\bibnamefont {Thomas}}, \bibinfo {author} {\bibfnamefont
  {D.}~\bibnamefont {Tom}}, \bibinfo {author} {\bibfnamefont {E.}~\bibnamefont
  {Toomey}}, \bibinfo {author} {\bibfnamefont {J.}~\bibnamefont {Tracy}},
  \bibinfo {author} {\bibfnamefont {M.}~\bibnamefont {Troyer}}, \bibinfo
  {author} {\bibfnamefont {M.}~\bibnamefont {Turley}}, \bibinfo {author}
  {\bibfnamefont {M.~D.}\ \bibnamefont {Turner}}, \bibinfo {author}
  {\bibfnamefont {S.}~\bibnamefont {Upadhyay}}, \bibinfo {author}
  {\bibfnamefont {I.}~\bibnamefont {Urban}}, \bibinfo {author} {\bibfnamefont
  {A.}~\bibnamefont {Vaschillo}}, \bibinfo {author} {\bibfnamefont
  {D.}~\bibnamefont {Viazmitinov}}, \bibinfo {author} {\bibfnamefont
  {D.}~\bibnamefont {Vogel}}, \bibinfo {author} {\bibfnamefont
  {Z.}~\bibnamefont {Wang}}, \bibinfo {author} {\bibfnamefont {J.}~\bibnamefont
  {Watson}}, \bibinfo {author} {\bibfnamefont {A.}~\bibnamefont {Webster}},
  \bibinfo {author} {\bibfnamefont {J.}~\bibnamefont {Weston}}, \bibinfo
  {author} {\bibfnamefont {T.}~\bibnamefont {Williamson}}, \bibinfo {author}
  {\bibfnamefont {G.~W.}\ \bibnamefont {Winkler}}, \bibinfo {author}
  {\bibfnamefont {D.~J.}\ \bibnamefont {van Woerkom}}, \bibinfo {author}
  {\bibfnamefont {B.~P.}\ \bibnamefont {W{\"u}tz}}, \bibinfo {author}
  {\bibfnamefont {C.~K.}\ \bibnamefont {Yang}}, \bibinfo {author}
  {\bibfnamefont {R.}~\bibnamefont {Yu}}, \bibinfo {author} {\bibfnamefont
  {E.}~\bibnamefont {Yucelen}}, \bibinfo {author} {\bibfnamefont {J.~H.}\
  \bibnamefont {Zamorano}}, \bibinfo {author} {\bibfnamefont {R.}~\bibnamefont
  {Zeisel}}, \bibinfo {author} {\bibfnamefont {G.}~\bibnamefont {Zheng}},
  \bibinfo {author} {\bibfnamefont {J.}~\bibnamefont {Zilke}},\ and\ \bibinfo
  {author} {\bibfnamefont {A.}~\bibnamefont {Zimmerman}},\ }\bibfield  {title}
  {\bibinfo {title} {Blueprint for fault-tolerant quantum computation with
  topological qubit arrays},\ }\href {https://doi.org/10.1103/qx36-4rv1}
  {\bibfield  {journal} {\bibinfo  {journal} {Phys. Rev. Res.}\ }\textbf
  {\bibinfo {volume} {7}},\ \bibinfo {pages} {041002} (\bibinfo {year}
  {2025})}\BibitemShut {NoStop}%
\bibitem [{\citenamefont {Aghaee}\ \emph {et~al.}(2023)\citenamefont {Aghaee},
  \citenamefont {Akkala}, \citenamefont {Alam}, \citenamefont {Ali},
  \citenamefont {Alcaraz~Ramirez}, \citenamefont {Andrzejczuk}, \citenamefont
  {Antipov}, \citenamefont {Aseev}, \citenamefont {Astafev}, \citenamefont
  {Bauer}, \citenamefont {Becker}, \citenamefont {Boddapati}, \citenamefont
  {Boekhout}, \citenamefont {Bommer}, \citenamefont {Bosma}, \citenamefont
  {Bourdet}, \citenamefont {Boutin}, \citenamefont {Caroff}, \citenamefont
  {Casparis}, \citenamefont {Cassidy}, \citenamefont {Chatoor}, \citenamefont
  {Christensen}, \citenamefont {Clay}, \citenamefont {Cole}, \citenamefont
  {Corsetti}, \citenamefont {Cui}, \citenamefont {Dalampiras}, \citenamefont
  {Dokania}, \citenamefont {{de Lange}}, \citenamefont {{de Moor}},
  \citenamefont {Estrada~Salda{\~n}a}, \citenamefont {Fallahi}, \citenamefont
  {Fathabad}, \citenamefont {Gamble}, \citenamefont {Gardner}, \citenamefont
  {Govender}, \citenamefont {Griggio}, \citenamefont {Grigoryan}, \citenamefont
  {Gronin}, \citenamefont {Gukelberger}, \citenamefont {Hansen}, \citenamefont
  {Heedt}, \citenamefont {Herranz~Zamorano}, \citenamefont {Ho}, \citenamefont
  {Holgaard}, \citenamefont {Ingerslev}, \citenamefont {Johansson},
  \citenamefont {Jones}, \citenamefont {Kallaher}, \citenamefont {Karimi},
  \citenamefont {Karzig}, \citenamefont {King}, \citenamefont {Kloster},
  \citenamefont {Knapp}, \citenamefont {Kocon}, \citenamefont {Koski},
  \citenamefont {Kostamo}, \citenamefont {Krogstrup}, \citenamefont {Kumar},
  \citenamefont {Laeven}, \citenamefont {Larsen}, \citenamefont {Li},
  \citenamefont {Lindemann}, \citenamefont {Love}, \citenamefont {Lutchyn},
  \citenamefont {Madsen}, \citenamefont {Manfra}, \citenamefont {Markussen},
  \citenamefont {Martinez}, \citenamefont {McNeil}, \citenamefont {Memisevic},
  \citenamefont {Morgan}, \citenamefont {Mullally}, \citenamefont {Nayak},
  \citenamefont {Nielsen}, \citenamefont {Nielsen}, \citenamefont {Nijholt},
  \citenamefont {Nurmohamed}, \citenamefont {O'Farrell}, \citenamefont {Otani},
  \citenamefont {Pauka}, \citenamefont {Petersson}, \citenamefont {Petit},
  \citenamefont {Pikulin}, \citenamefont {Preiss}, \citenamefont
  {{Quintero-Perez}}, \citenamefont {Rajpalke}, \citenamefont {Rasmussen},
  \citenamefont {Razmadze}, \citenamefont {Reentila}, \citenamefont {Reilly},
  \citenamefont {Rouse}, \citenamefont {Sadovskyy}, \citenamefont {Sainiemi},
  \citenamefont {Schreppler}, \citenamefont {Sidorkin}, \citenamefont {Singh},
  \citenamefont {Singh}, \citenamefont {Sinha}, \citenamefont {Sohr},
  \citenamefont {Stankevi{\v c}}, \citenamefont {Stek}, \citenamefont
  {Suominen}, \citenamefont {Suter}, \citenamefont {Svidenko}, \citenamefont
  {Teicher}, \citenamefont {Temuerhan}, \citenamefont {Thiyagarajah},
  \citenamefont {Tholapi}, \citenamefont {Thomas}, \citenamefont {Toomey},
  \citenamefont {Upadhyay}, \citenamefont {Urban}, \citenamefont
  {Vaitiek{\.e}nas}, \citenamefont {Van~Hoogdalem}, \citenamefont
  {Van~Woerkom}, \citenamefont {Viazmitinov}, \citenamefont {Vogel},
  \citenamefont {Waddy}, \citenamefont {Watson}, \citenamefont {Weston},
  \citenamefont {Winkler}, \citenamefont {Yang}, \citenamefont {Yau},
  \citenamefont {Yi}, \citenamefont {Yucelen}, \citenamefont {Webster},
  \citenamefont {Zeisel},\ and\ \citenamefont
  {Zhao}}]{microsoftquantum2023inasal}%
  \BibitemOpen
  \bibfield  {author} {\bibinfo {author} {\bibfnamefont {M.}~\bibnamefont
  {Aghaee}}, \bibinfo {author} {\bibfnamefont {A.}~\bibnamefont {Akkala}},
  \bibinfo {author} {\bibfnamefont {Z.}~\bibnamefont {Alam}}, \bibinfo {author}
  {\bibfnamefont {R.}~\bibnamefont {Ali}}, \bibinfo {author} {\bibfnamefont
  {A.}~\bibnamefont {Alcaraz~Ramirez}}, \bibinfo {author} {\bibfnamefont
  {M.}~\bibnamefont {Andrzejczuk}}, \bibinfo {author} {\bibfnamefont {A.~E.}\
  \bibnamefont {Antipov}}, \bibinfo {author} {\bibfnamefont {P.}~\bibnamefont
  {Aseev}}, \bibinfo {author} {\bibfnamefont {M.}~\bibnamefont {Astafev}},
  \bibinfo {author} {\bibfnamefont {B.}~\bibnamefont {Bauer}}, \bibinfo
  {author} {\bibfnamefont {J.}~\bibnamefont {Becker}}, \bibinfo {author}
  {\bibfnamefont {S.}~\bibnamefont {Boddapati}}, \bibinfo {author}
  {\bibfnamefont {F.}~\bibnamefont {Boekhout}}, \bibinfo {author}
  {\bibfnamefont {J.}~\bibnamefont {Bommer}}, \bibinfo {author} {\bibfnamefont
  {T.}~\bibnamefont {Bosma}}, \bibinfo {author} {\bibfnamefont
  {L.}~\bibnamefont {Bourdet}}, \bibinfo {author} {\bibfnamefont
  {S.}~\bibnamefont {Boutin}}, \bibinfo {author} {\bibfnamefont
  {P.}~\bibnamefont {Caroff}}, \bibinfo {author} {\bibfnamefont
  {L.}~\bibnamefont {Casparis}}, \bibinfo {author} {\bibfnamefont
  {M.}~\bibnamefont {Cassidy}}, \bibinfo {author} {\bibfnamefont
  {S.}~\bibnamefont {Chatoor}}, \bibinfo {author} {\bibfnamefont {A.~W.}\
  \bibnamefont {Christensen}}, \bibinfo {author} {\bibfnamefont
  {N.}~\bibnamefont {Clay}}, \bibinfo {author} {\bibfnamefont {W.~S.}\
  \bibnamefont {Cole}}, \bibinfo {author} {\bibfnamefont {F.}~\bibnamefont
  {Corsetti}}, \bibinfo {author} {\bibfnamefont {A.}~\bibnamefont {Cui}},
  \bibinfo {author} {\bibfnamefont {P.}~\bibnamefont {Dalampiras}}, \bibinfo
  {author} {\bibfnamefont {A.}~\bibnamefont {Dokania}}, \bibinfo {author}
  {\bibfnamefont {G.}~\bibnamefont {{de Lange}}}, \bibinfo {author}
  {\bibfnamefont {M.}~\bibnamefont {{de Moor}}}, \bibinfo {author}
  {\bibfnamefont {J.~C.}\ \bibnamefont {Estrada~Salda{\~n}a}}, \bibinfo
  {author} {\bibfnamefont {S.}~\bibnamefont {Fallahi}}, \bibinfo {author}
  {\bibfnamefont {Z.~H.}\ \bibnamefont {Fathabad}}, \bibinfo {author}
  {\bibfnamefont {J.}~\bibnamefont {Gamble}}, \bibinfo {author} {\bibfnamefont
  {G.}~\bibnamefont {Gardner}}, \bibinfo {author} {\bibfnamefont
  {D.}~\bibnamefont {Govender}}, \bibinfo {author} {\bibfnamefont
  {F.}~\bibnamefont {Griggio}}, \bibinfo {author} {\bibfnamefont
  {R.}~\bibnamefont {Grigoryan}}, \bibinfo {author} {\bibfnamefont
  {S.}~\bibnamefont {Gronin}}, \bibinfo {author} {\bibfnamefont
  {J.}~\bibnamefont {Gukelberger}}, \bibinfo {author} {\bibfnamefont {E.~B.}\
  \bibnamefont {Hansen}}, \bibinfo {author} {\bibfnamefont {S.}~\bibnamefont
  {Heedt}}, \bibinfo {author} {\bibfnamefont {J.}~\bibnamefont
  {Herranz~Zamorano}}, \bibinfo {author} {\bibfnamefont {S.}~\bibnamefont
  {Ho}}, \bibinfo {author} {\bibfnamefont {U.~L.}\ \bibnamefont {Holgaard}},
  \bibinfo {author} {\bibfnamefont {H.}~\bibnamefont {Ingerslev}}, \bibinfo
  {author} {\bibfnamefont {L.}~\bibnamefont {Johansson}}, \bibinfo {author}
  {\bibfnamefont {J.}~\bibnamefont {Jones}}, \bibinfo {author} {\bibfnamefont
  {R.}~\bibnamefont {Kallaher}}, \bibinfo {author} {\bibfnamefont
  {F.}~\bibnamefont {Karimi}}, \bibinfo {author} {\bibfnamefont
  {T.}~\bibnamefont {Karzig}}, \bibinfo {author} {\bibfnamefont
  {C.}~\bibnamefont {King}}, \bibinfo {author} {\bibfnamefont {M.~E.}\
  \bibnamefont {Kloster}}, \bibinfo {author} {\bibfnamefont {C.}~\bibnamefont
  {Knapp}}, \bibinfo {author} {\bibfnamefont {D.}~\bibnamefont {Kocon}},
  \bibinfo {author} {\bibfnamefont {J.}~\bibnamefont {Koski}}, \bibinfo
  {author} {\bibfnamefont {P.}~\bibnamefont {Kostamo}}, \bibinfo {author}
  {\bibfnamefont {P.}~\bibnamefont {Krogstrup}}, \bibinfo {author}
  {\bibfnamefont {M.}~\bibnamefont {Kumar}}, \bibinfo {author} {\bibfnamefont
  {T.}~\bibnamefont {Laeven}}, \bibinfo {author} {\bibfnamefont
  {T.}~\bibnamefont {Larsen}}, \bibinfo {author} {\bibfnamefont
  {K.}~\bibnamefont {Li}}, \bibinfo {author} {\bibfnamefont {T.}~\bibnamefont
  {Lindemann}}, \bibinfo {author} {\bibfnamefont {J.}~\bibnamefont {Love}},
  \bibinfo {author} {\bibfnamefont {R.}~\bibnamefont {Lutchyn}}, \bibinfo
  {author} {\bibfnamefont {M.~H.}\ \bibnamefont {Madsen}}, \bibinfo {author}
  {\bibfnamefont {M.}~\bibnamefont {Manfra}}, \bibinfo {author} {\bibfnamefont
  {S.}~\bibnamefont {Markussen}}, \bibinfo {author} {\bibfnamefont
  {E.}~\bibnamefont {Martinez}}, \bibinfo {author} {\bibfnamefont
  {R.}~\bibnamefont {McNeil}}, \bibinfo {author} {\bibfnamefont
  {E.}~\bibnamefont {Memisevic}}, \bibinfo {author} {\bibfnamefont
  {T.}~\bibnamefont {Morgan}}, \bibinfo {author} {\bibfnamefont
  {A.}~\bibnamefont {Mullally}}, \bibinfo {author} {\bibfnamefont
  {C.}~\bibnamefont {Nayak}}, \bibinfo {author} {\bibfnamefont
  {J.}~\bibnamefont {Nielsen}}, \bibinfo {author} {\bibfnamefont {W.~H.~P.}\
  \bibnamefont {Nielsen}}, \bibinfo {author} {\bibfnamefont {B.}~\bibnamefont
  {Nijholt}}, \bibinfo {author} {\bibfnamefont {A.}~\bibnamefont {Nurmohamed}},
  \bibinfo {author} {\bibfnamefont {E.}~\bibnamefont {O'Farrell}}, \bibinfo
  {author} {\bibfnamefont {K.}~\bibnamefont {Otani}}, \bibinfo {author}
  {\bibfnamefont {S.}~\bibnamefont {Pauka}}, \bibinfo {author} {\bibfnamefont
  {K.}~\bibnamefont {Petersson}}, \bibinfo {author} {\bibfnamefont
  {L.}~\bibnamefont {Petit}}, \bibinfo {author} {\bibfnamefont {D.~I.}\
  \bibnamefont {Pikulin}}, \bibinfo {author} {\bibfnamefont {F.}~\bibnamefont
  {Preiss}}, \bibinfo {author} {\bibfnamefont {M.}~\bibnamefont
  {{Quintero-Perez}}}, \bibinfo {author} {\bibfnamefont {M.}~\bibnamefont
  {Rajpalke}}, \bibinfo {author} {\bibfnamefont {K.}~\bibnamefont {Rasmussen}},
  \bibinfo {author} {\bibfnamefont {D.}~\bibnamefont {Razmadze}}, \bibinfo
  {author} {\bibfnamefont {O.}~\bibnamefont {Reentila}}, \bibinfo {author}
  {\bibfnamefont {D.}~\bibnamefont {Reilly}}, \bibinfo {author} {\bibfnamefont
  {R.}~\bibnamefont {Rouse}}, \bibinfo {author} {\bibfnamefont
  {I.}~\bibnamefont {Sadovskyy}}, \bibinfo {author} {\bibfnamefont
  {L.}~\bibnamefont {Sainiemi}}, \bibinfo {author} {\bibfnamefont
  {S.}~\bibnamefont {Schreppler}}, \bibinfo {author} {\bibfnamefont
  {V.}~\bibnamefont {Sidorkin}}, \bibinfo {author} {\bibfnamefont
  {A.}~\bibnamefont {Singh}}, \bibinfo {author} {\bibfnamefont
  {S.}~\bibnamefont {Singh}}, \bibinfo {author} {\bibfnamefont
  {S.}~\bibnamefont {Sinha}}, \bibinfo {author} {\bibfnamefont
  {P.}~\bibnamefont {Sohr}}, \bibinfo {author} {\bibfnamefont {T.}~\bibnamefont
  {Stankevi{\v c}}}, \bibinfo {author} {\bibfnamefont {L.}~\bibnamefont
  {Stek}}, \bibinfo {author} {\bibfnamefont {H.}~\bibnamefont {Suominen}},
  \bibinfo {author} {\bibfnamefont {J.}~\bibnamefont {Suter}}, \bibinfo
  {author} {\bibfnamefont {V.}~\bibnamefont {Svidenko}}, \bibinfo {author}
  {\bibfnamefont {S.}~\bibnamefont {Teicher}}, \bibinfo {author} {\bibfnamefont
  {M.}~\bibnamefont {Temuerhan}}, \bibinfo {author} {\bibfnamefont
  {N.}~\bibnamefont {Thiyagarajah}}, \bibinfo {author} {\bibfnamefont
  {R.}~\bibnamefont {Tholapi}}, \bibinfo {author} {\bibfnamefont
  {M.}~\bibnamefont {Thomas}}, \bibinfo {author} {\bibfnamefont
  {E.}~\bibnamefont {Toomey}}, \bibinfo {author} {\bibfnamefont
  {S.}~\bibnamefont {Upadhyay}}, \bibinfo {author} {\bibfnamefont
  {I.}~\bibnamefont {Urban}}, \bibinfo {author} {\bibfnamefont
  {S.}~\bibnamefont {Vaitiek{\.e}nas}}, \bibinfo {author} {\bibfnamefont
  {K.}~\bibnamefont {Van~Hoogdalem}}, \bibinfo {author} {\bibfnamefont
  {D.}~\bibnamefont {Van~Woerkom}}, \bibinfo {author} {\bibfnamefont {D.~V.}\
  \bibnamefont {Viazmitinov}}, \bibinfo {author} {\bibfnamefont
  {D.}~\bibnamefont {Vogel}}, \bibinfo {author} {\bibfnamefont
  {S.}~\bibnamefont {Waddy}}, \bibinfo {author} {\bibfnamefont
  {J.}~\bibnamefont {Watson}}, \bibinfo {author} {\bibfnamefont
  {J.}~\bibnamefont {Weston}}, \bibinfo {author} {\bibfnamefont {G.~W.}\
  \bibnamefont {Winkler}}, \bibinfo {author} {\bibfnamefont {C.~K.}\
  \bibnamefont {Yang}}, \bibinfo {author} {\bibfnamefont {S.}~\bibnamefont
  {Yau}}, \bibinfo {author} {\bibfnamefont {D.}~\bibnamefont {Yi}}, \bibinfo
  {author} {\bibfnamefont {E.}~\bibnamefont {Yucelen}}, \bibinfo {author}
  {\bibfnamefont {A.}~\bibnamefont {Webster}}, \bibinfo {author} {\bibfnamefont
  {R.}~\bibnamefont {Zeisel}},\ and\ \bibinfo {author} {\bibfnamefont
  {R.}~\bibnamefont {Zhao}},\ }\bibfield  {title} {\bibinfo {title}
  {{{InAs-Al}} hybrid devices passing the topological gap protocol},\ }\href
  {https://link.aps.org/doi/10.1103/PhysRevB.107.245423} {\bibfield  {journal}
  {\bibinfo  {journal} {Phys. Rev. B}\ }\textbf {\bibinfo {volume} {107}},\
  \bibinfo {pages} {245423} (\bibinfo {year} {2023})}\BibitemShut {NoStop}%
\bibitem [{\citenamefont {Aghaee}\ \emph
  {et~al.}(2025{\natexlab{a}})\citenamefont {Aghaee}, \citenamefont
  {Alcaraz~Ramirez}, \citenamefont {Alam}, \citenamefont {Ali}, \citenamefont
  {Andrzejczuk}, \citenamefont {Antipov}, \citenamefont {Astafev},
  \citenamefont {Barzegar}, \citenamefont {Bauer}, \citenamefont {Becker},
  \citenamefont {Bhaskar}, \citenamefont {Bocharov}, \citenamefont {Boddapati},
  \citenamefont {Bohn}, \citenamefont {Bommer}, \citenamefont {Bourdet},
  \citenamefont {Bousquet}, \citenamefont {Boutin}, \citenamefont {Casparis},
  \citenamefont {Chapman}, \citenamefont {Chatoor}, \citenamefont
  {Christensen}, \citenamefont {Chua}, \citenamefont {Codd}, \citenamefont
  {Cole}, \citenamefont {Cooper}, \citenamefont {Corsetti}, \citenamefont
  {Cui}, \citenamefont {Dalpasso}, \citenamefont {Dehollain}, \citenamefont
  {{de Lange}}, \citenamefont {{de Moor}}, \citenamefont {Ekefj{\"a}rd},
  \citenamefont {El~Dandachi}, \citenamefont {Estrada~Salda{\~n}a},
  \citenamefont {Fallahi}, \citenamefont {Galletti}, \citenamefont {Gardner},
  \citenamefont {Govender}, \citenamefont {Griggio}, \citenamefont {Grigoryan},
  \citenamefont {Grijalva}, \citenamefont {Gronin}, \citenamefont
  {Gukelberger}, \citenamefont {Hamdast}, \citenamefont {Hamze}, \citenamefont
  {Hansen}, \citenamefont {Heedt}, \citenamefont {Heidarnia}, \citenamefont
  {Herranz~Zamorano}, \citenamefont {Ho}, \citenamefont {Holgaard},
  \citenamefont {Hornibrook}, \citenamefont {Indrapiromkul}, \citenamefont
  {Ingerslev}, \citenamefont {Ivancevic}, \citenamefont {Jensen}, \citenamefont
  {Jhoja}, \citenamefont {Jones}, \citenamefont {Kalashnikov}, \citenamefont
  {Kallaher}, \citenamefont {Kalra}, \citenamefont {Karimi}, \citenamefont
  {Karzig}, \citenamefont {King}, \citenamefont {Kloster}, \citenamefont
  {Knapp}, \citenamefont {Kocon}, \citenamefont {Koski}, \citenamefont
  {Kostamo}, \citenamefont {Kumar}, \citenamefont {Laeven}, \citenamefont
  {Larsen}, \citenamefont {Lee}, \citenamefont {Lee}, \citenamefont {Leum},
  \citenamefont {Li}, \citenamefont {Lindemann}, \citenamefont {Looij},
  \citenamefont {Love}, \citenamefont {Lucas}, \citenamefont {Lutchyn},
  \citenamefont {Madsen}, \citenamefont {Madulid}, \citenamefont {Malmros},
  \citenamefont {Manfra}, \citenamefont {Mantri}, \citenamefont {Markussen},
  \citenamefont {Martinez}, \citenamefont {Mattila}, \citenamefont {McNeil},
  \citenamefont {Mei}, \citenamefont {Mishmash}, \citenamefont {Mohandas},
  \citenamefont {Mollgaard}, \citenamefont {Morgan}, \citenamefont {Moussa},
  \citenamefont {Nayak}, \citenamefont {Nielsen}, \citenamefont {Nielsen},
  \citenamefont {Nielsen}, \citenamefont {Nijholt}, \citenamefont {Nystrom},
  \citenamefont {O'Farrell}, \citenamefont {Ohki}, \citenamefont {Otani},
  \citenamefont {Paquelet~W{\"u}tz}, \citenamefont {Pauka}, \citenamefont
  {Petersson}, \citenamefont {Petit}, \citenamefont {Pikulin}, \citenamefont
  {Prawiroatmodjo}, \citenamefont {Preiss}, \citenamefont {Puchol~Morejon},
  \citenamefont {Rajpalke}, \citenamefont {Ranta}, \citenamefont {Rasmussen},
  \citenamefont {Razmadze}, \citenamefont {Reentila}, \citenamefont {Reilly},
  \citenamefont {Ren}, \citenamefont {Reneris}, \citenamefont {Rouse},
  \citenamefont {Sadovskyy}, \citenamefont {Sainiemi}, \citenamefont
  {Sanlorenzo}, \citenamefont {Schmidgall}, \citenamefont {Sfiligoj},
  \citenamefont {Shah}, \citenamefont {Simoes}, \citenamefont {Singh},
  \citenamefont {Sinha}, \citenamefont {Soerensen}, \citenamefont {Sohr},
  \citenamefont {Stankevic}, \citenamefont {Stek}, \citenamefont {Stuppard},
  \citenamefont {Suominen}, \citenamefont {Suter}, \citenamefont {Teicher},
  \citenamefont {Thiyagarajah}, \citenamefont {Tholapi}, \citenamefont
  {Thomas}, \citenamefont {Toomey}, \citenamefont {Tracy}, \citenamefont
  {Turley}, \citenamefont {Upadhyay}, \citenamefont {Urban}, \citenamefont
  {Van~Hoogdalem}, \citenamefont {Van~Woerkom}, \citenamefont {Viazmitinov},
  \citenamefont {Vogel}, \citenamefont {Watson}, \citenamefont {Webster},
  \citenamefont {Weston}, \citenamefont {Winkler}, \citenamefont {Xu},
  \citenamefont {Yang}, \citenamefont {Yucelen}, \citenamefont {Zeisel},
  \citenamefont {Zheng},\ and\ \citenamefont
  {Zilke}}]{aghaee2025interferometric}%
  \BibitemOpen
  \bibfield  {author} {\bibinfo {author} {\bibfnamefont {M.}~\bibnamefont
  {Aghaee}}, \bibinfo {author} {\bibfnamefont {A.}~\bibnamefont
  {Alcaraz~Ramirez}}, \bibinfo {author} {\bibfnamefont {Z.}~\bibnamefont
  {Alam}}, \bibinfo {author} {\bibfnamefont {R.}~\bibnamefont {Ali}}, \bibinfo
  {author} {\bibfnamefont {M.}~\bibnamefont {Andrzejczuk}}, \bibinfo {author}
  {\bibfnamefont {A.}~\bibnamefont {Antipov}}, \bibinfo {author} {\bibfnamefont
  {M.}~\bibnamefont {Astafev}}, \bibinfo {author} {\bibfnamefont
  {A.}~\bibnamefont {Barzegar}}, \bibinfo {author} {\bibfnamefont
  {B.}~\bibnamefont {Bauer}}, \bibinfo {author} {\bibfnamefont
  {J.}~\bibnamefont {Becker}}, \bibinfo {author} {\bibfnamefont {U.~K.}\
  \bibnamefont {Bhaskar}}, \bibinfo {author} {\bibfnamefont {A.}~\bibnamefont
  {Bocharov}}, \bibinfo {author} {\bibfnamefont {S.}~\bibnamefont {Boddapati}},
  \bibinfo {author} {\bibfnamefont {D.}~\bibnamefont {Bohn}}, \bibinfo {author}
  {\bibfnamefont {J.}~\bibnamefont {Bommer}}, \bibinfo {author} {\bibfnamefont
  {L.}~\bibnamefont {Bourdet}}, \bibinfo {author} {\bibfnamefont
  {A.}~\bibnamefont {Bousquet}}, \bibinfo {author} {\bibfnamefont
  {S.}~\bibnamefont {Boutin}}, \bibinfo {author} {\bibfnamefont
  {L.}~\bibnamefont {Casparis}}, \bibinfo {author} {\bibfnamefont {B.~J.}\
  \bibnamefont {Chapman}}, \bibinfo {author} {\bibfnamefont {S.}~\bibnamefont
  {Chatoor}}, \bibinfo {author} {\bibfnamefont {A.~W.}\ \bibnamefont
  {Christensen}}, \bibinfo {author} {\bibfnamefont {C.}~\bibnamefont {Chua}},
  \bibinfo {author} {\bibfnamefont {P.}~\bibnamefont {Codd}}, \bibinfo {author}
  {\bibfnamefont {W.}~\bibnamefont {Cole}}, \bibinfo {author} {\bibfnamefont
  {P.}~\bibnamefont {Cooper}}, \bibinfo {author} {\bibfnamefont
  {F.}~\bibnamefont {Corsetti}}, \bibinfo {author} {\bibfnamefont
  {A.}~\bibnamefont {Cui}}, \bibinfo {author} {\bibfnamefont {P.}~\bibnamefont
  {Dalpasso}}, \bibinfo {author} {\bibfnamefont {J.~P.}\ \bibnamefont
  {Dehollain}}, \bibinfo {author} {\bibfnamefont {G.}~\bibnamefont {{de
  Lange}}}, \bibinfo {author} {\bibfnamefont {M.}~\bibnamefont {{de Moor}}},
  \bibinfo {author} {\bibfnamefont {A.}~\bibnamefont {Ekefj{\"a}rd}}, \bibinfo
  {author} {\bibfnamefont {T.}~\bibnamefont {El~Dandachi}}, \bibinfo {author}
  {\bibfnamefont {J.~C.}\ \bibnamefont {Estrada~Salda{\~n}a}}, \bibinfo
  {author} {\bibfnamefont {S.}~\bibnamefont {Fallahi}}, \bibinfo {author}
  {\bibfnamefont {L.}~\bibnamefont {Galletti}}, \bibinfo {author}
  {\bibfnamefont {G.}~\bibnamefont {Gardner}}, \bibinfo {author} {\bibfnamefont
  {D.}~\bibnamefont {Govender}}, \bibinfo {author} {\bibfnamefont
  {F.}~\bibnamefont {Griggio}}, \bibinfo {author} {\bibfnamefont
  {R.}~\bibnamefont {Grigoryan}}, \bibinfo {author} {\bibfnamefont
  {S.}~\bibnamefont {Grijalva}}, \bibinfo {author} {\bibfnamefont
  {S.}~\bibnamefont {Gronin}}, \bibinfo {author} {\bibfnamefont
  {J.}~\bibnamefont {Gukelberger}}, \bibinfo {author} {\bibfnamefont
  {M.}~\bibnamefont {Hamdast}}, \bibinfo {author} {\bibfnamefont
  {F.}~\bibnamefont {Hamze}}, \bibinfo {author} {\bibfnamefont {E.~B.}\
  \bibnamefont {Hansen}}, \bibinfo {author} {\bibfnamefont {S.}~\bibnamefont
  {Heedt}}, \bibinfo {author} {\bibfnamefont {Z.}~\bibnamefont {Heidarnia}},
  \bibinfo {author} {\bibfnamefont {J.}~\bibnamefont {Herranz~Zamorano}},
  \bibinfo {author} {\bibfnamefont {S.}~\bibnamefont {Ho}}, \bibinfo {author}
  {\bibfnamefont {L.}~\bibnamefont {Holgaard}}, \bibinfo {author}
  {\bibfnamefont {J.}~\bibnamefont {Hornibrook}}, \bibinfo {author}
  {\bibfnamefont {J.}~\bibnamefont {Indrapiromkul}}, \bibinfo {author}
  {\bibfnamefont {H.}~\bibnamefont {Ingerslev}}, \bibinfo {author}
  {\bibfnamefont {L.}~\bibnamefont {Ivancevic}}, \bibinfo {author}
  {\bibfnamefont {T.}~\bibnamefont {Jensen}}, \bibinfo {author} {\bibfnamefont
  {J.}~\bibnamefont {Jhoja}}, \bibinfo {author} {\bibfnamefont
  {J.}~\bibnamefont {Jones}}, \bibinfo {author} {\bibfnamefont {K.~V.}\
  \bibnamefont {Kalashnikov}}, \bibinfo {author} {\bibfnamefont
  {R.}~\bibnamefont {Kallaher}}, \bibinfo {author} {\bibfnamefont
  {R.}~\bibnamefont {Kalra}}, \bibinfo {author} {\bibfnamefont
  {F.}~\bibnamefont {Karimi}}, \bibinfo {author} {\bibfnamefont
  {T.}~\bibnamefont {Karzig}}, \bibinfo {author} {\bibfnamefont
  {E.}~\bibnamefont {King}}, \bibinfo {author} {\bibfnamefont {M.~E.}\
  \bibnamefont {Kloster}}, \bibinfo {author} {\bibfnamefont {C.}~\bibnamefont
  {Knapp}}, \bibinfo {author} {\bibfnamefont {D.}~\bibnamefont {Kocon}},
  \bibinfo {author} {\bibfnamefont {J.~V.}\ \bibnamefont {Koski}}, \bibinfo
  {author} {\bibfnamefont {P.}~\bibnamefont {Kostamo}}, \bibinfo {author}
  {\bibfnamefont {M.}~\bibnamefont {Kumar}}, \bibinfo {author} {\bibfnamefont
  {T.}~\bibnamefont {Laeven}}, \bibinfo {author} {\bibfnamefont
  {T.}~\bibnamefont {Larsen}}, \bibinfo {author} {\bibfnamefont
  {J.}~\bibnamefont {Lee}}, \bibinfo {author} {\bibfnamefont {K.}~\bibnamefont
  {Lee}}, \bibinfo {author} {\bibfnamefont {G.}~\bibnamefont {Leum}}, \bibinfo
  {author} {\bibfnamefont {K.}~\bibnamefont {Li}}, \bibinfo {author}
  {\bibfnamefont {T.}~\bibnamefont {Lindemann}}, \bibinfo {author}
  {\bibfnamefont {M.}~\bibnamefont {Looij}}, \bibinfo {author} {\bibfnamefont
  {J.}~\bibnamefont {Love}}, \bibinfo {author} {\bibfnamefont {M.}~\bibnamefont
  {Lucas}}, \bibinfo {author} {\bibfnamefont {R.}~\bibnamefont {Lutchyn}},
  \bibinfo {author} {\bibfnamefont {M.~H.}\ \bibnamefont {Madsen}}, \bibinfo
  {author} {\bibfnamefont {N.}~\bibnamefont {Madulid}}, \bibinfo {author}
  {\bibfnamefont {A.}~\bibnamefont {Malmros}}, \bibinfo {author} {\bibfnamefont
  {M.}~\bibnamefont {Manfra}}, \bibinfo {author} {\bibfnamefont
  {D.}~\bibnamefont {Mantri}}, \bibinfo {author} {\bibfnamefont {S.~B.}\
  \bibnamefont {Markussen}}, \bibinfo {author} {\bibfnamefont {E.}~\bibnamefont
  {Martinez}}, \bibinfo {author} {\bibfnamefont {M.}~\bibnamefont {Mattila}},
  \bibinfo {author} {\bibfnamefont {R.}~\bibnamefont {McNeil}}, \bibinfo
  {author} {\bibfnamefont {A.~B.}\ \bibnamefont {Mei}}, \bibinfo {author}
  {\bibfnamefont {R.~V.}\ \bibnamefont {Mishmash}}, \bibinfo {author}
  {\bibfnamefont {G.}~\bibnamefont {Mohandas}}, \bibinfo {author}
  {\bibfnamefont {C.}~\bibnamefont {Mollgaard}}, \bibinfo {author}
  {\bibfnamefont {T.}~\bibnamefont {Morgan}}, \bibinfo {author} {\bibfnamefont
  {G.}~\bibnamefont {Moussa}}, \bibinfo {author} {\bibfnamefont
  {C.}~\bibnamefont {Nayak}}, \bibinfo {author} {\bibfnamefont {J.~H.}\
  \bibnamefont {Nielsen}}, \bibinfo {author} {\bibfnamefont {J.~M.}\
  \bibnamefont {Nielsen}}, \bibinfo {author} {\bibfnamefont {W.~H.~P.}\
  \bibnamefont {Nielsen}}, \bibinfo {author} {\bibfnamefont {B.}~\bibnamefont
  {Nijholt}}, \bibinfo {author} {\bibfnamefont {M.}~\bibnamefont {Nystrom}},
  \bibinfo {author} {\bibfnamefont {E.}~\bibnamefont {O'Farrell}}, \bibinfo
  {author} {\bibfnamefont {T.}~\bibnamefont {Ohki}}, \bibinfo {author}
  {\bibfnamefont {K.}~\bibnamefont {Otani}}, \bibinfo {author} {\bibfnamefont
  {B.}~\bibnamefont {Paquelet~W{\"u}tz}}, \bibinfo {author} {\bibfnamefont
  {S.}~\bibnamefont {Pauka}}, \bibinfo {author} {\bibfnamefont
  {K.}~\bibnamefont {Petersson}}, \bibinfo {author} {\bibfnamefont
  {L.}~\bibnamefont {Petit}}, \bibinfo {author} {\bibfnamefont
  {D.}~\bibnamefont {Pikulin}}, \bibinfo {author} {\bibfnamefont
  {G.}~\bibnamefont {Prawiroatmodjo}}, \bibinfo {author} {\bibfnamefont
  {F.}~\bibnamefont {Preiss}}, \bibinfo {author} {\bibfnamefont
  {E.}~\bibnamefont {Puchol~Morejon}}, \bibinfo {author} {\bibfnamefont
  {M.}~\bibnamefont {Rajpalke}}, \bibinfo {author} {\bibfnamefont
  {C.}~\bibnamefont {Ranta}}, \bibinfo {author} {\bibfnamefont
  {K.}~\bibnamefont {Rasmussen}}, \bibinfo {author} {\bibfnamefont
  {D.}~\bibnamefont {Razmadze}}, \bibinfo {author} {\bibfnamefont
  {O.}~\bibnamefont {Reentila}}, \bibinfo {author} {\bibfnamefont {D.~J.}\
  \bibnamefont {Reilly}}, \bibinfo {author} {\bibfnamefont {Y.}~\bibnamefont
  {Ren}}, \bibinfo {author} {\bibfnamefont {K.}~\bibnamefont {Reneris}},
  \bibinfo {author} {\bibfnamefont {R.}~\bibnamefont {Rouse}}, \bibinfo
  {author} {\bibfnamefont {I.}~\bibnamefont {Sadovskyy}}, \bibinfo {author}
  {\bibfnamefont {L.}~\bibnamefont {Sainiemi}}, \bibinfo {author}
  {\bibfnamefont {I.}~\bibnamefont {Sanlorenzo}}, \bibinfo {author}
  {\bibfnamefont {E.}~\bibnamefont {Schmidgall}}, \bibinfo {author}
  {\bibfnamefont {C.}~\bibnamefont {Sfiligoj}}, \bibinfo {author}
  {\bibfnamefont {M.~B.}\ \bibnamefont {Shah}}, \bibinfo {author}
  {\bibfnamefont {K.}~\bibnamefont {Simoes}}, \bibinfo {author} {\bibfnamefont
  {S.}~\bibnamefont {Singh}}, \bibinfo {author} {\bibfnamefont
  {S.}~\bibnamefont {Sinha}}, \bibinfo {author} {\bibfnamefont
  {T.}~\bibnamefont {Soerensen}}, \bibinfo {author} {\bibfnamefont
  {P.}~\bibnamefont {Sohr}}, \bibinfo {author} {\bibfnamefont {T.}~\bibnamefont
  {Stankevic}}, \bibinfo {author} {\bibfnamefont {L.}~\bibnamefont {Stek}},
  \bibinfo {author} {\bibfnamefont {E.}~\bibnamefont {Stuppard}}, \bibinfo
  {author} {\bibfnamefont {H.}~\bibnamefont {Suominen}}, \bibinfo {author}
  {\bibfnamefont {J.}~\bibnamefont {Suter}}, \bibinfo {author} {\bibfnamefont
  {S.}~\bibnamefont {Teicher}}, \bibinfo {author} {\bibfnamefont
  {N.}~\bibnamefont {Thiyagarajah}}, \bibinfo {author} {\bibfnamefont
  {R.}~\bibnamefont {Tholapi}}, \bibinfo {author} {\bibfnamefont
  {M.}~\bibnamefont {Thomas}}, \bibinfo {author} {\bibfnamefont
  {E.}~\bibnamefont {Toomey}}, \bibinfo {author} {\bibfnamefont
  {J.}~\bibnamefont {Tracy}}, \bibinfo {author} {\bibfnamefont
  {M.}~\bibnamefont {Turley}}, \bibinfo {author} {\bibfnamefont
  {S.}~\bibnamefont {Upadhyay}}, \bibinfo {author} {\bibfnamefont
  {I.}~\bibnamefont {Urban}}, \bibinfo {author} {\bibfnamefont
  {K.}~\bibnamefont {Van~Hoogdalem}}, \bibinfo {author} {\bibfnamefont {D.~J.}\
  \bibnamefont {Van~Woerkom}}, \bibinfo {author} {\bibfnamefont {D.~V.}\
  \bibnamefont {Viazmitinov}}, \bibinfo {author} {\bibfnamefont
  {D.}~\bibnamefont {Vogel}}, \bibinfo {author} {\bibfnamefont
  {J.}~\bibnamefont {Watson}}, \bibinfo {author} {\bibfnamefont
  {A.}~\bibnamefont {Webster}}, \bibinfo {author} {\bibfnamefont
  {J.}~\bibnamefont {Weston}}, \bibinfo {author} {\bibfnamefont {G.~W.}\
  \bibnamefont {Winkler}}, \bibinfo {author} {\bibfnamefont {D.}~\bibnamefont
  {Xu}}, \bibinfo {author} {\bibfnamefont {C.~K.}\ \bibnamefont {Yang}},
  \bibinfo {author} {\bibfnamefont {E.}~\bibnamefont {Yucelen}}, \bibinfo
  {author} {\bibfnamefont {R.}~\bibnamefont {Zeisel}}, \bibinfo {author}
  {\bibfnamefont {G.}~\bibnamefont {Zheng}},\ and\ \bibinfo {author}
  {\bibfnamefont {J.}~\bibnamefont {Zilke}},\ }\bibfield  {title} {\bibinfo
  {title} {Interferometric single-shot parity measurement in {{InAs}}--{{Al}}
  hybrid devices},\ }\href {https://www.nature.com/articles/s41586-024-08445-2}
  {\bibfield  {journal} {\bibinfo  {journal} {Nature}\ }\textbf {\bibinfo
  {volume} {638}},\ \bibinfo {pages} {651} (\bibinfo {year}
  {2025}{\natexlab{a}})}\BibitemShut {NoStop}%
\bibitem [{\citenamefont {Akhmerov}\ \emph {et~al.}(2009)\citenamefont
  {Akhmerov}, \citenamefont {Nilsson},\ and\ \citenamefont
  {Beenakker}}]{akhmerov2009electrically}%
  \BibitemOpen
  \bibfield  {author} {\bibinfo {author} {\bibfnamefont {A.~R.}\ \bibnamefont
  {Akhmerov}}, \bibinfo {author} {\bibfnamefont {J.}~\bibnamefont {Nilsson}},\
  and\ \bibinfo {author} {\bibfnamefont {C.~W.~J.}\ \bibnamefont {Beenakker}},\
  }\bibfield  {title} {\bibinfo {title} {Electrically detected interferometry
  of {Majorana} fermions in a topological insulator},\ }\href
  {https://link.aps.org/doi/10.1103/PhysRevLett.102.216404} {\bibfield
  {journal} {\bibinfo  {journal} {Phys. Rev. Lett.}\ }\textbf {\bibinfo
  {volume} {102}},\ \bibinfo {pages} {216404} (\bibinfo {year}
  {2009})}\BibitemShut {NoStop}%
\bibitem [{\citenamefont {Fu}(2010)}]{fu2010electron}%
  \BibitemOpen
  \bibfield  {author} {\bibinfo {author} {\bibfnamefont {L.}~\bibnamefont
  {Fu}},\ }\bibfield  {title} {\bibinfo {title} {Electron teleportation via
  {Majorana} bound states in a mesoscopic superconductor},\ }\href
  {https://link.aps.org/doi/10.1103/PhysRevLett.104.056402} {\bibfield
  {journal} {\bibinfo  {journal} {Phys. Rev. Lett.}\ }\textbf {\bibinfo
  {volume} {104}},\ \bibinfo {pages} {056402} (\bibinfo {year}
  {2010})}\BibitemShut {NoStop}%
\bibitem [{\citenamefont {Karzig}\ \emph {et~al.}(2017)\citenamefont {Karzig},
  \citenamefont {Knapp}, \citenamefont {Lutchyn}, \citenamefont {Bonderson},
  \citenamefont {Hastings}, \citenamefont {Nayak}, \citenamefont {Alicea},
  \citenamefont {Flensberg}, \citenamefont {Plugge}, \citenamefont {Oreg},
  \citenamefont {Marcus},\ and\ \citenamefont {Freedman}}]{karzig2017scalable}%
  \BibitemOpen
  \bibfield  {author} {\bibinfo {author} {\bibfnamefont {T.}~\bibnamefont
  {Karzig}}, \bibinfo {author} {\bibfnamefont {C.}~\bibnamefont {Knapp}},
  \bibinfo {author} {\bibfnamefont {R.~M.}\ \bibnamefont {Lutchyn}}, \bibinfo
  {author} {\bibfnamefont {P.}~\bibnamefont {Bonderson}}, \bibinfo {author}
  {\bibfnamefont {M.~B.}\ \bibnamefont {Hastings}}, \bibinfo {author}
  {\bibfnamefont {C.}~\bibnamefont {Nayak}}, \bibinfo {author} {\bibfnamefont
  {J.}~\bibnamefont {Alicea}}, \bibinfo {author} {\bibfnamefont
  {K.}~\bibnamefont {Flensberg}}, \bibinfo {author} {\bibfnamefont
  {S.}~\bibnamefont {Plugge}}, \bibinfo {author} {\bibfnamefont
  {Y.}~\bibnamefont {Oreg}}, \bibinfo {author} {\bibfnamefont {C.~M.}\
  \bibnamefont {Marcus}},\ and\ \bibinfo {author} {\bibfnamefont {M.~H.}\
  \bibnamefont {Freedman}},\ }\bibfield  {title} {\bibinfo {title} {Scalable
  designs for quasiparticle-poisoning-protected topological quantum computation
  with {{Majorana}} zero modes},\ }\href
  {https://link.aps.org/doi/10.1103/PhysRevB.95.235305} {\bibfield  {journal}
  {\bibinfo  {journal} {Phys. Rev. B}\ }\textbf {\bibinfo {volume} {95}},\
  \bibinfo {pages} {235305} (\bibinfo {year} {2017})}\BibitemShut {NoStop}%
\bibitem [{\citenamefont {Plugge}\ \emph {et~al.}(2017)\citenamefont {Plugge},
  \citenamefont {Rasmussen}, \citenamefont {Egger},\ and\ \citenamefont
  {Flensberg}}]{plugge2017majorana}%
  \BibitemOpen
  \bibfield  {author} {\bibinfo {author} {\bibfnamefont {S.}~\bibnamefont
  {Plugge}}, \bibinfo {author} {\bibfnamefont {A.}~\bibnamefont {Rasmussen}},
  \bibinfo {author} {\bibfnamefont {R.}~\bibnamefont {Egger}},\ and\ \bibinfo
  {author} {\bibfnamefont {K.}~\bibnamefont {Flensberg}},\ }\bibfield  {title}
  {\bibinfo {title} {Majorana box qubits},\ }\href
  {http://stacks.iop.org/1367-2630/19/i=1/a=012001} {\bibfield  {journal}
  {\bibinfo  {journal} {New J. Phys.}\ }\textbf {\bibinfo {volume} {19}},\
  \bibinfo {pages} {012001} (\bibinfo {year} {2017})}\BibitemShut {NoStop}%
\bibitem [{\citenamefont {Aghaee}\ \emph
  {et~al.}(2025{\natexlab{b}})\citenamefont {Aghaee}, \citenamefont {Alam},
  \citenamefont {Andersen}, \citenamefont {Andrzejczuk}, \citenamefont
  {Antipov}, \citenamefont {Astafev}, \citenamefont {Avilovas}, \citenamefont
  {Azizimanesh}, \citenamefont {Banek}, \citenamefont {Bauer}, \citenamefont
  {Becker}, \citenamefont {Bhaskar}, \citenamefont {Boa}, \citenamefont
  {Boddapati}, \citenamefont {Bohac}, \citenamefont {Bommer}, \citenamefont
  {Borovsky}, \citenamefont {Bourdet}, \citenamefont {Boutin}, \citenamefont
  {Casparis}, \citenamefont {Chakravarthi}, \citenamefont {Chalabi},
  \citenamefont {Chapman}, \citenamefont {Chatzaras}, \citenamefont {Chien},
  \citenamefont {Cho}, \citenamefont {Codd}, \citenamefont {Cole},
  \citenamefont {Cooper}, \citenamefont {Corsetti}, \citenamefont {Cui},
  \citenamefont {Dandachi}, \citenamefont {Dinesen}, \citenamefont {Ekefjärd},
  \citenamefont {Fallahi}, \citenamefont {Galletti}, \citenamefont {Gardner},
  \citenamefont {Gonzalez}, \citenamefont {Govender}, \citenamefont {Griggio},
  \citenamefont {Grigoryan}, \citenamefont {Grijalva}, \citenamefont {Gronin},
  \citenamefont {Gukelberger}, \citenamefont {Hamdast}, \citenamefont {Hamida},
  \citenamefont {Hansen}, \citenamefont {Hansen}, \citenamefont {Heedt},
  \citenamefont {Ho}, \citenamefont {Holgaard}, \citenamefont {van Hoogdalem},
  \citenamefont {Hornibrook}, \citenamefont {Ingerslev}, \citenamefont
  {Ivancevic}, \citenamefont {Jamo}, \citenamefont {Jantos}, \citenamefont
  {Jensen}, \citenamefont {Jhoja}, \citenamefont {Jones}, \citenamefont
  {Joshi}, \citenamefont {Kalashnikov}, \citenamefont {Kallaher}, \citenamefont
  {Kalra}, \citenamefont {Karimi}, \citenamefont {Karzig}, \citenamefont
  {Kimes}, \citenamefont {King}, \citenamefont {Kloster}, \citenamefont
  {Knapp}, \citenamefont {Koski}, \citenamefont {Kostamo}, \citenamefont
  {Laeven}, \citenamefont {Lai}, \citenamefont {de~Lange}, \citenamefont
  {Larsen}, \citenamefont {Lee}, \citenamefont {Li}, \citenamefont {Li},
  \citenamefont {Liang}, \citenamefont {Lindemann}, \citenamefont {Looij},
  \citenamefont {Lucas}, \citenamefont {Lutchyn}, \citenamefont {Madsen},
  \citenamefont {Madulid}, \citenamefont {Manfra}, \citenamefont {Manjunath},
  \citenamefont {Markussen}, \citenamefont {Martinez}, \citenamefont {Mattila},
  \citenamefont {Mattinson}, \citenamefont {McNeil}, \citenamefont {Millan},
  \citenamefont {Mishmash}, \citenamefont {Mittal}, \citenamefont {Møllgaard},
  \citenamefont {de~Moor}, \citenamefont {Morejon}, \citenamefont {Morgan},
  \citenamefont {Moussa}, \citenamefont {Nabar}, \citenamefont {Narla},
  \citenamefont {Nayak}, \citenamefont {Nielsen}, \citenamefont {Nielsen},
  \citenamefont {Nolet}, \citenamefont {Nystrom}, \citenamefont {O'Farrell},
  \citenamefont {Ohki}, \citenamefont {Otani}, \citenamefont {Papon},
  \citenamefont {Petersson}, \citenamefont {Petit}, \citenamefont {Pikulin},
  \citenamefont {Rajpalke}, \citenamefont {Ramirez}, \citenamefont {Razmadze},
  \citenamefont {Ren}, \citenamefont {Sadovskyy}, \citenamefont {Sainiemi},
  \citenamefont {Saldaña}, \citenamefont {Sanlorenzo}, \citenamefont {dos
  Santos}, \citenamefont {Schaal}, \citenamefont {Schack}, \citenamefont
  {Schmidgall}, \citenamefont {Sfetsou}, \citenamefont {Sfiligoj},
  \citenamefont {Sinha}, \citenamefont {Sohr}, \citenamefont {Sørensen},
  \citenamefont {Spiegelhauer}, \citenamefont {Stankević}, \citenamefont
  {Stek}, \citenamefont {Strøm-Hansen}, \citenamefont {Suominen},
  \citenamefont {Suter}, \citenamefont {Teicher}, \citenamefont {Tholapi},
  \citenamefont {Thomas}, \citenamefont {Tom}, \citenamefont {Toomey},
  \citenamefont {Tracy}, \citenamefont {Turley}, \citenamefont {Turner},
  \citenamefont {Upadhyay}, \citenamefont {Urban}, \citenamefont {Viazmitinov},
  \citenamefont {Viazmitinova}, \citenamefont {Viegas}, \citenamefont {Vogel},
  \citenamefont {Watson}, \citenamefont {Webster}, \citenamefont {Weston},
  \citenamefont {Williamson}, \citenamefont {Winkler}, \citenamefont {van
  Woerkom}, \citenamefont {Wuetz}, \citenamefont {Yang}, \citenamefont
  {Shang-Jyun}, \citenamefont {Yu}, \citenamefont {Yucelen}, \citenamefont
  {Zamorano}, \citenamefont {Zeisel}, \citenamefont {Zheng},\ and\
  \citenamefont {Zimmerman}}]{aghaee2025distinctlifetimesxz}%
  \BibitemOpen
  \bibfield  {author} {\bibinfo {author} {\bibfnamefont {M.}~\bibnamefont
  {Aghaee}}, \bibinfo {author} {\bibfnamefont {Z.}~\bibnamefont {Alam}},
  \bibinfo {author} {\bibfnamefont {R.}~\bibnamefont {Andersen}}, \bibinfo
  {author} {\bibfnamefont {M.}~\bibnamefont {Andrzejczuk}}, \bibinfo {author}
  {\bibfnamefont {A.}~\bibnamefont {Antipov}}, \bibinfo {author} {\bibfnamefont
  {M.}~\bibnamefont {Astafev}}, \bibinfo {author} {\bibfnamefont
  {L.}~\bibnamefont {Avilovas}}, \bibinfo {author} {\bibfnamefont
  {A.}~\bibnamefont {Azizimanesh}}, \bibinfo {author} {\bibfnamefont
  {E.}~\bibnamefont {Banek}}, \bibinfo {author} {\bibfnamefont
  {B.}~\bibnamefont {Bauer}}, \bibinfo {author} {\bibfnamefont
  {J.}~\bibnamefont {Becker}}, \bibinfo {author} {\bibfnamefont {U.~K.}\
  \bibnamefont {Bhaskar}}, \bibinfo {author} {\bibfnamefont {A.~G.}\
  \bibnamefont {Boa}}, \bibinfo {author} {\bibfnamefont {S.}~\bibnamefont
  {Boddapati}}, \bibinfo {author} {\bibfnamefont {N.}~\bibnamefont {Bohac}},
  \bibinfo {author} {\bibfnamefont {J.~D.~S.}\ \bibnamefont {Bommer}}, \bibinfo
  {author} {\bibfnamefont {J.}~\bibnamefont {Borovsky}}, \bibinfo {author}
  {\bibfnamefont {L.}~\bibnamefont {Bourdet}}, \bibinfo {author} {\bibfnamefont
  {S.}~\bibnamefont {Boutin}}, \bibinfo {author} {\bibfnamefont
  {L.}~\bibnamefont {Casparis}}, \bibinfo {author} {\bibfnamefont
  {S.}~\bibnamefont {Chakravarthi}}, \bibinfo {author} {\bibfnamefont
  {H.}~\bibnamefont {Chalabi}}, \bibinfo {author} {\bibfnamefont {B.~J.}\
  \bibnamefont {Chapman}}, \bibinfo {author} {\bibfnamefont {N.}~\bibnamefont
  {Chatzaras}}, \bibinfo {author} {\bibfnamefont {T.-C.}\ \bibnamefont
  {Chien}}, \bibinfo {author} {\bibfnamefont {J.}~\bibnamefont {Cho}}, \bibinfo
  {author} {\bibfnamefont {P.}~\bibnamefont {Codd}}, \bibinfo {author}
  {\bibfnamefont {W.}~\bibnamefont {Cole}}, \bibinfo {author} {\bibfnamefont
  {P.~W.}\ \bibnamefont {Cooper}}, \bibinfo {author} {\bibfnamefont
  {F.}~\bibnamefont {Corsetti}}, \bibinfo {author} {\bibfnamefont
  {A.}~\bibnamefont {Cui}}, \bibinfo {author} {\bibfnamefont {T.~E.}\
  \bibnamefont {Dandachi}}, \bibinfo {author} {\bibfnamefont {C.}~\bibnamefont
  {Dinesen}}, \bibinfo {author} {\bibfnamefont {A.}~\bibnamefont {Ekefjärd}},
  \bibinfo {author} {\bibfnamefont {S.}~\bibnamefont {Fallahi}}, \bibinfo
  {author} {\bibfnamefont {L.}~\bibnamefont {Galletti}}, \bibinfo {author}
  {\bibfnamefont {G.~C.}\ \bibnamefont {Gardner}}, \bibinfo {author}
  {\bibfnamefont {G.~L.}\ \bibnamefont {Gonzalez}}, \bibinfo {author}
  {\bibfnamefont {D.}~\bibnamefont {Govender}}, \bibinfo {author}
  {\bibfnamefont {F.}~\bibnamefont {Griggio}}, \bibinfo {author} {\bibfnamefont
  {R.}~\bibnamefont {Grigoryan}}, \bibinfo {author} {\bibfnamefont
  {S.}~\bibnamefont {Grijalva}}, \bibinfo {author} {\bibfnamefont
  {S.}~\bibnamefont {Gronin}}, \bibinfo {author} {\bibfnamefont
  {J.}~\bibnamefont {Gukelberger}}, \bibinfo {author} {\bibfnamefont
  {M.}~\bibnamefont {Hamdast}}, \bibinfo {author} {\bibfnamefont {A.~B.}\
  \bibnamefont {Hamida}}, \bibinfo {author} {\bibfnamefont {E.~B.}\
  \bibnamefont {Hansen}}, \bibinfo {author} {\bibfnamefont {C.~T.}\
  \bibnamefont {Hansen}}, \bibinfo {author} {\bibfnamefont {S.}~\bibnamefont
  {Heedt}}, \bibinfo {author} {\bibfnamefont {S.}~\bibnamefont {Ho}}, \bibinfo
  {author} {\bibfnamefont {L.}~\bibnamefont {Holgaard}}, \bibinfo {author}
  {\bibfnamefont {K.}~\bibnamefont {van Hoogdalem}}, \bibinfo {author}
  {\bibfnamefont {J.}~\bibnamefont {Hornibrook}}, \bibinfo {author}
  {\bibfnamefont {H.}~\bibnamefont {Ingerslev}}, \bibinfo {author}
  {\bibfnamefont {L.}~\bibnamefont {Ivancevic}}, \bibinfo {author}
  {\bibfnamefont {S.}~\bibnamefont {Jamo}}, \bibinfo {author} {\bibfnamefont
  {M.}~\bibnamefont {Jantos}}, \bibinfo {author} {\bibfnamefont
  {T.}~\bibnamefont {Jensen}}, \bibinfo {author} {\bibfnamefont {J.~S.}\
  \bibnamefont {Jhoja}}, \bibinfo {author} {\bibfnamefont {J.~C.}\ \bibnamefont
  {Jones}}, \bibinfo {author} {\bibfnamefont {V.}~\bibnamefont {Joshi}},
  \bibinfo {author} {\bibfnamefont {K.~V.}\ \bibnamefont {Kalashnikov}},
  \bibinfo {author} {\bibfnamefont {R.}~\bibnamefont {Kallaher}}, \bibinfo
  {author} {\bibfnamefont {R.}~\bibnamefont {Kalra}}, \bibinfo {author}
  {\bibfnamefont {F.}~\bibnamefont {Karimi}}, \bibinfo {author} {\bibfnamefont
  {T.}~\bibnamefont {Karzig}}, \bibinfo {author} {\bibfnamefont
  {S.}~\bibnamefont {Kimes}}, \bibinfo {author} {\bibfnamefont
  {E.}~\bibnamefont {King}}, \bibinfo {author} {\bibfnamefont {M.~E.}\
  \bibnamefont {Kloster}}, \bibinfo {author} {\bibfnamefont {C.}~\bibnamefont
  {Knapp}}, \bibinfo {author} {\bibfnamefont {J.~V.}\ \bibnamefont {Koski}},
  \bibinfo {author} {\bibfnamefont {P.}~\bibnamefont {Kostamo}}, \bibinfo
  {author} {\bibfnamefont {T.}~\bibnamefont {Laeven}}, \bibinfo {author}
  {\bibfnamefont {J.}~\bibnamefont {Lai}}, \bibinfo {author} {\bibfnamefont
  {G.}~\bibnamefont {de~Lange}}, \bibinfo {author} {\bibfnamefont {T.~W.}\
  \bibnamefont {Larsen}}, \bibinfo {author} {\bibfnamefont {K.}~\bibnamefont
  {Lee}}, \bibinfo {author} {\bibfnamefont {K.}~\bibnamefont {Li}}, \bibinfo
  {author} {\bibfnamefont {G.}~\bibnamefont {Li}}, \bibinfo {author}
  {\bibfnamefont {S.}~\bibnamefont {Liang}}, \bibinfo {author} {\bibfnamefont
  {T.}~\bibnamefont {Lindemann}}, \bibinfo {author} {\bibfnamefont
  {M.}~\bibnamefont {Looij}}, \bibinfo {author} {\bibfnamefont
  {M.}~\bibnamefont {Lucas}}, \bibinfo {author} {\bibfnamefont
  {R.}~\bibnamefont {Lutchyn}}, \bibinfo {author} {\bibfnamefont {M.~H.}\
  \bibnamefont {Madsen}}, \bibinfo {author} {\bibfnamefont {N.}~\bibnamefont
  {Madulid}}, \bibinfo {author} {\bibfnamefont {M.~J.}\ \bibnamefont {Manfra}},
  \bibinfo {author} {\bibfnamefont {L.}~\bibnamefont {Manjunath}}, \bibinfo
  {author} {\bibfnamefont {S.}~\bibnamefont {Markussen}}, \bibinfo {author}
  {\bibfnamefont {E.}~\bibnamefont {Martinez}}, \bibinfo {author}
  {\bibfnamefont {M.}~\bibnamefont {Mattila}}, \bibinfo {author} {\bibfnamefont
  {J.~R.}\ \bibnamefont {Mattinson}}, \bibinfo {author} {\bibfnamefont
  {R.~P.~G.}\ \bibnamefont {McNeil}}, \bibinfo {author} {\bibfnamefont {A.~P.}\
  \bibnamefont {Millan}}, \bibinfo {author} {\bibfnamefont {R.~V.}\
  \bibnamefont {Mishmash}}, \bibinfo {author} {\bibfnamefont {S.}~\bibnamefont
  {Mittal}}, \bibinfo {author} {\bibfnamefont {C.}~\bibnamefont {Møllgaard}},
  \bibinfo {author} {\bibfnamefont {M.~W.~A.}\ \bibnamefont {de~Moor}},
  \bibinfo {author} {\bibfnamefont {E.~P.}\ \bibnamefont {Morejon}}, \bibinfo
  {author} {\bibfnamefont {T.}~\bibnamefont {Morgan}}, \bibinfo {author}
  {\bibfnamefont {G.}~\bibnamefont {Moussa}}, \bibinfo {author} {\bibfnamefont
  {B.~P.}\ \bibnamefont {Nabar}}, \bibinfo {author} {\bibfnamefont
  {A.}~\bibnamefont {Narla}}, \bibinfo {author} {\bibfnamefont
  {C.}~\bibnamefont {Nayak}}, \bibinfo {author} {\bibfnamefont {J.~H.}\
  \bibnamefont {Nielsen}}, \bibinfo {author} {\bibfnamefont {W.~H.~P.}\
  \bibnamefont {Nielsen}}, \bibinfo {author} {\bibfnamefont {F.}~\bibnamefont
  {Nolet}}, \bibinfo {author} {\bibfnamefont {M.~J.}\ \bibnamefont {Nystrom}},
  \bibinfo {author} {\bibfnamefont {E.}~\bibnamefont {O'Farrell}}, \bibinfo
  {author} {\bibfnamefont {T.~A.}\ \bibnamefont {Ohki}}, \bibinfo {author}
  {\bibfnamefont {K.}~\bibnamefont {Otani}}, \bibinfo {author} {\bibfnamefont
  {C.}~\bibnamefont {Papon}}, \bibinfo {author} {\bibfnamefont {K.~D.}\
  \bibnamefont {Petersson}}, \bibinfo {author} {\bibfnamefont {L.}~\bibnamefont
  {Petit}}, \bibinfo {author} {\bibfnamefont {D.}~\bibnamefont {Pikulin}},
  \bibinfo {author} {\bibfnamefont {M.}~\bibnamefont {Rajpalke}}, \bibinfo
  {author} {\bibfnamefont {A.~A.}\ \bibnamefont {Ramirez}}, \bibinfo {author}
  {\bibfnamefont {D.}~\bibnamefont {Razmadze}}, \bibinfo {author}
  {\bibfnamefont {Y.}~\bibnamefont {Ren}}, \bibinfo {author} {\bibfnamefont
  {I.}~\bibnamefont {Sadovskyy}}, \bibinfo {author} {\bibfnamefont
  {L.}~\bibnamefont {Sainiemi}}, \bibinfo {author} {\bibfnamefont {J.~C.~E.}\
  \bibnamefont {Saldaña}}, \bibinfo {author} {\bibfnamefont {I.}~\bibnamefont
  {Sanlorenzo}}, \bibinfo {author} {\bibfnamefont {T.~P.}\ \bibnamefont {dos
  Santos}}, \bibinfo {author} {\bibfnamefont {S.}~\bibnamefont {Schaal}},
  \bibinfo {author} {\bibfnamefont {J.}~\bibnamefont {Schack}}, \bibinfo
  {author} {\bibfnamefont {E.~R.}\ \bibnamefont {Schmidgall}}, \bibinfo
  {author} {\bibfnamefont {C.}~\bibnamefont {Sfetsou}}, \bibinfo {author}
  {\bibfnamefont {C.}~\bibnamefont {Sfiligoj}}, \bibinfo {author}
  {\bibfnamefont {S.}~\bibnamefont {Sinha}}, \bibinfo {author} {\bibfnamefont
  {P.}~\bibnamefont {Sohr}}, \bibinfo {author} {\bibfnamefont {T.~L.}\
  \bibnamefont {Sørensen}}, \bibinfo {author} {\bibfnamefont {K.}~\bibnamefont
  {Spiegelhauer}}, \bibinfo {author} {\bibfnamefont {T.}~\bibnamefont
  {Stankević}}, \bibinfo {author} {\bibfnamefont {L.~J.}\ \bibnamefont
  {Stek}}, \bibinfo {author} {\bibfnamefont {P.}~\bibnamefont {Strøm-Hansen}},
  \bibinfo {author} {\bibfnamefont {H.~J.}\ \bibnamefont {Suominen}}, \bibinfo
  {author} {\bibfnamefont {J.}~\bibnamefont {Suter}}, \bibinfo {author}
  {\bibfnamefont {S.~M.~L.}\ \bibnamefont {Teicher}}, \bibinfo {author}
  {\bibfnamefont {R.}~\bibnamefont {Tholapi}}, \bibinfo {author} {\bibfnamefont
  {M.}~\bibnamefont {Thomas}}, \bibinfo {author} {\bibfnamefont {D.~W.}\
  \bibnamefont {Tom}}, \bibinfo {author} {\bibfnamefont {E.}~\bibnamefont
  {Toomey}}, \bibinfo {author} {\bibfnamefont {J.}~\bibnamefont {Tracy}},
  \bibinfo {author} {\bibfnamefont {M.}~\bibnamefont {Turley}}, \bibinfo
  {author} {\bibfnamefont {M.~D.}\ \bibnamefont {Turner}}, \bibinfo {author}
  {\bibfnamefont {S.}~\bibnamefont {Upadhyay}}, \bibinfo {author}
  {\bibfnamefont {I.}~\bibnamefont {Urban}}, \bibinfo {author} {\bibfnamefont
  {D.~V.}\ \bibnamefont {Viazmitinov}}, \bibinfo {author} {\bibfnamefont
  {A.~W.}\ \bibnamefont {Viazmitinova}}, \bibinfo {author} {\bibfnamefont
  {B.}~\bibnamefont {Viegas}}, \bibinfo {author} {\bibfnamefont {D.~J.}\
  \bibnamefont {Vogel}}, \bibinfo {author} {\bibfnamefont {J.}~\bibnamefont
  {Watson}}, \bibinfo {author} {\bibfnamefont {A.}~\bibnamefont {Webster}},
  \bibinfo {author} {\bibfnamefont {J.}~\bibnamefont {Weston}}, \bibinfo
  {author} {\bibfnamefont {T.}~\bibnamefont {Williamson}}, \bibinfo {author}
  {\bibfnamefont {G.~W.}\ \bibnamefont {Winkler}}, \bibinfo {author}
  {\bibfnamefont {D.~J.}\ \bibnamefont {van Woerkom}}, \bibinfo {author}
  {\bibfnamefont {B.~P.}\ \bibnamefont {Wuetz}}, \bibinfo {author}
  {\bibfnamefont {C.-K.}\ \bibnamefont {Yang}}, \bibinfo {author} {\bibnamefont
  {Shang-Jyun}}, \bibinfo {author} {\bibnamefont {Yu}}, \bibinfo {author}
  {\bibfnamefont {E.}~\bibnamefont {Yucelen}}, \bibinfo {author} {\bibfnamefont
  {J.~H.}\ \bibnamefont {Zamorano}}, \bibinfo {author} {\bibfnamefont
  {R.}~\bibnamefont {Zeisel}}, \bibinfo {author} {\bibfnamefont
  {G.}~\bibnamefont {Zheng}},\ and\ \bibinfo {author} {\bibfnamefont {A.~M.}\
  \bibnamefont {Zimmerman}},\ }\bibfield  {title} {\bibinfo {title} {Distinct
  lifetimes for {$X$} and {$Z$} loop measurements in a {Majorana} tetron
  device},\ }\href {https://arxiv.org/abs/2507.08795} {\bibfield  {journal}
  {\bibinfo  {journal} {arXiv:2507.08795}\ } (\bibinfo {year}
  {2025}{\natexlab{b}})}\BibitemShut {NoStop}%
\bibitem [{\citenamefont {Aghaee}\ \emph {et~al.}(2026)\citenamefont {Aghaee},
  \citenamefont {Alam}, \citenamefont {Andrzejczuk}, \citenamefont {Antipov},
  \citenamefont {Asimakidis}, \citenamefont {Astafev}, \citenamefont
  {Avilovas}, \citenamefont {Azizimanesh}, \citenamefont {Barzegar},
  \citenamefont {Bauer}, \citenamefont {Becker}, \citenamefont {Bhaskar},
  \citenamefont {Boa}, \citenamefont {Boddapati}, \citenamefont {Bohac},
  \citenamefont {Bommer}, \citenamefont {Borovsky}, \citenamefont {Bourdet},
  \citenamefont {Boutin}, \citenamefont {Chakravarthi}, \citenamefont
  {Chapman}, \citenamefont {Chatzaras}, \citenamefont {Chien}, \citenamefont
  {Cho}, \citenamefont {Codd}, \citenamefont {Cole}, \citenamefont {Cooper},
  \citenamefont {Corsetti}, \citenamefont {Cui}, \citenamefont {Dandachi},
  \citenamefont {Divanis}, \citenamefont {Doyle}, \citenamefont {Ekefjard},
  \citenamefont {Falcon}, \citenamefont {Fallahi}, \citenamefont {Galletti},
  \citenamefont {Gardner}, \citenamefont {Gavranovic}, \citenamefont {Gomes},
  \citenamefont {Govender}, \citenamefont {Griggio}, \citenamefont {Grigoryan},
  \citenamefont {Grijalva}, \citenamefont {Gronin}, \citenamefont
  {Gukelberger}, \citenamefont {Hamdast}, \citenamefont {Hansen}, \citenamefont
  {Heedt}, \citenamefont {Ho}, \citenamefont {Holgaard}, \citenamefont {van
  Hoogdalem}, \citenamefont {Indrapiromkul}, \citenamefont {Ingerslev},
  \citenamefont {Ivancevic}, \citenamefont {Jantos}, \citenamefont {Jensen},
  \citenamefont {Jhoja}, \citenamefont {Joshi}, \citenamefont {Kalashnikov},
  \citenamefont {Kallaher}, \citenamefont {Kalra}, \citenamefont {Karimi},
  \citenamefont {Karzig}, \citenamefont {Kloster}, \citenamefont {Knapp},
  \citenamefont {Knoblauch}, \citenamefont {Koski}, \citenamefont {Kringhøj},
  \citenamefont {Laeven}, \citenamefont {Lai}, \citenamefont {de~Lange},
  \citenamefont {Larsen}, \citenamefont {Lee}, \citenamefont {Li},
  \citenamefont {Liang}, \citenamefont {Lindemann}, \citenamefont {Lochmatter},
  \citenamefont {Lucas}, \citenamefont {Lutchyn}, \citenamefont {Madsen},
  \citenamefont {Madulid}, \citenamefont {Maliyov}, \citenamefont {Mampaey},
  \citenamefont {Manfra}, \citenamefont {Markussen}, \citenamefont {Martinez},
  \citenamefont {Mattinson}, \citenamefont {Meira}, \citenamefont {Mikolas},
  \citenamefont {Mittal}, \citenamefont {Mohandas}, \citenamefont {Mollgaard},
  \citenamefont {de~Moor}, \citenamefont {Moore}, \citenamefont {Moussa},
  \citenamefont {Nabar}, \citenamefont {Narla}, \citenamefont {Naseri},
  \citenamefont {Nayak}, \citenamefont {Nielsen}, \citenamefont {Nielsen},
  \citenamefont {Nystrom}, \citenamefont {O'Farrell}, \citenamefont {Ohtani},
  \citenamefont {Orth}, \citenamefont {Paolo}, \citenamefont {Papon},
  \citenamefont {Petit}, \citenamefont {Pikulin}, \citenamefont {Rajpalke},
  \citenamefont {Ramirez}, \citenamefont {Rasmussen}, \citenamefont {Razmadze},
  \citenamefont {Ren}, \citenamefont {Romanova}, \citenamefont {Roure},
  \citenamefont {Sadovskyy}, \citenamefont {Sainiemi}, \citenamefont
  {Saldaña}, \citenamefont {Sanlorenzo}, \citenamefont {dos Santos},
  \citenamefont {Saruda}, \citenamefont {Schaal}, \citenamefont {Schack},
  \citenamefont {Schmidgall}, \citenamefont {Sfetsou}, \citenamefont
  {Sfiligoj}, \citenamefont {Shekason}, \citenamefont {Sinha}, \citenamefont
  {Sohr}, \citenamefont {de~Sousa}, \citenamefont {Spiegelhauer}, \citenamefont
  {Stankevic}, \citenamefont {Suominen}, \citenamefont {Suter}, \citenamefont
  {Szénási}, \citenamefont {Teicher}, \citenamefont {Thiyagarajah},
  \citenamefont {Tholapi}, \citenamefont {Thomas}, \citenamefont {Tom},
  \citenamefont {Toomey}, \citenamefont {Tracy}, \citenamefont {Turley},
  \citenamefont {Turner}, \citenamefont {Urban}, \citenamefont {Valliappan},
  \citenamefont {Viazmitinov}, \citenamefont {Viazmitinova}, \citenamefont
  {Vogel}, \citenamefont {Wang}, \citenamefont {Watson}, \citenamefont
  {Watson}, \citenamefont {Webster}, \citenamefont {Weston}, \citenamefont
  {Williamson}, \citenamefont {Winkler}, \citenamefont {van Woerkom},
  \citenamefont {Wuetz}, \citenamefont {Yu}, \citenamefont {Yucelen},
  \citenamefont {Zamorano}, \citenamefont {Zeisel}, \citenamefont {Zheng},\
  and\ \citenamefont {Zimmerman}}]{aghaee202620secondparitylifetime}%
  \BibitemOpen
  \bibfield  {author} {\bibinfo {author} {\bibfnamefont {M.}~\bibnamefont
  {Aghaee}}, \bibinfo {author} {\bibfnamefont {Z.}~\bibnamefont {Alam}},
  \bibinfo {author} {\bibfnamefont {M.}~\bibnamefont {Andrzejczuk}}, \bibinfo
  {author} {\bibfnamefont {A.}~\bibnamefont {Antipov}}, \bibinfo {author}
  {\bibfnamefont {T.}~\bibnamefont {Asimakidis}}, \bibinfo {author}
  {\bibfnamefont {M.}~\bibnamefont {Astafev}}, \bibinfo {author} {\bibfnamefont
  {L.}~\bibnamefont {Avilovas}}, \bibinfo {author} {\bibfnamefont
  {A.}~\bibnamefont {Azizimanesh}}, \bibinfo {author} {\bibfnamefont
  {A.}~\bibnamefont {Barzegar}}, \bibinfo {author} {\bibfnamefont
  {B.}~\bibnamefont {Bauer}}, \bibinfo {author} {\bibfnamefont
  {J.}~\bibnamefont {Becker}}, \bibinfo {author} {\bibfnamefont {U.~K.}\
  \bibnamefont {Bhaskar}}, \bibinfo {author} {\bibfnamefont {A.~G.}\
  \bibnamefont {Boa}}, \bibinfo {author} {\bibfnamefont {S.}~\bibnamefont
  {Boddapati}}, \bibinfo {author} {\bibfnamefont {N.}~\bibnamefont {Bohac}},
  \bibinfo {author} {\bibfnamefont {J.}~\bibnamefont {Bommer}}, \bibinfo
  {author} {\bibfnamefont {J.}~\bibnamefont {Borovsky}}, \bibinfo {author}
  {\bibfnamefont {L.}~\bibnamefont {Bourdet}}, \bibinfo {author} {\bibfnamefont
  {S.}~\bibnamefont {Boutin}}, \bibinfo {author} {\bibfnamefont
  {S.}~\bibnamefont {Chakravarthi}}, \bibinfo {author} {\bibfnamefont {B.~J.}\
  \bibnamefont {Chapman}}, \bibinfo {author} {\bibfnamefont {N.}~\bibnamefont
  {Chatzaras}}, \bibinfo {author} {\bibfnamefont {T.-C.}\ \bibnamefont
  {Chien}}, \bibinfo {author} {\bibfnamefont {J.}~\bibnamefont {Cho}}, \bibinfo
  {author} {\bibfnamefont {P.~T.}\ \bibnamefont {Codd}}, \bibinfo {author}
  {\bibfnamefont {W.}~\bibnamefont {Cole}}, \bibinfo {author} {\bibfnamefont
  {P.~W.}\ \bibnamefont {Cooper}}, \bibinfo {author} {\bibfnamefont
  {F.}~\bibnamefont {Corsetti}}, \bibinfo {author} {\bibfnamefont
  {A.}~\bibnamefont {Cui}}, \bibinfo {author} {\bibfnamefont {T.~E.}\
  \bibnamefont {Dandachi}}, \bibinfo {author} {\bibfnamefont {K.}~\bibnamefont
  {Divanis}}, \bibinfo {author} {\bibfnamefont {C.}~\bibnamefont {Doyle}},
  \bibinfo {author} {\bibfnamefont {A.}~\bibnamefont {Ekefjard}}, \bibinfo
  {author} {\bibfnamefont {J.~A.}\ \bibnamefont {Falcon}}, \bibinfo {author}
  {\bibfnamefont {S.}~\bibnamefont {Fallahi}}, \bibinfo {author} {\bibfnamefont
  {L.}~\bibnamefont {Galletti}}, \bibinfo {author} {\bibfnamefont {G.~C.}\
  \bibnamefont {Gardner}}, \bibinfo {author} {\bibfnamefont {H.}~\bibnamefont
  {Gavranovic}}, \bibinfo {author} {\bibfnamefont {J.~P.~M.}\ \bibnamefont
  {Gomes}}, \bibinfo {author} {\bibfnamefont {D.}~\bibnamefont {Govender}},
  \bibinfo {author} {\bibfnamefont {F.}~\bibnamefont {Griggio}}, \bibinfo
  {author} {\bibfnamefont {R.}~\bibnamefont {Grigoryan}}, \bibinfo {author}
  {\bibfnamefont {S.}~\bibnamefont {Grijalva}}, \bibinfo {author}
  {\bibfnamefont {S.}~\bibnamefont {Gronin}}, \bibinfo {author} {\bibfnamefont
  {J.}~\bibnamefont {Gukelberger}}, \bibinfo {author} {\bibfnamefont
  {M.}~\bibnamefont {Hamdast}}, \bibinfo {author} {\bibfnamefont {E.~B.}\
  \bibnamefont {Hansen}}, \bibinfo {author} {\bibfnamefont {S.}~\bibnamefont
  {Heedt}}, \bibinfo {author} {\bibfnamefont {S.}~\bibnamefont {Ho}}, \bibinfo
  {author} {\bibfnamefont {L.}~\bibnamefont {Holgaard}}, \bibinfo {author}
  {\bibfnamefont {K.}~\bibnamefont {van Hoogdalem}}, \bibinfo {author}
  {\bibfnamefont {J.}~\bibnamefont {Indrapiromkul}}, \bibinfo {author}
  {\bibfnamefont {H.}~\bibnamefont {Ingerslev}}, \bibinfo {author}
  {\bibfnamefont {L.}~\bibnamefont {Ivancevic}}, \bibinfo {author}
  {\bibfnamefont {M.}~\bibnamefont {Jantos}}, \bibinfo {author} {\bibfnamefont
  {T.}~\bibnamefont {Jensen}}, \bibinfo {author} {\bibfnamefont {J.~S.}\
  \bibnamefont {Jhoja}}, \bibinfo {author} {\bibfnamefont {V.~R.}\ \bibnamefont
  {Joshi}}, \bibinfo {author} {\bibfnamefont {K.~V.}\ \bibnamefont
  {Kalashnikov}}, \bibinfo {author} {\bibfnamefont {R.}~\bibnamefont
  {Kallaher}}, \bibinfo {author} {\bibfnamefont {R.}~\bibnamefont {Kalra}},
  \bibinfo {author} {\bibfnamefont {F.}~\bibnamefont {Karimi}}, \bibinfo
  {author} {\bibfnamefont {T.}~\bibnamefont {Karzig}}, \bibinfo {author}
  {\bibfnamefont {M.~E.}\ \bibnamefont {Kloster}}, \bibinfo {author}
  {\bibfnamefont {C.}~\bibnamefont {Knapp}}, \bibinfo {author} {\bibfnamefont
  {J.}~\bibnamefont {Knoblauch}}, \bibinfo {author} {\bibfnamefont
  {J.}~\bibnamefont {Koski}}, \bibinfo {author} {\bibfnamefont
  {A.}~\bibnamefont {Kringhøj}}, \bibinfo {author} {\bibfnamefont
  {T.}~\bibnamefont {Laeven}}, \bibinfo {author} {\bibfnamefont
  {J.}~\bibnamefont {Lai}}, \bibinfo {author} {\bibfnamefont {G.}~\bibnamefont
  {de~Lange}}, \bibinfo {author} {\bibfnamefont {T.~W.}\ \bibnamefont
  {Larsen}}, \bibinfo {author} {\bibfnamefont {K.}~\bibnamefont {Lee}},
  \bibinfo {author} {\bibfnamefont {K.}~\bibnamefont {Li}}, \bibinfo {author}
  {\bibfnamefont {S.}~\bibnamefont {Liang}}, \bibinfo {author} {\bibfnamefont
  {T.}~\bibnamefont {Lindemann}}, \bibinfo {author} {\bibfnamefont
  {L.}~\bibnamefont {Lochmatter}}, \bibinfo {author} {\bibfnamefont
  {M.}~\bibnamefont {Lucas}}, \bibinfo {author} {\bibfnamefont
  {R.}~\bibnamefont {Lutchyn}}, \bibinfo {author} {\bibfnamefont {M.~H.}\
  \bibnamefont {Madsen}}, \bibinfo {author} {\bibfnamefont {N.}~\bibnamefont
  {Madulid}}, \bibinfo {author} {\bibfnamefont {I.}~\bibnamefont {Maliyov}},
  \bibinfo {author} {\bibfnamefont {Y.}~\bibnamefont {Mampaey}}, \bibinfo
  {author} {\bibfnamefont {M.}~\bibnamefont {Manfra}}, \bibinfo {author}
  {\bibfnamefont {S.~B.}\ \bibnamefont {Markussen}}, \bibinfo {author}
  {\bibfnamefont {E.~A.}\ \bibnamefont {Martinez}}, \bibinfo {author}
  {\bibfnamefont {J.~R.}\ \bibnamefont {Mattinson}}, \bibinfo {author}
  {\bibfnamefont {M.}~\bibnamefont {Meira}}, \bibinfo {author} {\bibfnamefont
  {C.~A.}\ \bibnamefont {Mikolas}}, \bibinfo {author} {\bibfnamefont
  {S.}~\bibnamefont {Mittal}}, \bibinfo {author} {\bibfnamefont
  {G.}~\bibnamefont {Mohandas}}, \bibinfo {author} {\bibfnamefont
  {C.}~\bibnamefont {Mollgaard}}, \bibinfo {author} {\bibfnamefont {M.~W.~A.}\
  \bibnamefont {de~Moor}}, \bibinfo {author} {\bibfnamefont {C.}~\bibnamefont
  {Moore}}, \bibinfo {author} {\bibfnamefont {G.}~\bibnamefont {Moussa}},
  \bibinfo {author} {\bibfnamefont {B.}~\bibnamefont {Nabar}}, \bibinfo
  {author} {\bibfnamefont {A.}~\bibnamefont {Narla}}, \bibinfo {author}
  {\bibfnamefont {A.}~\bibnamefont {Naseri}}, \bibinfo {author} {\bibfnamefont
  {C.}~\bibnamefont {Nayak}}, \bibinfo {author} {\bibfnamefont {B.~F.~S.}\
  \bibnamefont {Nielsen}}, \bibinfo {author} {\bibfnamefont {J.~H.}\
  \bibnamefont {Nielsen}}, \bibinfo {author} {\bibfnamefont {M.~J.}\
  \bibnamefont {Nystrom}}, \bibinfo {author} {\bibfnamefont {E.}~\bibnamefont
  {O'Farrell}}, \bibinfo {author} {\bibfnamefont {K.}~\bibnamefont {Ohtani}},
  \bibinfo {author} {\bibfnamefont {T.~R.}\ \bibnamefont {Orth}}, \bibinfo
  {author} {\bibfnamefont {S.~D.}\ \bibnamefont {Paolo}}, \bibinfo {author}
  {\bibfnamefont {C.}~\bibnamefont {Papon}}, \bibinfo {author} {\bibfnamefont
  {L.}~\bibnamefont {Petit}}, \bibinfo {author} {\bibfnamefont
  {D.}~\bibnamefont {Pikulin}}, \bibinfo {author} {\bibfnamefont
  {M.}~\bibnamefont {Rajpalke}}, \bibinfo {author} {\bibfnamefont {A.~A.}\
  \bibnamefont {Ramirez}}, \bibinfo {author} {\bibfnamefont {K.}~\bibnamefont
  {Rasmussen}}, \bibinfo {author} {\bibfnamefont {D.}~\bibnamefont {Razmadze}},
  \bibinfo {author} {\bibfnamefont {Y.}~\bibnamefont {Ren}}, \bibinfo {author}
  {\bibfnamefont {M.}~\bibnamefont {Romanova}}, \bibinfo {author}
  {\bibfnamefont {L.}~\bibnamefont {Roure}}, \bibinfo {author} {\bibfnamefont
  {I.}~\bibnamefont {Sadovskyy}}, \bibinfo {author} {\bibfnamefont
  {L.}~\bibnamefont {Sainiemi}}, \bibinfo {author} {\bibfnamefont {J.~C.~E.}\
  \bibnamefont {Saldaña}}, \bibinfo {author} {\bibfnamefont {I.}~\bibnamefont
  {Sanlorenzo}}, \bibinfo {author} {\bibfnamefont {T.~P.}\ \bibnamefont {dos
  Santos}}, \bibinfo {author} {\bibfnamefont {C.~V.~C.}\ \bibnamefont
  {Saruda}}, \bibinfo {author} {\bibfnamefont {S.}~\bibnamefont {Schaal}},
  \bibinfo {author} {\bibfnamefont {J.}~\bibnamefont {Schack}}, \bibinfo
  {author} {\bibfnamefont {E.~R.}\ \bibnamefont {Schmidgall}}, \bibinfo
  {author} {\bibfnamefont {C.}~\bibnamefont {Sfetsou}}, \bibinfo {author}
  {\bibfnamefont {C.}~\bibnamefont {Sfiligoj}}, \bibinfo {author}
  {\bibfnamefont {Z.}~\bibnamefont {Shekason}}, \bibinfo {author}
  {\bibfnamefont {S.~S.}\ \bibnamefont {Sinha}}, \bibinfo {author}
  {\bibfnamefont {P.}~\bibnamefont {Sohr}}, \bibinfo {author} {\bibfnamefont
  {M.~J.~L.}\ \bibnamefont {de~Sousa}}, \bibinfo {author} {\bibfnamefont
  {K.~R.}\ \bibnamefont {Spiegelhauer}}, \bibinfo {author} {\bibfnamefont
  {T.}~\bibnamefont {Stankevic}}, \bibinfo {author} {\bibfnamefont {H.~J.}\
  \bibnamefont {Suominen}}, \bibinfo {author} {\bibfnamefont {J.}~\bibnamefont
  {Suter}}, \bibinfo {author} {\bibfnamefont {A.}~\bibnamefont {Szénási}},
  \bibinfo {author} {\bibfnamefont {S.~M.~L.}\ \bibnamefont {Teicher}},
  \bibinfo {author} {\bibfnamefont {N.}~\bibnamefont {Thiyagarajah}}, \bibinfo
  {author} {\bibfnamefont {R.}~\bibnamefont {Tholapi}}, \bibinfo {author}
  {\bibfnamefont {M.}~\bibnamefont {Thomas}}, \bibinfo {author} {\bibfnamefont
  {D.}~\bibnamefont {Tom}}, \bibinfo {author} {\bibfnamefont {E.}~\bibnamefont
  {Toomey}}, \bibinfo {author} {\bibfnamefont {J.}~\bibnamefont {Tracy}},
  \bibinfo {author} {\bibfnamefont {M.}~\bibnamefont {Turley}}, \bibinfo
  {author} {\bibfnamefont {M.~D.}\ \bibnamefont {Turner}}, \bibinfo {author}
  {\bibfnamefont {I.}~\bibnamefont {Urban}}, \bibinfo {author} {\bibfnamefont
  {A.}~\bibnamefont {Valliappan}}, \bibinfo {author} {\bibfnamefont {D.~V.}\
  \bibnamefont {Viazmitinov}}, \bibinfo {author} {\bibfnamefont {A.~W.}\
  \bibnamefont {Viazmitinova}}, \bibinfo {author} {\bibfnamefont {D.~J.}\
  \bibnamefont {Vogel}}, \bibinfo {author} {\bibfnamefont {W.}~\bibnamefont
  {Wang}}, \bibinfo {author} {\bibfnamefont {C.~A.}\ \bibnamefont {Watson}},
  \bibinfo {author} {\bibfnamefont {J.}~\bibnamefont {Watson}}, \bibinfo
  {author} {\bibfnamefont {A.}~\bibnamefont {Webster}}, \bibinfo {author}
  {\bibfnamefont {J.}~\bibnamefont {Weston}}, \bibinfo {author} {\bibfnamefont
  {T.}~\bibnamefont {Williamson}}, \bibinfo {author} {\bibfnamefont {G.~W.}\
  \bibnamefont {Winkler}}, \bibinfo {author} {\bibfnamefont {D.~J.}\
  \bibnamefont {van Woerkom}}, \bibinfo {author} {\bibfnamefont {B.~P.}\
  \bibnamefont {Wuetz}}, \bibinfo {author} {\bibfnamefont {C.~X.}\ \bibnamefont
  {Yu}}, \bibinfo {author} {\bibfnamefont {E.}~\bibnamefont {Yucelen}},
  \bibinfo {author} {\bibfnamefont {J.~H.}\ \bibnamefont {Zamorano}}, \bibinfo
  {author} {\bibfnamefont {R.}~\bibnamefont {Zeisel}}, \bibinfo {author}
  {\bibfnamefont {G.}~\bibnamefont {Zheng}},\ and\ \bibinfo {author}
  {\bibfnamefont {A.~M.}\ \bibnamefont {Zimmerman}},\ }\bibfield  {title}
  {\bibinfo {title} {20 second parity lifetime in an {InAs--Pb} tetron
  device},\ }\href {https://arxiv.org/abs/2606.03884} {\bibfield  {journal}
  {\bibinfo  {journal} {arXiv preprint arXiv:2606.03884}\ } (\bibinfo {year}
  {2026})}\BibitemShut {NoStop}%
\bibitem [{\citenamefont {Zatelli}\ \emph {et~al.}(2026)\citenamefont
  {Zatelli}, \citenamefont {Roovers}, \citenamefont {van Loo}, \citenamefont
  {Lombardi}, \citenamefont {Luna}, \citenamefont {Miles}, \citenamefont
  {Sietses}, \citenamefont {Evertsz'}, \citenamefont {Fariña}, \citenamefont
  {Bordin}, \citenamefont {Badawy}, \citenamefont {Bakkers}, \citenamefont
  {Wimmer},\ and\ \citenamefont
  {Kouwenhoven}}]{zatelli2026majoranaparityqubitcoupled}%
  \BibitemOpen
  \bibfield  {author} {\bibinfo {author} {\bibfnamefont {F.}~\bibnamefont
  {Zatelli}}, \bibinfo {author} {\bibfnamefont {B.}~\bibnamefont {Roovers}},
  \bibinfo {author} {\bibfnamefont {N.}~\bibnamefont {van Loo}}, \bibinfo
  {author} {\bibfnamefont {A.}~\bibnamefont {Lombardi}}, \bibinfo {author}
  {\bibfnamefont {J.~D.~T.}\ \bibnamefont {Luna}}, \bibinfo {author}
  {\bibfnamefont {S.}~\bibnamefont {Miles}}, \bibinfo {author} {\bibfnamefont
  {V.~P.~M.}\ \bibnamefont {Sietses}}, \bibinfo {author} {\bibfnamefont
  {F.~J.~B.}\ \bibnamefont {Evertsz'}}, \bibinfo {author} {\bibfnamefont
  {P.~C.}\ \bibnamefont {Fariña}}, \bibinfo {author} {\bibfnamefont
  {A.}~\bibnamefont {Bordin}}, \bibinfo {author} {\bibfnamefont
  {G.}~\bibnamefont {Badawy}}, \bibinfo {author} {\bibfnamefont {E.~P. A.~M.}\
  \bibnamefont {Bakkers}}, \bibinfo {author} {\bibfnamefont {M.}~\bibnamefont
  {Wimmer}},\ and\ \bibinfo {author} {\bibfnamefont {L.~P.}\ \bibnamefont
  {Kouwenhoven}},\ }\bibfield  {title} {\bibinfo {title} {Majorana parity qubit
  in coupled minimal {Kitaev} chains},\ }\href
  {https://arxiv.org/abs/2607.09511} {\bibfield  {journal} {\bibinfo  {journal}
  {arXiv preprint arXiv:2607.09511}\ } (\bibinfo {year} {2026})}\BibitemShut
  {NoStop}%
\bibitem [{\citenamefont {van Loo}\ \emph {et~al.}(2026)\citenamefont {van
  Loo}, \citenamefont {Zatelli}, \citenamefont {Steffensen}, \citenamefont
  {Roovers}, \citenamefont {Wang}, \citenamefont {Van~Caekenberghe},
  \citenamefont {Bordin}, \citenamefont {van Driel}, \citenamefont {Zhang},
  \citenamefont {Huisman}, \citenamefont {Badawy}, \citenamefont {Bakkers},
  \citenamefont {Mazur}, \citenamefont {Aguado},\ and\ \citenamefont
  {Kouwenhoven}}]{vanLoo2026}%
  \BibitemOpen
  \bibfield  {author} {\bibinfo {author} {\bibfnamefont {N.}~\bibnamefont {van
  Loo}}, \bibinfo {author} {\bibfnamefont {F.}~\bibnamefont {Zatelli}},
  \bibinfo {author} {\bibfnamefont {G.~O.}\ \bibnamefont {Steffensen}},
  \bibinfo {author} {\bibfnamefont {B.}~\bibnamefont {Roovers}}, \bibinfo
  {author} {\bibfnamefont {G.}~\bibnamefont {Wang}}, \bibinfo {author}
  {\bibfnamefont {T.}~\bibnamefont {Van~Caekenberghe}}, \bibinfo {author}
  {\bibfnamefont {A.}~\bibnamefont {Bordin}}, \bibinfo {author} {\bibfnamefont
  {D.}~\bibnamefont {van Driel}}, \bibinfo {author} {\bibfnamefont
  {Y.}~\bibnamefont {Zhang}}, \bibinfo {author} {\bibfnamefont {W.~D.}\
  \bibnamefont {Huisman}}, \bibinfo {author} {\bibfnamefont {G.}~\bibnamefont
  {Badawy}}, \bibinfo {author} {\bibfnamefont {E.~P. A.~M.}\ \bibnamefont
  {Bakkers}}, \bibinfo {author} {\bibfnamefont {G.~P.}\ \bibnamefont {Mazur}},
  \bibinfo {author} {\bibfnamefont {R.}~\bibnamefont {Aguado}},\ and\ \bibinfo
  {author} {\bibfnamefont {L.~P.}\ \bibnamefont {Kouwenhoven}},\ }\bibfield
  {title} {\bibinfo {title} {Single-shot parity readout of a minimal {Kitaev}
  chain},\ }\href {https://doi.org/10.1038/s41586-025-09927-7} {\bibfield
  {journal} {\bibinfo  {journal} {Nature}\ }\textbf {\bibinfo {volume} {650}},\
  \bibinfo {pages} {334} (\bibinfo {year} {2026})}\BibitemShut {NoStop}%
\bibitem [{\citenamefont {Knapp}\ \emph {et~al.}(2018)\citenamefont {Knapp},
  \citenamefont {Karzig}, \citenamefont {Lutchyn},\ and\ \citenamefont
  {Nayak}}]{Knapp2018}%
  \BibitemOpen
  \bibfield  {author} {\bibinfo {author} {\bibfnamefont {C.}~\bibnamefont
  {Knapp}}, \bibinfo {author} {\bibfnamefont {T.}~\bibnamefont {Karzig}},
  \bibinfo {author} {\bibfnamefont {R.~M.}\ \bibnamefont {Lutchyn}},\ and\
  \bibinfo {author} {\bibfnamefont {C.}~\bibnamefont {Nayak}},\ }\bibfield
  {title} {\bibinfo {title} {Dephasing of {Majorana-based} qubits},\ }\href
  {https://doi.org/10.1103/PhysRevB.97.125404} {\bibfield  {journal} {\bibinfo
  {journal} {Phys. Rev. B}\ }\textbf {\bibinfo {volume} {97}},\ \bibinfo
  {pages} {125404} (\bibinfo {year} {2018})}\BibitemShut {NoStop}%
\bibitem [{\citenamefont {Mishmash}\ \emph {et~al.}(2020)\citenamefont
  {Mishmash}, \citenamefont {Bauer}, \citenamefont {{von Oppen}},\ and\
  \citenamefont {Alicea}}]{mishmash2020dephasing}%
  \BibitemOpen
  \bibfield  {author} {\bibinfo {author} {\bibfnamefont {R.~V.}\ \bibnamefont
  {Mishmash}}, \bibinfo {author} {\bibfnamefont {B.}~\bibnamefont {Bauer}},
  \bibinfo {author} {\bibfnamefont {F.}~\bibnamefont {{von Oppen}}},\ and\
  \bibinfo {author} {\bibfnamefont {J.}~\bibnamefont {Alicea}},\ }\bibfield
  {title} {\bibinfo {title} {Dephasing and leakage dynamics of noisy
  {{Majorana-based}} qubits: {{Topological}} versus {{Andreev}}},\ }\href
  {http://arxiv.org/abs/1911.02582} {\bibfield  {journal} {\bibinfo  {journal}
  {Phys. Rev. B}\ }\textbf {\bibinfo {volume} {101}},\ \bibinfo {pages}
  {075404} (\bibinfo {year} {2020})}\BibitemShut {NoStop}%
\bibitem [{\citenamefont {Das~Sarma}\ and\ \citenamefont
  {Pan}(2023)}]{dassarma2023density}%
  \BibitemOpen
  \bibfield  {author} {\bibinfo {author} {\bibfnamefont {S.}~\bibnamefont
  {Das~Sarma}}\ and\ \bibinfo {author} {\bibfnamefont {H.}~\bibnamefont
  {Pan}},\ }\bibfield  {title} {\bibinfo {title} {Density of states, transport,
  and topology in disordered {{Majorana}} nanowires},\ }\href
  {https://link.aps.org/doi/10.1103/PhysRevB.108.085415} {\bibfield  {journal}
  {\bibinfo  {journal} {Phys. Rev. B}\ }\textbf {\bibinfo {volume} {108}},\
  \bibinfo {pages} {085415} (\bibinfo {year} {2023})}\BibitemShut {NoStop}%
\bibitem [{\citenamefont {Das~Sarma}\ \emph {et~al.}(2023)\citenamefont
  {Das~Sarma}, \citenamefont {Sau},\ and\ \citenamefont
  {Stanescu}}]{dassarma2023spectral}%
  \BibitemOpen
  \bibfield  {author} {\bibinfo {author} {\bibfnamefont {S.}~\bibnamefont
  {Das~Sarma}}, \bibinfo {author} {\bibfnamefont {J.~D.}\ \bibnamefont {Sau}},\
  and\ \bibinfo {author} {\bibfnamefont {T.~D.}\ \bibnamefont {Stanescu}},\
  }\bibfield  {title} {\bibinfo {title} {Spectral properties, topological
  patches, and effective phase diagrams of finite disordered {{Majorana}}
  nanowires},\ }\href {https://link.aps.org/doi/10.1103/PhysRevB.108.085416}
  {\bibfield  {journal} {\bibinfo  {journal} {Phys. Rev. B}\ }\textbf {\bibinfo
  {volume} {108}},\ \bibinfo {pages} {085416} (\bibinfo {year}
  {2023})}\BibitemShut {NoStop}%
\bibitem [{\citenamefont {Pan}\ and\ \citenamefont
  {Das~Sarma}(2026)}]{Pan2026}%
  \BibitemOpen
  \bibfield  {author} {\bibinfo {author} {\bibfnamefont {H.}~\bibnamefont
  {Pan}}\ and\ \bibinfo {author} {\bibfnamefont {S.}~\bibnamefont
  {Das~Sarma}},\ }\bibfield  {title} {\bibinfo {title} {Majorana zero modes in
  semiconductor-superconductor hybrid structures: Defining topology in short
  and disordered nanowires through {Majorana} splitting},\ }\href
  {https://doi.org/10.1103/2v41-yvs1} {\bibfield  {journal} {\bibinfo
  {journal} {Phys. Rev. B}\ }\textbf {\bibinfo {volume} {113}},\ \bibinfo
  {pages} {165420} (\bibinfo {year} {2026})}\BibitemShut {NoStop}%
\bibitem [{\citenamefont {Taylor}\ \emph {et~al.}(2026)\citenamefont {Taylor},
  \citenamefont {Pan},\ and\ \citenamefont {Das~Sarma}}]{Taylor2026}%
  \BibitemOpen
  \bibfield  {author} {\bibinfo {author} {\bibfnamefont {J.~R.}\ \bibnamefont
  {Taylor}}, \bibinfo {author} {\bibfnamefont {H.}~\bibnamefont {Pan}},\ and\
  \bibinfo {author} {\bibfnamefont {S.}~\bibnamefont {Das~Sarma}},\ }\bibfield
  {title} {\bibinfo {title} {Machine learning {Majorana} topology using
  unsupervised and supervised learning},\ }\href
  {https://link.aps.org/doi/10.1103/ywsw-29xc} {\bibfield  {journal} {\bibinfo
  {journal} {Phys. Rev. B}\ }\textbf {\bibinfo {volume} {114}},\ \bibinfo
  {pages} {165418} (\bibinfo {year} {2026})}\BibitemShut {NoStop}%
\bibitem [{\citenamefont {Sau}\ and\ \citenamefont
  {Das~Sarma}(2025)}]{Sau2025}%
  \BibitemOpen
  \bibfield  {author} {\bibinfo {author} {\bibfnamefont {J.~D.}\ \bibnamefont
  {Sau}}\ and\ \bibinfo {author} {\bibfnamefont {S.}~\bibnamefont
  {Das~Sarma}},\ }\bibfield  {title} {\bibinfo {title} {Capacitance-based
  fermion parity readout and predicted {Rabi} oscillations in a {Majorana}
  nanowire},\ }\href {https://doi.org/10.1103/p8qg-p1fp} {\bibfield  {journal}
  {\bibinfo  {journal} {Phys. Rev. B}\ }\textbf {\bibinfo {volume} {111}},\
  \bibinfo {pages} {224509} (\bibinfo {year} {2025})}\BibitemShut {NoStop}%
\bibitem [{\citenamefont {Stanescu}\ and\ \citenamefont
  {Tewari}(2026)}]{stanescu2025fermionparityquantumcapacitance}%
  \BibitemOpen
  \bibfield  {author} {\bibinfo {author} {\bibfnamefont {T.~D.}\ \bibnamefont
  {Stanescu}}\ and\ \bibinfo {author} {\bibfnamefont {S.}~\bibnamefont
  {Tewari}},\ }\bibfield  {title} {\bibinfo {title} {Fermion parity and quantum
  capacitance oscillation with partially separated majorana and quasi-majorana
  modes},\ }\href {https://doi.org/10.1103/q488-97dp} {\bibfield  {journal}
  {\bibinfo  {journal} {Phys. Rev. B}\ }\textbf {\bibinfo {volume} {113}},\
  \bibinfo {pages} {165422} (\bibinfo {year} {2026})}\BibitemShut {NoStop}%
\bibitem [{\citenamefont {Boutin}\ \emph {et~al.}(2025)\citenamefont {Boutin},
  \citenamefont {Karzig}, \citenamefont {Dandachi}, \citenamefont {Mishmash},
  \citenamefont {Gukelberger}, \citenamefont {Lutchyn},\ and\ \citenamefont
  {Bauer}}]{Boutin2025}%
  \BibitemOpen
  \bibfield  {author} {\bibinfo {author} {\bibfnamefont {S.}~\bibnamefont
  {Boutin}}, \bibinfo {author} {\bibfnamefont {T.}~\bibnamefont {Karzig}},
  \bibinfo {author} {\bibfnamefont {T.~E.}\ \bibnamefont {Dandachi}}, \bibinfo
  {author} {\bibfnamefont {R.~V.}\ \bibnamefont {Mishmash}}, \bibinfo {author}
  {\bibfnamefont {J.}~\bibnamefont {Gukelberger}}, \bibinfo {author}
  {\bibfnamefont {R.~M.}\ \bibnamefont {Lutchyn}},\ and\ \bibinfo {author}
  {\bibfnamefont {B.}~\bibnamefont {Bauer}},\ }\bibfield  {title} {\bibinfo
  {title} {Predictive simulations of the dynamical response of mesoscopic
  devices},\ }\href {https://arxiv.org/abs/2502.12960} {\bibfield  {journal}
  {\bibinfo  {journal} {arXiv preprint arXiv:2502.12960}\ } (\bibinfo {year}
  {2025})}\BibitemShut {NoStop}%
\bibitem [{Note1()}]{Note1}%
  \BibitemOpen
  \bibinfo {note} {Adapting the noise-amplitude estimate of Ref.~\cite
  {Boutin2025} to our convention for the noise spectral density yields $A_D\sim
  (1~\mu {\protect \rm eV})^2/2\pi \sim 1.6\times 10^{-13}~{\protect \rm
  eV}^2$.}\BibitemShut {Stop}%
\bibitem [{Note2()}]{Note2}%
  \BibitemOpen
  \bibinfo {note} {We estimate $A_z$ by combining a representative
  dimensionless charge sensitivity $D\equiv |\partial \varepsilon _M/\partial
  \mu |\sim 0.1$ of the Majorana quasiparticle energy $\varepsilon _M=2E_M$ to
  the local electrostatic potential $\mu $, inferred from the decoherence rates
  reported in Ref.~\cite {Sau2025}, with the electric-field noise model of
  Ref.~\cite {Knapp2018}. In our convention, the intrawire splitting $E_M$ is
  half the corresponding BdG excitation energy, giving $A_M\simeq \alpha
  _E(eDd_{\protect \rm eff})^2/4$ for a single wire, where $\alpha _E$ is the
  amplitude of the electric-field noise spectrum and $d_{\protect \rm eff}$ is
  an effective electrostatic length scale converting electric-field
  fluctuations into potential fluctuations. Assuming comparable independent
  fluctuations in the two wires then gives $A_z=A_M+A'_M\simeq \alpha
  _E(eDd_{\protect \rm eff})^2/2$. Using $\alpha _E\sim 10~({\protect \rm
  V/m})^2$, $D\sim 0.1$, and $d_{\protect \rm eff}\sim 10-20~{\protect \rm nm}$
  gives $A_z\sim 5\times 10^{-18}-2\times 10^{-17}~{\protect \rm eV}^2$,
  motivating the representative value $A_z=10^{-17}~{\protect \rm eV}^2$ used
  here. Since $d_{\protect \rm eff}$ and the microscopic charge sensitivity are
  device dependent, this should be regarded only as an order of magnitude
  estimate. For comparison, the decoherence rate reported for the ideal
  clean-wire case in the same work corresponds to an effective charge
  sensitivity $D\sim 10^{-4}$, illustrating that the electrostatic sensitivity
  can increase strongly away from the ideal clean-wire regime.}\BibitemShut
  {Stop}%
\bibitem [{\citenamefont {Bagrets}\ and\ \citenamefont
  {Altland}(2012)}]{bagrets2012class}%
  \BibitemOpen
  \bibfield  {author} {\bibinfo {author} {\bibfnamefont {D.}~\bibnamefont
  {Bagrets}}\ and\ \bibinfo {author} {\bibfnamefont {A.}~\bibnamefont
  {Altland}},\ }\bibfield  {title} {\bibinfo {title} {Class {$D$} spectral peak
  in {Majorana} quantum wires},\ }\href
  {https://link.aps.org/doi/10.1103/PhysRevLett.109.227005} {\bibfield
  {journal} {\bibinfo  {journal} {Phys. Rev. Lett.}\ }\textbf {\bibinfo
  {volume} {109}},\ \bibinfo {pages} {227005} (\bibinfo {year}
  {2012})}\BibitemShut {NoStop}%
\bibitem [{\citenamefont {Kells}\ \emph {et~al.}(2012)\citenamefont {Kells},
  \citenamefont {Meidan},\ and\ \citenamefont
  {Brouwer}}]{kells2012nearzeroenergy}%
  \BibitemOpen
  \bibfield  {author} {\bibinfo {author} {\bibfnamefont {G.}~\bibnamefont
  {Kells}}, \bibinfo {author} {\bibfnamefont {D.}~\bibnamefont {Meidan}},\ and\
  \bibinfo {author} {\bibfnamefont {P.~W.}\ \bibnamefont {Brouwer}},\
  }\bibfield  {title} {\bibinfo {title} {Near-zero-energy end states in
  topologically trivial spin-orbit coupled superconducting nanowires with a
  smooth confinement},\ }\href
  {https://link.aps.org/doi/10.1103/PhysRevB.86.100503} {\bibfield  {journal}
  {\bibinfo  {journal} {Phys. Rev. B}\ }\textbf {\bibinfo {volume} {86}},\
  \bibinfo {pages} {100503} (\bibinfo {year} {2012})}\BibitemShut {NoStop}%
\bibitem [{\citenamefont {Prada}\ \emph {et~al.}(2012)\citenamefont {Prada},
  \citenamefont {{San-Jose}},\ and\ \citenamefont
  {Aguado}}]{prada2012transport}%
  \BibitemOpen
  \bibfield  {author} {\bibinfo {author} {\bibfnamefont {E.}~\bibnamefont
  {Prada}}, \bibinfo {author} {\bibfnamefont {P.}~\bibnamefont {{San-Jose}}},\
  and\ \bibinfo {author} {\bibfnamefont {R.}~\bibnamefont {Aguado}},\
  }\bibfield  {title} {\bibinfo {title} {Transport spectroscopy of {$NS$}
  nanowire junctions with {{Majorana}} fermions},\ }\href
  {https://link.aps.org/doi/10.1103/PhysRevB.86.180503} {\bibfield  {journal}
  {\bibinfo  {journal} {Phys. Rev. B}\ }\textbf {\bibinfo {volume} {86}},\
  \bibinfo {pages} {180503} (\bibinfo {year} {2012})}\BibitemShut {NoStop}%
\bibitem [{\citenamefont {Reeg}\ \emph {et~al.}(2018)\citenamefont {Reeg},
  \citenamefont {Dmytruk}, \citenamefont {Chevallier}, \citenamefont {Loss},\
  and\ \citenamefont {Klinovaja}}]{Reeg2018}%
  \BibitemOpen
  \bibfield  {author} {\bibinfo {author} {\bibfnamefont {C.}~\bibnamefont
  {Reeg}}, \bibinfo {author} {\bibfnamefont {O.}~\bibnamefont {Dmytruk}},
  \bibinfo {author} {\bibfnamefont {D.}~\bibnamefont {Chevallier}}, \bibinfo
  {author} {\bibfnamefont {D.}~\bibnamefont {Loss}},\ and\ \bibinfo {author}
  {\bibfnamefont {J.}~\bibnamefont {Klinovaja}},\ }\bibfield  {title} {\bibinfo
  {title} {Zero-energy {Andreev} bound states from quantum dots in proximitized
  {Rashba} nanowires},\ }\href
  {https://link.aps.org/doi/10.1103/PhysRevB.98.245407} {\bibfield  {journal}
  {\bibinfo  {journal} {Phys. Rev. B}\ }\textbf {\bibinfo {volume} {98}},\
  \bibinfo {pages} {245407} (\bibinfo {year} {2018})}\BibitemShut {NoStop}%
\bibitem [{\citenamefont {Vuik}\ \emph {et~al.}(2019)\citenamefont {Vuik},
  \citenamefont {Nijholt}, \citenamefont {Akhmerov},\ and\ \citenamefont
  {Wimmer}}]{vuik2019reproducing}%
  \BibitemOpen
  \bibfield  {author} {\bibinfo {author} {\bibfnamefont {A.}~\bibnamefont
  {Vuik}}, \bibinfo {author} {\bibfnamefont {B.}~\bibnamefont {Nijholt}},
  \bibinfo {author} {\bibfnamefont {A.}~\bibnamefont {Akhmerov}},\ and\
  \bibinfo {author} {\bibfnamefont {M.}~\bibnamefont {Wimmer}},\ }\bibfield
  {title} {\bibinfo {title} {Reproducing topological properties with
  quasi-{{Majorana}} states},\ }\href
  {https://scipost.org/10.21468/SciPostPhys.7.5.061} {\bibfield  {journal}
  {\bibinfo  {journal} {SciPost Physics}\ }\textbf {\bibinfo {volume} {7}},\
  \bibinfo {pages} {061} (\bibinfo {year} {2019})}\BibitemShut {NoStop}%
\bibitem [{\citenamefont {Pan}\ and\ \citenamefont
  {Das~Sarma}(2020)}]{pan2020physical}%
  \BibitemOpen
  \bibfield  {author} {\bibinfo {author} {\bibfnamefont {H.}~\bibnamefont
  {Pan}}\ and\ \bibinfo {author} {\bibfnamefont {S.}~\bibnamefont
  {Das~Sarma}},\ }\bibfield  {title} {\bibinfo {title} {Physical mechanisms for
  zero-bias conductance peaks in {{Majorana}} nanowires},\ }\href
  {https://link.aps.org/doi/10.1103/PhysRevResearch.2.013377} {\bibfield
  {journal} {\bibinfo  {journal} {Phys. Rev. Research}\ }\textbf {\bibinfo
  {volume} {2}},\ \bibinfo {pages} {013377} (\bibinfo {year}
  {2020})}\BibitemShut {NoStop}%
\bibitem [{\citenamefont {Das~Sarma}\ and\ \citenamefont
  {Pan}(2021)}]{dassarma2021disorderinduced}%
  \BibitemOpen
  \bibfield  {author} {\bibinfo {author} {\bibfnamefont {S.}~\bibnamefont
  {Das~Sarma}}\ and\ \bibinfo {author} {\bibfnamefont {H.}~\bibnamefont
  {Pan}},\ }\bibfield  {title} {\bibinfo {title} {Disorder-induced zero-bias
  peaks in {{Majorana}} nanowires},\ }\href
  {https://link.aps.org/doi/10.1103/PhysRevB.103.195158} {\bibfield  {journal}
  {\bibinfo  {journal} {Phys. Rev. B}\ }\textbf {\bibinfo {volume} {103}},\
  \bibinfo {pages} {195158} (\bibinfo {year} {2021})}\BibitemShut {NoStop}%
\bibitem [{\citenamefont {Pan}\ \emph {et~al.}(2021)\citenamefont {Pan},
  \citenamefont {Liu}, \citenamefont {Wimmer},\ and\ \citenamefont
  {Das~Sarma}}]{pan2021quantized}%
  \BibitemOpen
  \bibfield  {author} {\bibinfo {author} {\bibfnamefont {H.}~\bibnamefont
  {Pan}}, \bibinfo {author} {\bibfnamefont {C.-X.}\ \bibnamefont {Liu}},
  \bibinfo {author} {\bibfnamefont {M.}~\bibnamefont {Wimmer}},\ and\ \bibinfo
  {author} {\bibfnamefont {S.}~\bibnamefont {Das~Sarma}},\ }\bibfield  {title}
  {\bibinfo {title} {Quantized and unquantized zero-bias tunneling conductance
  peaks in {{Majorana}} nanowires: {{Conductance}} below and above
  {$2{e}^{2}/h$}},\ }\href
  {https://link.aps.org/doi/10.1103/PhysRevB.103.214502} {\bibfield  {journal}
  {\bibinfo  {journal} {Phys. Rev. B}\ }\textbf {\bibinfo {volume} {103}},\
  \bibinfo {pages} {214502} (\bibinfo {year} {2021})}\BibitemShut {NoStop}%
\bibitem [{\citenamefont {Prada}\ \emph {et~al.}(2020)\citenamefont {Prada},
  \citenamefont {{San-Jose}}, \citenamefont {{de Moor}}, \citenamefont
  {Geresdi}, \citenamefont {Lee}, \citenamefont {Klinovaja}, \citenamefont
  {Loss}, \citenamefont {Nyg{\aa}rd}, \citenamefont {Aguado},\ and\
  \citenamefont {Kouwenhoven}}]{prada2020andreev}%
  \BibitemOpen
  \bibfield  {author} {\bibinfo {author} {\bibfnamefont {E.}~\bibnamefont
  {Prada}}, \bibinfo {author} {\bibfnamefont {P.}~\bibnamefont {{San-Jose}}},
  \bibinfo {author} {\bibfnamefont {M.~W.~A.}\ \bibnamefont {{de Moor}}},
  \bibinfo {author} {\bibfnamefont {A.}~\bibnamefont {Geresdi}}, \bibinfo
  {author} {\bibfnamefont {E.~J.~H.}\ \bibnamefont {Lee}}, \bibinfo {author}
  {\bibfnamefont {J.}~\bibnamefont {Klinovaja}}, \bibinfo {author}
  {\bibfnamefont {D.}~\bibnamefont {Loss}}, \bibinfo {author} {\bibfnamefont
  {J.}~\bibnamefont {Nyg{\aa}rd}}, \bibinfo {author} {\bibfnamefont
  {R.}~\bibnamefont {Aguado}},\ and\ \bibinfo {author} {\bibfnamefont {L.~P.}\
  \bibnamefont {Kouwenhoven}},\ }\bibfield  {title} {\bibinfo {title} {From
  {{Andreev}} to {{Majorana}} bound states in hybrid
  superconductor--semiconductor nanowires},\ }\href
  {http://www.nature.com/articles/s42254-020-0228-y} {\bibfield  {journal}
  {\bibinfo  {journal} {Nature Reviews Physics}\ }\textbf {\bibinfo {volume}
  {2}},\ \bibinfo {pages} {575} (\bibinfo {year} {2020})}\BibitemShut {NoStop}%
\end{thebibliography}%

\end{document}